\documentclass{elsarticle}
\usepackage{fullpage}

\usepackage{amsfonts,amssymb,amsbsy,amsmath,mathtools,multirow,cases}
\usepackage{graphicx}
\usepackage{pgfplots}
\usepackage{amsthm}
\usepackage{stmaryrd}
\theoremstyle{remark}

\DeclareMathAlphabet\mathbfcal{OMS}{cmsy}{b}{n}

\usepgfplotslibrary{external}  
\usepackage{calligra,calrsfs,mathrsfs}
\usepackage{ulem,cancel}
\usepackage{subcaption}
\usepackage{tkz-euclide}
\usetikzlibrary[patterns]
\usepackage{hyperref}
\usepackage[ruled,vlined]{algorithm2e}
\usepackage{algpseudocode}
\usepackage{dsfont}
\usepackage{caption}
\usepackage{epstopdf}
\usepackage{epsfig}
\usepackage{natbib}
\usepackage{grffile}
\usepackage{amsmath,tabularx}
\usepackage{enumerate}

\usepackage[version=4]{mhchem}

\usepackage{longtable,tabularx}
\SetKwInput{KwInput}{Input}                
\SetKwInput{KwOutput}{Output}              

\usepackage{diagbox}
\usepackage{booktabs}
\usepackage{float}

\DeclareMathOperator*{\argmin}{arg\,min}

\newcommand\ub{\mbox{$\bar{u}$}}

\newcommand{\ifcomment}{\iffalse}

\newdefinition{rem}{Remark}

\allowdisplaybreaks

\begin{document}
\begin{frontmatter}
\title{
Enhancing Dynamical System Modeling through Interpretable  Machine Learning Augmentations: A Case Study in Cathodic Electrophoretic Deposition}

\cortext[cor1]{~Corresponding author}
\address[a]{Department of Mechanical Engineering, University of Michigan, Ann Arbor, MI, 48109, USA}
\address[b]{Department of Aerospace Engineering, University of Michigan, Ann Arbor, MI, 48109, USA}
\address[d]{Ford Research \& Innovation Center, Dearborn, MI, 48121, USA}
\address[c]{The Dow Chemical Company, Core R\&D, Engineering and Process Science, Lake Jackson, TX 77566, USA}
\author[b]{Christian~Jacobsen\corref{cor1}}
\ead{csjacobs@umich.edu}
\author[a]{Jiayuan~Dong}
\ead{jiayuand@umich.edu}
\author[a,c]{Mehdi~Khalloufi}
\ead{khalloufi.mehdi@gmail.com}
\author[a]{Xun~Huan}
\ead{xhuan@umich.edu}
\author[b]{Karthik~Duraisamy}
\ead{kdur@umich.edu}
\author[d]{Maryam~Akram}
\ead{makram13@ford.com}
\author[d]{Wanjiao~Liu}
\ead{lwanjiao@ford.com}

\begin{abstract}
We introduce a comprehensive data-driven framework aimed at enhancing the modeling of physical systems, employing inference techniques and machine learning enhancements. As a demonstrative application, we pursue the modeling of cathodic electrophoretic deposition (EPD), commonly known as e-coating. Our approach illustrates a systematic procedure for enhancing physical models by identifying their limitations through inference on experimental data and introducing adaptable model enhancements to address these shortcomings.
We begin by tackling the issue of model parameter identifiability, which reveals aspects of the model 
that require improvement. To address generalizability , we introduce modifications which also enhance identifiability. However, these modifications do not fully capture essential experimental behaviors.
To overcome this limitation, we incorporate interpretable yet flexible augmentations into the baseline model. These augmentations are parameterized by simple fully-connected neural networks (FNNs), and we leverage machine learning tools, particularly Neural Ordinary Differential Equations (Neural ODEs), to learn these augmentations. Our simulations demonstrate that the machine learning-augmented model more accurately captures observed behaviors and improves predictive accuracy.
Nevertheless, we contend that while the model updates offer superior performance and capture the relevant physics, we can reduce off-line computational costs by eliminating certain dynamics without compromising accuracy or interpretability in downstream predictions of quantities of interest, particularly film thickness predictions.
The entire process outlined here provides a structured approach to leverage data-driven methods. Firstly, it helps us comprehend the root causes of model inaccuracies, and secondly, it offers a principled method for enhancing model performance.
\end{abstract}
\end{frontmatter}
%
%

\section{Introduction}

Cathodic electrophoretic deposition (EPD), commonly known as e-coating, is a pivotal technique used in industries such as automotive and manufacturing to apply protective coatings to various surfaces, preventing corrosion and ensuring durability. Achieving optimal coating properties and process efficiency requires accurate modeling of EPD dynamics, which presents significant challenges due to the complex electrochemical interactions involved.

The primary challenges in modeling the e-coat process is the uncertainty surrounding the physical properties of the coating film during the deposition and a limited understanding of the underlying physics. Various models proposed in the literature have attempted to address these issues~\cite{pierce1981physical,RASTEGAR200817,BOYD199625,mivskovic2002mechanism,31b27711-89b9-3bd0-9dc3-5864cb482b97}, but they often require empirical calibration based on measurements for film initiation and growth or fail to capture experimental behaviors. Additionally, the deposition onset of the coating film, characterized by threshold parameters, remains an open question, leading to discontinuities in model outputs and rendering parameter inference difficult. An extensive study on the modeling of the e-coat process and chemistry involved for constant current and constant voltage can be found in~\cite{pierce1981physical,BECK19761}.

E-coating in the automotive industry typically involves employing Sand's equation~\cite{SandEquation} to compute the induction time under constant current conditions. However, the application of a linearly increasing voltage in time is also common in automotive practices to ensure good throw-power and prevent defects associated with high voltage~\cite{Marlar2020}. To address non-constant current and voltage scenarios, an equivalent to Sand's equation is derived using the Laplace transform, assuming a constant diffusion coefficient of the bath solution.

To enhance the modeling of EPD dynamics, specifically in the context of e-coating, this study aims to modify a baseline model to improve model performance with respect to experimentally obtained data. There exist two major types of uncertainty in the baseline model: parametric and model form uncertainties. Parametric uncertainties refer to the uncertainty of lack of precise knowledge about the values of certain parameters in a computational model while model form uncertainty refers to the uncertainty associated with the choice of computational model itself. Parametric uncertainty is often addressed by leveraging experimental data to perform parameter estimation (maximum likelihood estimation, Bayesian inference, etc.), and model form is typically addressed through physical insight and intuition along with data-driven modeling techniques, the latter of which is highlighted in this work.

Variational inference methods are employed to estimate the parameters of this model in an effort to address the parametric uncertainty associated with the model. However, inherent unidentifiability issues in the model hinder accurate parameter estimation. To overcome these challenges, modifications are introduced to the baseline model to ensure that all parameters are identifiable and more generalizable across experimental conditions, attempting to address model form uncertainty. Nonetheless, the model's inability to capture observed behaviors in experimental data prompts the use of machine learning (ML) tools, particularly NeuralODE~\cite{NeuralODE}, to introduce and learn physically relevant and interpretable modifications.

By leveraging NeuralODE and carefully crafting learnable augmentations using physical insights, the model's capability to match experimental data is improved while preserving physical interpretability.  We use data to enhance an existing full-order dynamical system, effectively creating a surrogate model in the form of model augmentations rather than entirely discovering the physical dynamics~\cite{doi:10.1073/pnas.1517384113, doi:10.1126/sciadv.1602614}. Our aim is that of improving predictive accuracy while maintaining an interpretable model. As the baseline model is computationally efficient, reduced order modelling (ROM) techniques~\cite{10.1063/5.0061577, 10.1063/1.5113494, San2018, Hasegawa2020} are not addressed in this work. To this end, simulations and comparisons with experimental results demonstrate the effectiveness of the improved model in capturing observed behaviors and enhancing the predictive accuracy of the baseline e-coat dynamics model, notably for film thickness prediction, while augmentation forms remain physically interpretable and flexible.
Accurately predicting the polymer film thickness over time is of utmost interest for time and cost savings in industrial applications, and this study argues that not all behaviors present in experimental data are necessary for accurate film thickness predictions. Removing unnecessary complex physical behavior from the model leads to a more efficient formulation of the e-coat process, striking a balance between interpretability, predictive accuracy, and efficiency.

A contribution of this work is to illustrate a principled process of clearly understanding and evaluating the root cause of model shortcomings followed by improving model performance using these insights along with experimental data. Together, our methods can be useful for addressing both parametric and model form uncertainties in computational models where experimental data is available. 

The prominent  aspects of our approach include the use of variational inference techniques to estimate model parameters from data, an in depth identifiability analysis of parameters in the baseline e-coat model, identifiability and generalizability improvements to the baseline model based on experiment type, and the introduction of interpretable and learnable ML augmentations to improve model accuracy.

We first illustrate issues with the baseline e-coat dynamics model during inference and prediction which inform the introduction of physically relevant augmentations to improve predictive accuracy. In particular, through variational Bayesian inference, we attempt to infer the parameters of the baseline model and show that identifiability issues hinder the process. Slightly altering the model results in an identifiable model which is more generalizable, but important physically relevant behaviour is found to be entirely absent from the model. Finally, we introduce interpretable ML augmentations to improve the accuracy of the baseline model on experimental data. The final model is assessed for both constant current and voltage ramp cases concerning experimental data while trained only on voltage ramp cases.

The organization of this manuscript is as follows. Sec.~\ref{sec:BaselineModel} introduces the baseline model for electro-deposition. Sec.~\ref{sec:inference} presents the Bayesian inference methods for estimating the parameters of the baseline model with quantified uncertainty, 
and illustrates
the issue of parameter unidentifiability. Sec.~\ref{sec:model_updates} introduces a new model based on a variation of Sand's equation, valid for any constant current/voltage ramp experiment, but we show that this model fails to capture all relevant physical behavior observed in experimental data. We then introduce an interpretable ML-based augmentations to the model later in Sec.~\ref{sec:model_updates} to better capture experimental data dynamics and show that removing some of the most complex physical behavior can greatly improve model efficiency without negatively impacting film thickness growth predictions much. Finally, the study concludes in Sec. \ref{sec:conclusion}, highlighting the contributions to e-coating modeling and offering insights into potential applications in diverse fields.

\section{Mathematical Formulation}\label{sec:BaselineModel}

In this section, we present the mathematical formulation of the baseline model~\cite{31b27711-89b9-3bd0-9dc3-5864cb482b97}, a computational framework for solving the baseline model, and the corresponding measurement data acquired from experiments.

\subsection{Baseline Model}

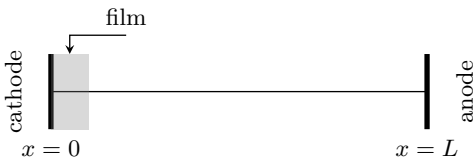
\begin{figure}[htb]
\centering
\tikzsetnextfilename{1Dsetup}
\begin{tikzpicture}[scale=5]
\tikzstyle{every node}=[font=\small]
\draw[line width=0.5pt] (0,0.0) -- (1,0.0); 
\draw[line width=2pt] (0.0,-0.1) -- (0.0,0.1); 
\draw[line width=2pt] (1.0,-0.1) -- (1.0,0.1); 

\node[draw=none,rotate=90] at (-0.1,0.0) {$\textrm{cathode}$};
\node[draw=none,rotate=90] at (1.1,0.0) {$\textrm{anode}$};
\node[draw=none] at (0,-0.15) {$x=0$};
\node[draw=none] at (1,-0.15) {$x=L$};
\fill[gray,opacity=0.3] (0.0,-0.1) rectangle (0.1,0.1);
\node[draw=none] at (0.2,0.2) {film};
\draw[-stealth,line width=0.5pt] (0.20,0.15) -- (0.05,0.15) -- (0.05,0.1); 
\end{tikzpicture}
\caption{Initial setup for the 1D case.\label{fig:1DSetup}}
\end{figure}

The baseline model is one-dimensional in space, consisting of a cathode and anode placed length $L$ apart and filled with a solution of suspended colloidal particles in between (see Fig.~\ref{fig:1DSetup}). 
%
%
After a voltage is applied and prior to film deposition, an electrochemical reaction takes place at the cathode with \ce{2H+ + 2e- -> H2} or \ce{2H2O + 2e- -> 2OH- + H2}~\cite{31b27711-89b9-3bd0-9dc3-5864cb482b97}. As the reaction proceeds, the bath solution basicity increases until a critical pH value is reached and film deposition starts. Film deposition is defined by suspended colloidal particles being deposited on the anode, increasing the circuit resistance.

The film deposition process is separated into three components. First, the electric field within the bath is computed using the conservation of current density given by
\begin{align}
\nabla \cdot \textbf{j}&=0 \label{eq:FluxConservation} \\ 
\textbf{j}&=-\sigma_{\rm{bath}}\nabla \phi \label{eq:CurrentDensity}\\ 
\left.\phi\right|_{\Gamma}&=R_{\textrm{film}} \,j_n \quad \textrm{ at the interface film-bath},\label{eq:RobinBC}
\end{align}
where \textbf{j} is the current density, $\sigma_{\rm{bath}}$ is the bath conductivity, $j_n=\textbf{j}\cdot \bf{n}$ is the normal component of the current density, $\phi$ is the electrical potential, $R_{\textrm{film}}$ is the film resistance and $\Gamma$ represents the interface between the film and the bath.
Second, once film deposition begins, deposition rate is defined as
\begin{align}
\frac{dh}{dt}=C_v j_n \label{FilmGrowth}, 
\end{align}
where $h$ is the film thickness and $C_v$ is the Coulombic efficiency. Third and finally, the film thickness and film resistance are related through
\begin{align}
\frac{dR_{\textrm{film}}}{dt}= \rho(j_n) \frac{dh}{dt}\label{ResistanceEquation},
\end{align}
where $ \rho(\textbf{j})$ is the film resistivity.
In particular, $\rho(\textbf{j})$ is a decreasing function of the current density, and we adopt an empirical estimate from~\cite{WanjiaoLiu} of 
\begin{equation}\label{eq:rho}
    \rho = \max(8\times 10^5 \exp( -0.1j), \; 2\times 10^6).
\end{equation}
Before deposition begins, the right hand side of Eqs.~\ref{ResistanceEquation} and~\ref{FilmGrowth} are both equal to 0. The model defines the deposition onset event according to two criteria: the minimum current density and minimum charge conditions. The onset criteria are critical for accurately predicting the film thickness growth in time. If both of the conditions are met, film deposition begins in the baseline model.



The first condition which determines deposition onset is a minimum current density condition. Once the current density at the cathode reaches a threshold value $j_{min}$, the film thickness begins to increase.
The onset condition parameter $j_{min}$ is unknown and estimated or inferred from experimental data.
%
The second onset condition is a minimum charge condition.
The minimum charge criterion assumes that the deposition does not start until the accumulated charge on the cathode reaches a threshold value $Q_{min}$,
with the electric charge $Q$  defined by
\begin{equation}
Q(t)=\int_t j_n dt.
\end{equation}
The deposition onset threshold $Q_{min}$ is also unknown and estimated or inferred from experimental data.
Once both the minimum charge and minimum current conditions are met, film thickness increase as
\begin{align}
\frac{dh}{dt} = C_v\,j_n \textrm{ for } Q> Q_{min} \; \textrm{and} \; j > j_{min} .
\end{align}

Additional model parameters have already been estimated from previous experimental data. We thus fix those parameter to values presented in Table~\ref{table:paramSim}, and focus only on estimating the key unknown parameters $C_v$, $j_{min}$, and $Q_{min}$ in the baseline model. 

\setlength{\tabcolsep}{0.35em}
\begin{table}[h]
\begin{center}
\caption{Summary of baseline model parameters.}
\begin{tabular}{lcclccccccc}
\hline
Name & Symbol & Value & Unit\\
\hline
Bath conductivity & $\sigma_{\rm{bath}}$ & 0.14 &$S/m$\\
Initial resistance at cathode& $R_0$ & 0.5 & $\Omega \cdot m^2$\\
\hline
\end{tabular}
\label{table:paramSim}
\end{center}
\end{table}

\subsection{Computational Formulation}


We consider two main types of experiments in our computational setup: voltage ramp (VR) and constant current (CC). In VR, the voltage is increased linearly with time at a rate of $V_R$ such that $\phi(t, x=L) = V(t, x=L) = V_Rt$. Electric potential is denoted $\phi$ or $V$ interchangeably in this work. In CC, the current density at the anode is held constant. Experimentally, the current can only be held constant until some maximum voltage $V_{max}$ determined by the available equipment. Numerically, once the maximum voltage is reached in CC, $V_{max}$ is enforced instead of constant current. 

Eqs.~\ref{eq:FluxConservation} through~\ref{eq:RobinBC} are the Poisson equation with Robin boundary condition on the film / bath interface. This model is imposed for both experiments, but the boundary condition at the anode is different for each. Both experiment types are modeled by
\begin{align}
\sigma_{\rm{bath}}\frac{\partial^2\phi}{\partial x^2}&=0 \qquad\textrm{ in the bath}\\
\phi-R_{\textrm{film}} \sigma_{\rm{bath}}\frac{\partial \phi}{\partial x}&=0 \qquad\textrm{ at the interface film-bath},
\end{align}
additionally with VR having an anode boundary condition
\begin{equation}\label{eq:bc_vr}
    \phi_{\rm{anode}}(t) = \phi_{t=0} + \phi_{\rm{ramp}}(t),
\end{equation}
and CC having an anode boundary condition
\begin{equation}\label{eq:bc_cc}
    \sigma_{\rm{bath}}\frac{\partial \phi}{\partial x}= j_0.
\end{equation}
These equations offer an analytic relationship between film resistance and current in 1D for VR:
\begin{equation}\label{eq:Rj_VR}
    j(t) = \frac{\sigma V(t, x=L)}{\sigma R_{film}(t) + L}.
\end{equation}
As $V(t, x=L)$ is set as the boundary condition, we can solve only the resistance dynamics in Eq.~\ref{ResistanceEquation} and compute $j(t)$ from $V$ and $R_{film}$. For CC, the current is set by the experimental conditions as $j_0$, and the voltage at the anode can be computed by 
\begin{equation}\label{eq:Rj_CC}
    \frac{\partial V}{\partial t} = \frac{\partial R_{film}}{\partial t}j_0 \; \textrm{for} \; V < V_{max},
\end{equation}
where $V_{max}$ is the maximum experimental voltage. If $V \geq V_{max}$ in CC, the relationship between voltage, current, and resistance is given by Eq.~\ref{eq:Rj_VR}. 

To simulate the baseline dynamics, any black box ODE solver can be implemented to solve Eq.~\ref{ResistanceEquation} forward in time, computing the time evolution of film resistance. Current, voltage, and charge are given as boundary conditions, computed analytically from Eq.~\ref{eq:Rj_VR}, or solved along with resistance dynamics using Eq.~\ref{eq:Rj_CC}.

\subsection{Experimental Setup and Data}\label{sec:experimental_setup}

\begin{figure}[htb]
    \centering
    \captionsetup{width=.75\linewidth}
    \includegraphics[width=.9\textwidth,angle=0,clip,trim=0pt 0pt 0pt 0pt]{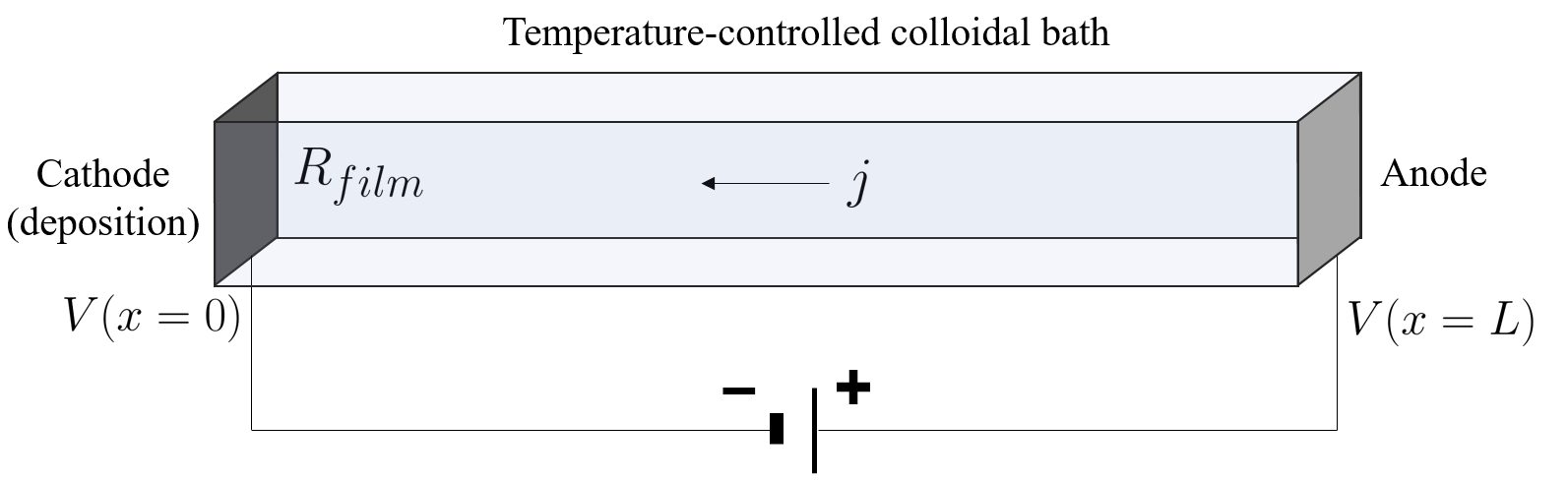}
    \caption{Experimental setup}
    \label{fig:expSetup}
\end{figure}

Experimental data is acquired from a laboratory setup that approximates the computational setting. In the experiment, a 16.0 cm$^2$ square anode and cathode are placed at the ends of a long e-coat bath and connected to a power source. A voltage is applied across the anode and the cathode according to either the VR or CC setup. During the course of each experiment, the voltage, current, and film resistance are all measured at a frequency of 10 Hz. For some experiments, thickness measurements are obtained by setting the voltage to zero at some time and measuring the thickness of the film at that time. Note that measuring the thickness terminates the experiment. A depiction of the experimental setup is given in Fig.~\ref{fig:expSetup}

\begin{figure}[htb]
    \centering
    \captionsetup{width=.7\linewidth}
    \includegraphics[width=.95\textwidth,angle=0,clip,trim=0pt 0pt 0pt 0pt]{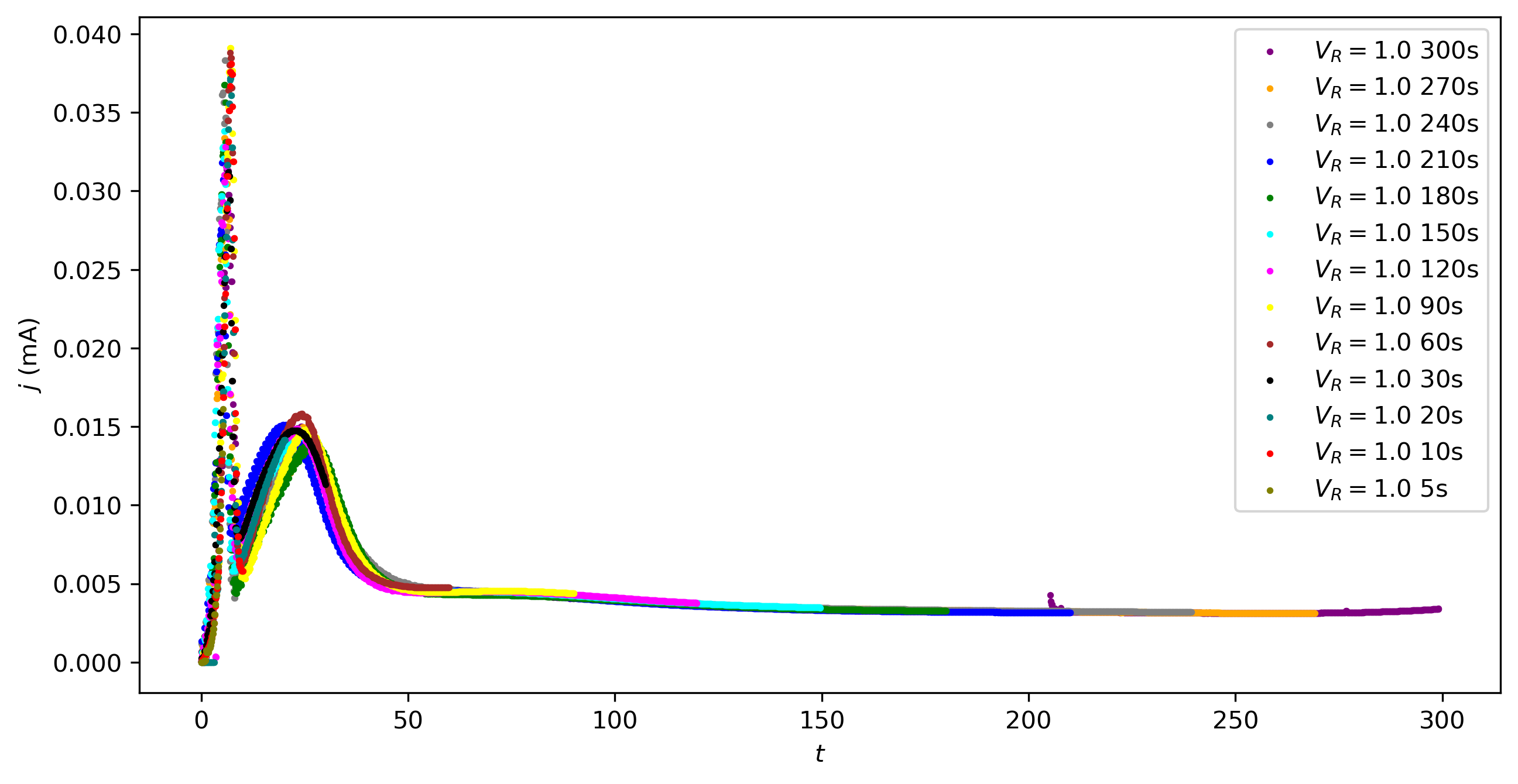}
    \caption{Visualization of the $\{j\}_1$ experimental data (all 13 trials) for configuration VR = 1.0. Each trial ends at a different time, and data is sampled at a rate of 10 Hz.}
    \label{fig:expData}
\end{figure}

Experiments are performed for both VR and CC at different experimental conditions, for a total of six configurations (see Table~\ref{table:expData} for an overview). Multiple trials are repeated for each configuration, where each trial is performed to a different final time in obtaining its thickness measurement. The experimental data is collected in $\mathcal{D} = \{\{j\}_i,\{R\}_i\}_{i=1}^6$ where $\{j\}_i, \{R\}_i$ respectively represent the sets of current and resistance measurements for each of the six  configurations. We do not use voltage data 
as they can be computed analytically from resistance and current. 
The data used during inference and learning is truncated for each experiment type $i$ to time $t_i$ such that data has been gathered for at least 3 trials at all times $t \leq t_i$. This is done to provide more accurate estimates of the variance during likelihood computation. An example of current measurements $\{j\}_1$ from the 13 trials under configuration VR, $V_R = 1V$ is shown in Fig~\ref{fig:expData}.

\begin{table}[htb]
\centering
\captionsetup{width=.9\linewidth}
 \begin{tabular}{|c c c c c|} 
 \hline
$i$ & Experiment & $n_i$ (\# trials) & $t_i$ (s) & Total Data Points ($j, R, V$) \\[.5ex] 
 \hline
 \hline
1 & VR, $V_R=1.0$ V/s & 13 & 239 & 31,870\\[.5ex]
\hline
2 & VR, $V_R=0.5$ V/s & 13 & 477 & 63,620\\[.5ex]
\hline
3 & VR, $V_R=0.125$ V/s & 12 & 639 & 85,178\\[.5ex]
\hline
4 & CC, $j_0=10.0$ mA & 10 & 80 & 10,420\\[.5ex]
\hline
5 & CC, $j_0=7.5$ mA & 10 & 160 & 20,820\\[.5ex]
\hline
6 & CC, $j_0=5.0$ mA & 10 & 240 & 31,216\\[.5ex]
\hline
\end{tabular}
\caption{Descriptions of the experimental data for each of the six configurations.}
\label{table:expData}
\end{table}


\section{Parameter Inference Using Experimental Data}\label{sec:inference}

Under the standard Bayesian inference framework, the initial knowledge of parameters of interest $\Theta : \omega \rightarrow \mathbb{R}^m$ is described by the prior $p(\theta)$. 
The observation of data $Y: \Omega \rightarrow \mathbb{R}^n$ given by realizations $y$ are assumed to be related to the parameters through a likelihood $p(y|\theta)$. The posterior $p(\theta |y)$ gives the updated knowledge about parameters $\theta$ after observing data $y$, and is given by Bayes rule:
\begin{equation}
p(\theta |y)= \frac{p(\theta) p(y|\theta)}{p(y)} \;,
\end{equation}
where $p(y)$ is known as the evidence and is typically intractable to compute. However, the evidence is constant for some data distribution as does not depend on the parameters $\theta$. In our applications, we are concerned with optimization techniques which require only the log-likelihood and log prior. Thus the evidence is neglected as a constant and not computed.

In the baseline e-coat model, our parameters of interest are $\theta = \{j_{min}, C_v, Q_{min}\}$. Our goal is to infer these parameters given the experimental data described in Sec.~\ref{sec:experimental_setup}. 
The truncated normal distributions are used for the prior to ensure the positive support while still enables the specification of the mode and spread regarding the parameters.
The data random variable $Y$ consists of both current $j$ and film resistance $R_{film}$. We assume a measurement model of 
\begin{equation}\label{eq:measurement_model}
 y(\theta, t; \eta) = G(\theta, t; \eta) + \epsilon(t, \eta)
\end{equation}
where $G(\theta, t; \eta) = \{R_{film}(\theta, t, \eta), j(\theta, t, \eta)\}$, computed by simulating the baseline model, and is a deterministic function of $\theta$, $t$, and the experimental configuration parameters $\eta$. The measurement noise $\epsilon(t, \eta) \sim \mathcal{N}(0, \Sigma(t, \eta))$ is a function of time and experimental configuration parameters, and we estimate the covariance matrix $\Sigma(t, \eta)$ from experimental data.

Considering our experimental data as $\mathcal{D}$, the posterior we seek to approximate is given by
\begin{equation}
    p(\theta|\mathcal{D}) = \frac{p(\mathcal{D}|\theta)p(\theta)}{p(\mathcal{D})} \; .
\end{equation}
The integration involved in directly computing $p(\mathcal{D}) = \int_{\theta}p(\mathcal{D}, \theta)d\theta$ (the ``brute force" approach) is expensive to evaluate, particularly in high dimensions. Therefore, numerical techniques have been widely used to approximate the posterior. Additionally, computing the posterior directly does not on its own necessarily allow samples to be easily drawn from the posterior. We discuss various inference techniques investigated to perform more efficient inference on the parameters of interest and easily sample from the inferred distributions. The brute force method of directly computing the posterior is referred to as the gridding approach hereafter due to computing the posterior on a discretized grid in the parameter space. Other methods investigated are all forms of variational inference in which the inference problem is transformed to an optimization problem
\cite{duraisamy2021VAE}
\cite{tran2021practical}
\cite{guo2017boosting}
\cite{pmlr-v37-rezende15}. 




\subsection{Likelihood}\label{sec:likelihood}
Each of the inference methods described require the computation of the (log) likelihood. 
Given the assumed measurement model in Eq.~\ref{eq:measurement_model}, the likelihood $p(\mathcal{D}|\theta)$ is a Gaussian distribution with mean $G(\theta, t; \eta)$ and covariance matrix $\Sigma(t, \eta)$. We assume that all experiments, trials, and samples within each trial are independent. While the later assumption may not be strictly true, it significantly decreases the computational cost of computing the log likelihood due to a diagonal covariance matrix. The log-likelihood is therefore given by
\begin{equation}\label{eq:loglike}
    \log p(\mathcal{D}|\theta) = -\frac{1}{2}\sum_{i=1}^6 \sum_{l=1}^{n_i} \sum_{r=1}^{10t_{l,i}} (G(\theta, 0.1r; \eta_i) - \mathcal{D}_{i, l, r})^T\Sigma(0.1r, \eta_i)(G(\theta, 0.1r; \eta_i) - \mathcal{D}_{i, l, r}) + Z \; , 
\end{equation}
where the time is given by $0.1r$ due to the constant sampling rate of 10Hz, and $\mathcal{D}_{i, l, r}$ denotes the 2D vector $[\{j\}_i^{(l)}(0.1r), \{R\}_i^{(l)}(0.1r)]^T$, the resistance and current for experiment $i$ in trial $l$ at time $0.1r$. Additionally, the matrix $\Sigma(0.1r, \eta_i)$ is diagonal and is computed for experiment $i$ at time $0.1r$ by computing the variance of current and resistance data over all trials which contain data at time $0.1r$. The time $t_{l,i}$ denotes the final experiment time for trial $l$ of experimental configuration $i$. Finally, $Z$ is the negative log of the likelihood normalization constant which does not depend on $\theta$. 


\subsection{Gridding approach}\label{sec:gridding}

Computing the Bayesian posterior is typically prohibitively expensive in practice. For our application, simulating the baseline model is the dominant bottleneck in terms of time complexity. Evaluating the forward model as little as possible will provide the greatest benefit for the efficiency of the inference process. More efficient methods of approximating the posterior are thus investigated to perform parameter inference. Each of these results is compared to the Bayesian posterior using a gridding approach, which is considered the `true' posterior. 

We employ a simple gridding approach to compute the Bayesian posterior in which the parameter space is discretized into a $d$-dimensional grid of uniform spacing. This grid is then sequentially refined with higher resolution near the maximum a posteriori (MAP) and any modes of the Bayesian posterior. 

It is often computationally more stable to compute the log-distributions first, rather than the distribution directly. Considering a single grid point $\theta_i$, we first compute the quantity
\begin{equation}\label{eq:grid}
    \log p(\theta_i|y) + \log p(y) = \log p(\theta_i) + \log p(y|\theta_i)
\end{equation}
at all grid points in the parameter space. Note that we ignore the \textit{evidence} term $p(y)$ in the gridding approach until Eq. \ref{eq:grid} is computed at each point. It is assumed that the grid bounds in parameter space are sufficiently large to capture the most significant aspects of the distribution. 

The final distribution is computed at each point in the grid by normalizing the quantity computed in Eq. \ref{eq:grid} using
\[
p(\theta_i|y) = \exp ( \log p(\theta_i) + \log p(y|\theta_i)) / Z \; ,
\]
where the normalization constant is given by
$
Z = \int_\Theta \exp( \log p(\theta) + \log p(y|\theta) ) d\theta$
and approximated using some numerical integration scheme based on the selected grid. 

We note that the gridding approach can provide a reasonable approximation to the posterior, but it does not provide a straightforward method of sampling from this posterior. 




\subsection{Variational Inference}


%
Variational inference methods~\cite{pmlr-v37-rezende15,Blei_2017,guo2017boosting,JMLR:v14:hoffman13a} are a powerful and versatile class of techniques used in probabilistic and Bayesian modeling. They have an advantage of being very efficient, in particular for high dimensional problems, compared to Markov chain monte carlo (MCMC) based methods~\cite{1953JChPh..21.1087M,10.5555/1051451,Robert_2011}. By formulating the posterior inference as an optimization problem, variational inference methods aim to approximate the Bayesian posterior with a tractable parametric distribution. Only the parameters of the parametric distribution are learned, rather than the typically intractable Bayesian posterior. An optimization problem is formed such that the parameters of the distribution are optimized by minimizing the Kullback-Leibler (KL) divergence~\cite{10.5555/1146355} between the variational approximation and the true posterior. This optimization problem is formulated as minimizing the evidence lower bound (ELBO).

Variational inference assumes a parametric distribution $q_{\phi}(\theta|y)$ which is used to approximate the true posterior distribution $p(\theta |y)$. The goal of the variational inference is to minimize the discrepancy between the true and approximate distributions by optimizing the distribution parameters. To achieve this, a measure of discrepancy or distance between two distributions, called the Kullback-Leibler divergence (KL divergence), is minimized between the two distributions. The KL divergence is given by
\begin{equation}
\rm{KL}\left[q_{\phi}(\theta|y) \| p(\theta |y)\right]
=\mathbb{E}_{\theta\sim q_{\phi}(\theta|y)}\left[\log\left(\dfrac{q_{\phi}(\theta|y)}{p(\theta |y)}\right)\right] \; .
\end{equation}


Variational inference seeks to minimize this by solving the optimization problem 
\begin{equation} \label{eq:variational_inference}
\phi^* = \argmin\limits_{\phi}\,\rm{KL}\left[q_{\phi}(\theta|y) \| p(\theta |y)\right] \; .
\end{equation}

Using Bayes' rule and defining the 
evidence lower bound (ELBO) $\mathcal{L}(q)$ as
\begin{equation}\label{eq:elbo}
\mathcal{L}(\theta,\phi)= \mathbb{E}_{Q}\left[\log\left(\dfrac{p(\theta ,y)}{q_{\phi}(\theta|y)}\right)\right]
= \mathbb{E}_{Q}\left[\log p(\theta ,y)\right] -  \mathbb{E}_{Q}\left[\log q_{\phi}(\theta|y)\right] \; ,
\end{equation}

the KL divergence can be written as a function of the ELBO by
\begin{equation}
\rm{KL}\left[q_{\phi}(\theta|y) \| p(\theta |y)\right]= -\mathcal{L}(\theta,	\phi) + \log\left(p(y)\right) \; .
\end{equation}
The optimization problem of Eq.~\ref{eq:variational_inference} is therefore equivalent to minimizing the negative ELBO:
\begin{equation*}
\phi^* = \argmin\limits_{\phi}\,-\mathcal{L}(\theta,\phi)
\end{equation*}

\subsubsection{Gaussian variational inference}

Fixed-form variational inference uses simple parametric distributions as the variational posterior. In our experiments, we choose an independent Gaussian distribution $q_\phi(\theta|y) = \mathcal{N}(\mu_\phi, \Sigma_\phi)$ for the fixed-form such that the parameters $\phi = [\mu_\phi, \Sigma_\phi]$ to be optimized consist of the mean $\mu_\phi \in \mathbb{R}^m$ and covariance $\Sigma_\phi \in \mathbb{R}^{m\times m}$ matrix. Note that the assumption of independence limits the covariance matrix to $m$ trainable parameters and $2m$ total trainable parameters. 

Gradient-free optimization techniques such as the Nelder-Mead algorithm~\cite{10.1093/comjnl/7.4.308} scale poorly in high dimensional problems and are often far less efficient than gradient-based optimization algorithms. This technique therefore requires the use of gradients to be effective, and gradients can be computed for the baseline model following the discussion in Sec.~\ref{sec:gradients}. 

\subsection{Gradients Through ODE Solve}\label{sec:gradients}

Our subsequent analysis will employ inference and ML methods that require gradient information. Therefore, ensuring that the ODE solve is differentiable and having an efficient method of computing gradients is critical.
To obtain gradients of the ODE forward solve with respect to model parameters, we employ an adjoint-based method known as NeuralODE~\cite{NeuralODE} using any black-box ODE solver. This method computes gradients through the ODE forward solve efficiently. It is also available as part of existing ML libraries such as Pytorch~\cite{NEURIPS2019_9015}, which allows its integration with other ML tools. 

Propagating gradients through the deposition onset criteria requires extra care. Before film deposition begins, the resistance and film thickness do not increase. Therefore, the following dynamical system is solved:
\begin{equation}\label{eq:baseline_part1}
     \frac{dh}{dt} = 0, \quad \frac{dR_{film}}{dt}= 0,
\end{equation}
with $V$, $j$, and $Q$ computed analytically according to the experiment type. When both $j > j_{min}$ and $Q > Q_{min}$, deposition begins and the model dynamics instantaneously switch to 
\begin{equation}\label{eq:baseline_part2}
    \frac{dh}{dt} = C_vj, \quad 
    \frac{dR_{film}}{dt} = \rho(j)\frac{dh}{dt}.
\end{equation}

As predicting the time of deposition onset is critical for predicting film growth, gradients of the output with respect to the onset condition parameters $j_{min}$ and $Q_{min}$ must be computed. However, the instantaneous change in model constitutes an in-place operation~\cite{NEURIPS2019_9015} and must be treated separately. We use ideas from an extension to NeuralODE which adds the ability the compute gradients through instantaneous event handling~\cite{chen2021learning}. Adding an additional \textit{switch state} $\xi$ to our model, we computationally solve the following equations instead of Eqs.~\ref{eq:baseline_part1} and~\ref{eq:baseline_part2}:
\begin{equation}\label{eq:baseline_event}
    \frac{dh}{dt} = \xi C_v j, \quad \frac{dR_{film}}{dt} = \xi \rho(j) \frac{dh}{dt}.
\end{equation}
The initial switch state $\xi(t=0) = 0$ is switched to $\xi(t=t_e) = 1$ at the event time, defined as the time $t_e$ such that either of the deposition onset criteria are met. This implementation detail allows computing the gradient of the forward solve with respect to the event time, and in turn with respect to the onset criteria parameters $j_{min}$ and $Q_{min}$, using the framework from~\cite{chen2021learning}.

\subsection{Parameter identifiability of baseline model}\label{sec:identifiability}
The process of parameter inference on the baseline model provides some insight into the shortcomings of the model, and results in proposed updates to the baseline model to address some of these shortcomings.

One of such shortcomings is parameter unidentifiability of the deposition onset criteria parameters, $Q_{min}$ and $j_{min}$. The nature of this double criteria for the onset of deposition creates situations in which one parameter or the other may not be identifiable. For example, consider the case in which the time $t_j$ at which $j>j_{min}$ is smaller than the time $t_Q$ at which $Q > Q_{min}$, such that $t_j < t_Q$. Changing $j_{min}$ over some range will thus not have any effect on the baseline model output because \textit{both} conditions $j>j_{min}$ and $Q>Q_{min}$ must be satisfied for deposition to begin. In this case, only $Q_{min}$ controls the deposition onset time and data may be uninformative about $j_{min}$. However, the opposite can occur and there also exist cases in which data may be uninformative about $Q_{min}$. 


\begin{figure}[h!]
    \centering
    \begin{subfigure}[b]{0.32\textwidth}
        \includegraphics[width=\linewidth]{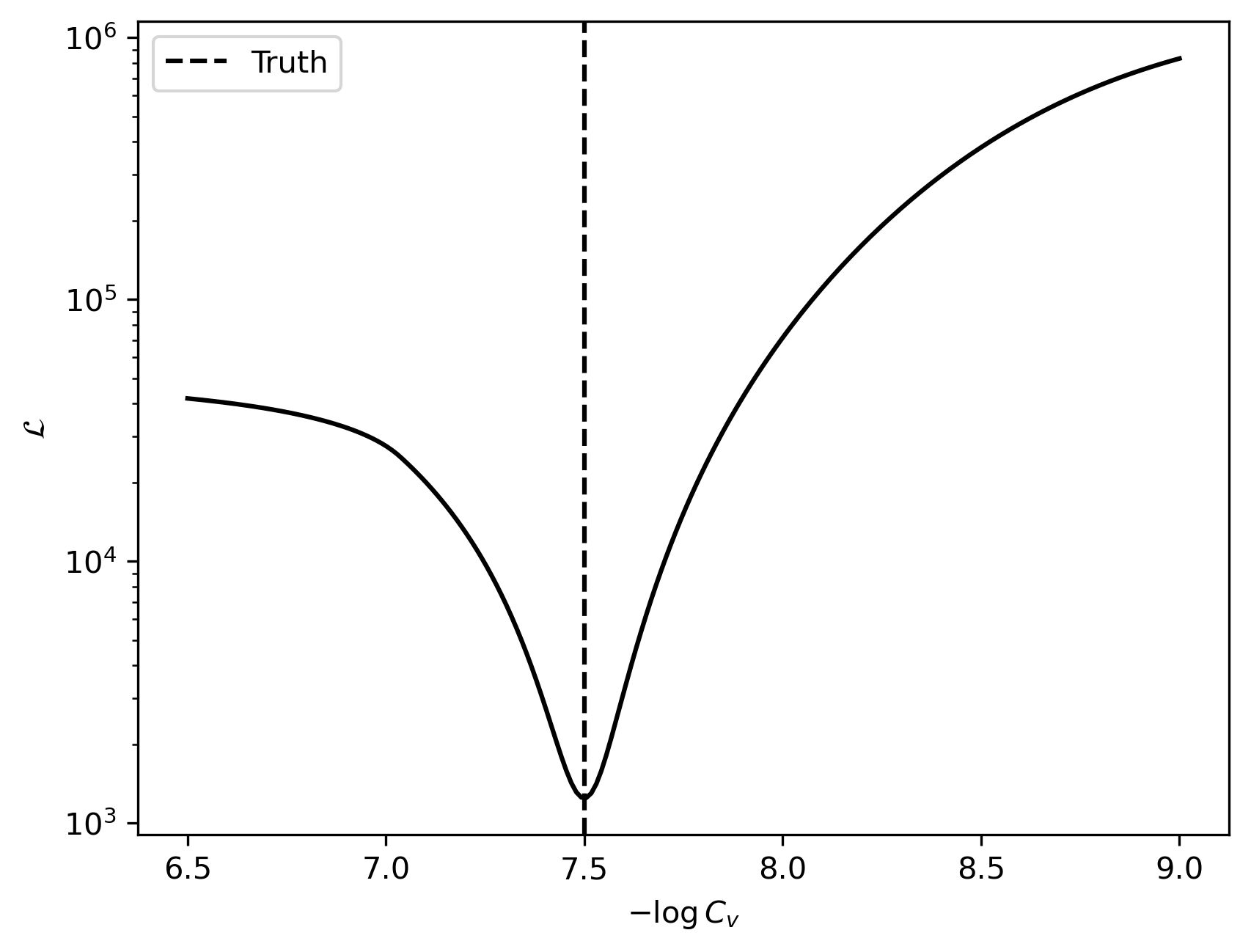}
        \caption{$-\log C_v$}
    \end{subfigure}
    \begin{subfigure}[b]{0.32\textwidth}
        \includegraphics[width=\linewidth]{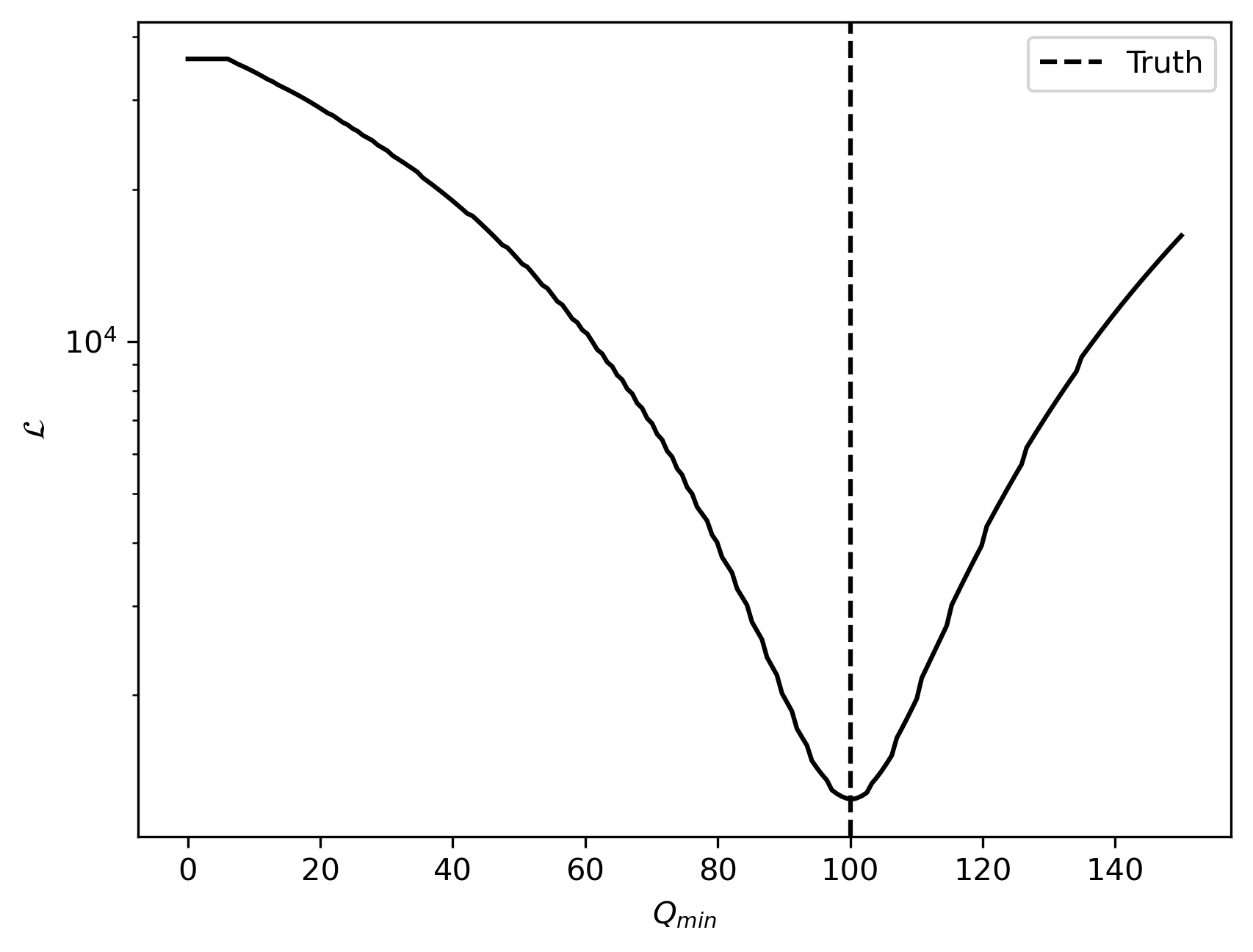}
        \caption{$Q_{min}$}
        \label{fig:Qmin_loglike}
    \end{subfigure}
    \begin{subfigure}[b]{0.32\textwidth}
        \includegraphics[width=\linewidth]{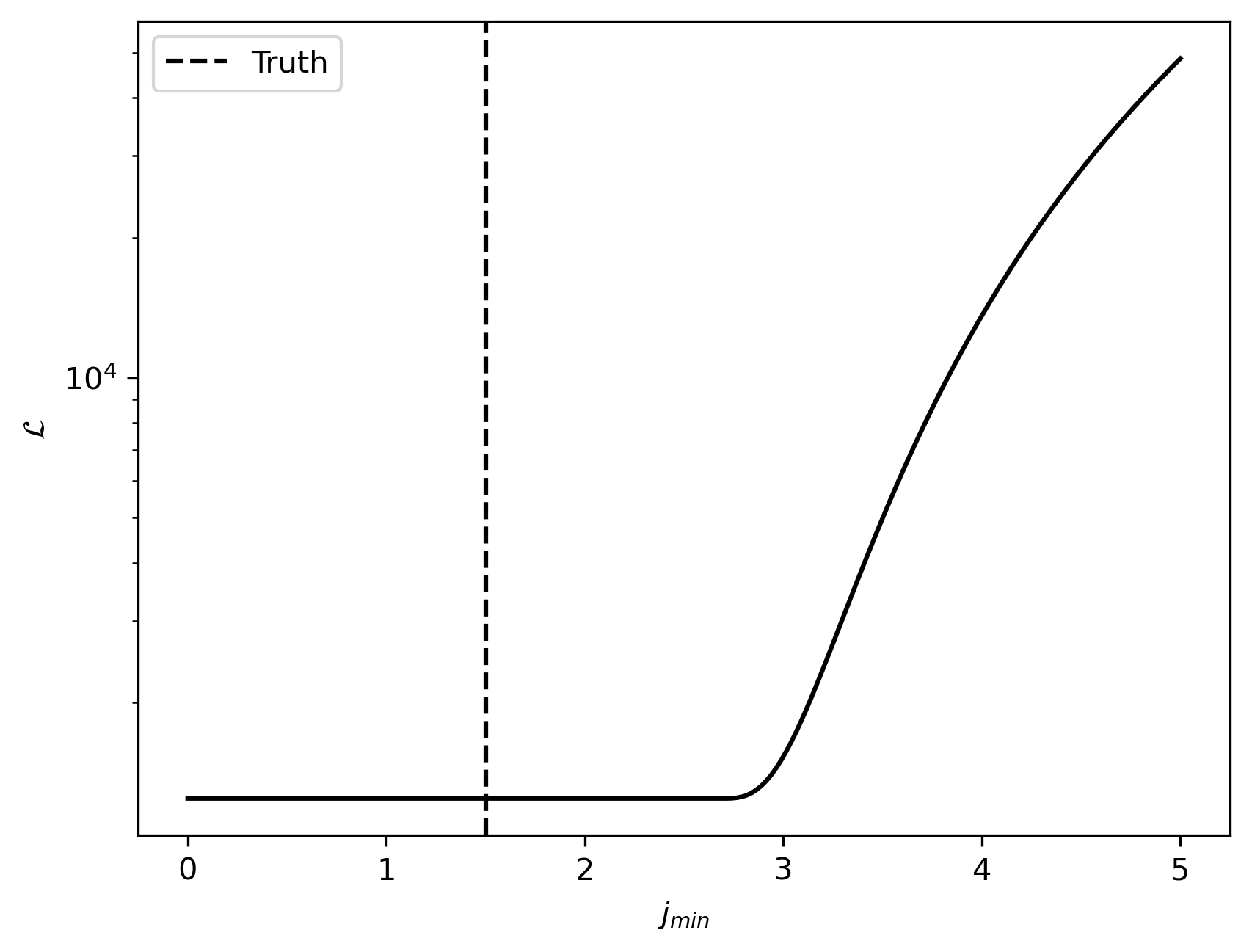}
        \caption{$j_{min}$}
        \label{fig:jmin_loglike}
    \end{subfigure}
    \caption{Negative log-likelihoods computed from simulated data on a voltage ramp experiment using the baseline model with experimental conditions $V_R = 0.125$, $\sigma=0.14$, $-\log C_v=8.5$, $Q_{min} = 100.0$, and $j_{min} = 1.5$.}
    \label{fig:unid_loglike}
\end{figure}

To illustrate this identifiability issue more clearly, we generate artificial data from the baseline forward model by taking the `true' parameters to be $[-\log C_v, Q_{min}, j_{min}] = [7.0, 150.0, 1.0]$. Other experimental parameters are given in Table \ref{table:paramSim}. A total of 10 samples are obtained by simulating a voltage ramp (VR) experiment with a small amount of noise added. Figure \ref{fig:unid_loglike} illustrates the negative log-likelihood of this data for each parameter to be inferred assuming that the other two parameters are fixed to their `true' values. For the first two parameters, a minimum exists at the `true' values, indicating that these parameters can be accurately inferred from the simulated data. However, the negative log-likelihood of the minimum current threshold parameter $j_{min}$ is flat over some region, indicating that the model is not effected by changes in $j_{min}$ over this region. 
If the `true' parameter value lies anywhere in the region in which the derivative of the log-likelihood is zero, that parameter is unidentifiable on that region. We use the term `unidentifiable' to mean that the first derivative of the likelihood is zero over some region in parameter space.

For voltage ramp experiments, there exists a set of boundaries in parameter space for which either $j_{min}$ or $Q_{min}$ will be unidentifiable, or both will be identifiable in baseline model. These boundaries change with the conditions of the voltage ramp experiment; gathering data from additional experiments could provide information on a parameter that is not informed by data from a different experiment. 

Consider a case in which data exists such that $t_Q<t_j$. This means that $t_j$ will control the deposition onset time. As $j(t)>0 \; \forall \; t>0$, then $Q(t>t_j) > Q(t_j) > Q_{min}$. Thus changing $Q_{min}$ on a range $0 \leq Q_{min} < Q(t_j)$ will have no effect on the output of the baseline model and the data will be uninformative about $Q_{min}$ on this range. Next consider a case in which data exists such that $t_j<t_Q$. Now $t_Q$ will control the deposition onset time. However, $j(t>t_Q) > j(t_Q)$ is \textit{not} guaranteed, which can be easily seen by taking the time derivative of Eq.~\ref{eq:Rj_VR}. If the voltage ramp $V_R < \rho(j)C_vj$, then $dj/dt < 0$ and it is not guaranteed that $j(t>t_Q) > j(t_Q)$. If $j(t>t_Q) < j(t_Q)$, then it is possible for $j(t>t_Q) < j_{min}$ and deposition stops, followed by an increase in current, restarting deposition, and the cycle repeats. Ultimately this indicates that the parameter $j_{min}$ will have an influence on the baseline model output, and data will be informative about $j_{min}$. Thus, if $t_j<t_Q$, data may or may not be informative about the parameter $j_{min}$, depending on the experimental conditions. 

\begin{figure}[h!]
    \centering
    \begin{subfigure}[b]{0.4\textwidth}
        \includegraphics[width=\linewidth]{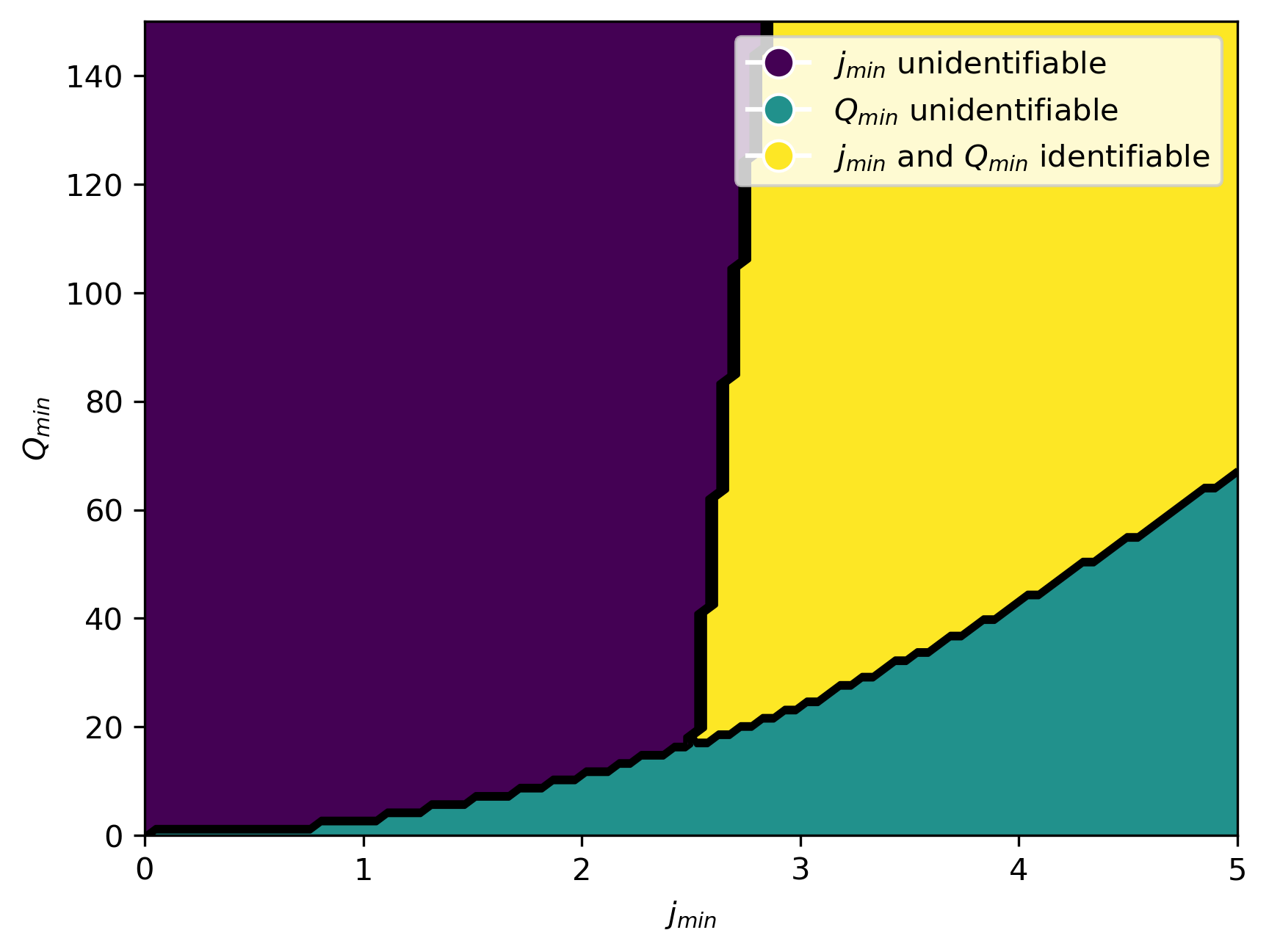}
        \caption{VR experiment, $V_R=0.125$, $\sigma=0.14$, $-\log C_v=8.5$}
        \label{fig:id_bound1}
    \end{subfigure}
    \hfill
    \begin{subfigure}[b]{0.4\textwidth}
        \includegraphics[width=\linewidth]{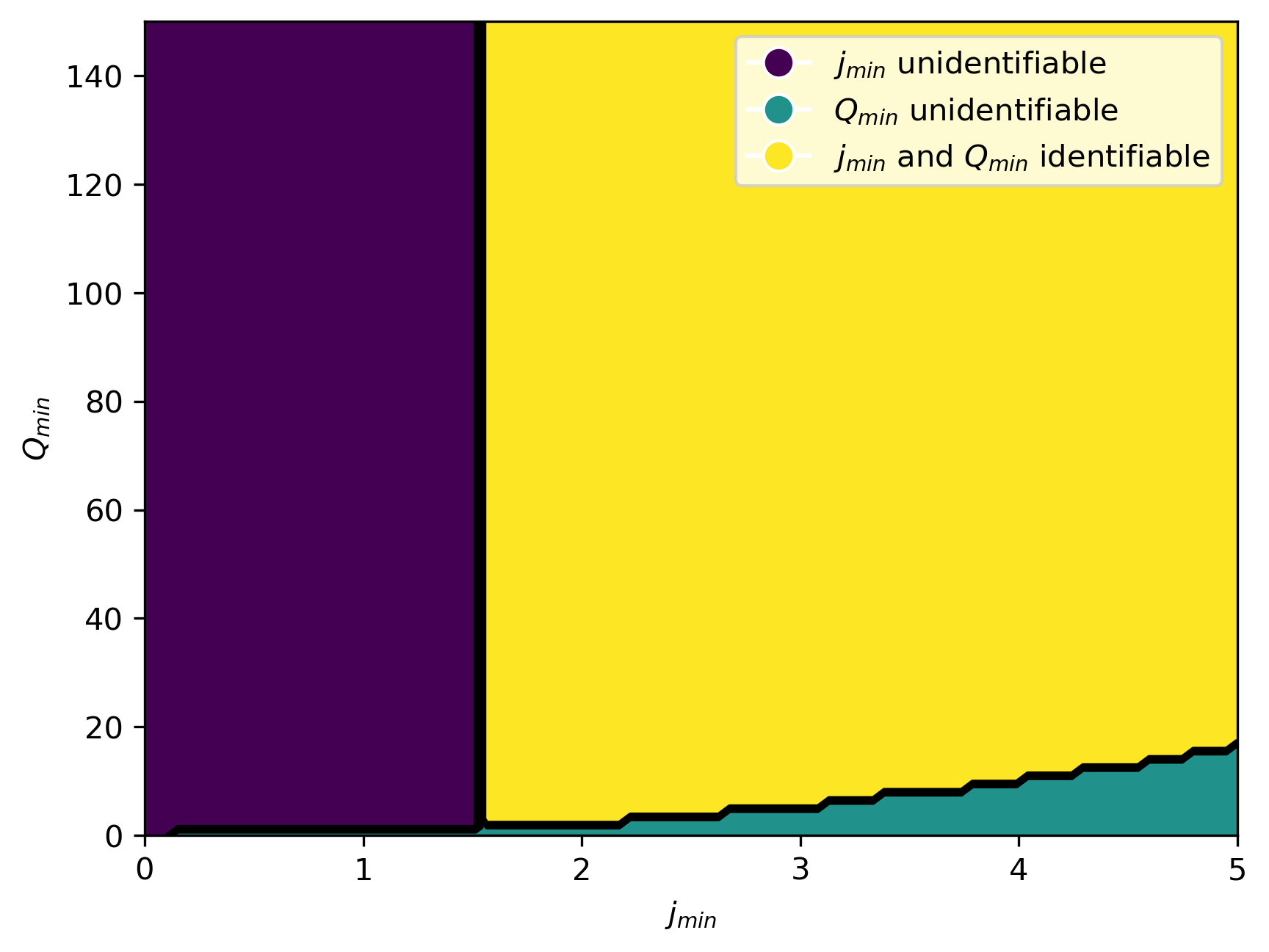}
        \caption{VR experiment, $V_R=0.5$, $\sigma=0.14$, $-\log C_v=7.5$}
    \end{subfigure}
    \caption{Identifiability regions of the baseline model for two different experimental conditions. The log-likelihood on the simulated experimental data will be constant in the purple and cyan regions, indicating that little information is gained about $j_{min}$ if the true value lies in the purple region or $Q_{min}$ if the true value lies in the cyan region. Note: the `stepping' behavior observed in the identifiability boundaries here are a product of discretizing the $j_{min}$ and $Q_{min}$ domains, but the boundaries are in fact smooth.}
    \label{fig:id_bounds}
\end{figure}

Two identifiability regions corresponding to different experiments are computed empirically and illustrated in Figure \ref{fig:id_bounds}. This regions are computed by simulating the baseline model according to the experimental conditions for a set of discretized points in the $j_{min}, Q_{min}$ space. For each simulation, identifiability is checked by checking $t_Q$ and $t_j$. Purple regions indicate that data from the experiment is not informative about $j_{min}$ if the true value of $j_{min}$ and $Q_{min}$ lie in the region. In other words, $\partial \mathcal{L}/\partial j_{min} = 0$ on the purple region. Cyan regions indicate that data from the experiment is not informative about $Q_{min}$ if the true values of $j_{min}$ and $Q_{min}$ lie in the region, or $\partial \mathcal{L}/\partial j_{min} = 0$. Yellow regions indicate that the experiment can inform both $j_{min}$ and $Q_{min}$. The boundary of the cyan region can be computed analytically, but the other is nontrivial and computed empirically. The times at which the deposition onset criteria are met in the voltage ramp experiment for the baseline model are given by
\begin{equation}\label{eq:tj}
    t_j = \frac{2Q_{min}}{\sigma V_R}(\sigma R_0+L), \quad t_Q = \left [\frac{2j_{min}}{\sigma V_R}(\sigma R_0+L)\right ]^{1/2} \; .
\end{equation}
Setting Eqs.~\ref{eq:tj} 
equal, we obtain a closed form solution to the $Q_{min}$ identifiability boundary as
\begin{equation}
    j_{min} = \left [ \frac{2Q_{min}\sigma V_R}{\sigma R_0 + L}\right ]^{1/2} \; ,
\end{equation}
which has been validated against the empirically computed boundaries shown in Fig.~\ref{fig:id_bounds}. We note that the log-likelihoods in Fig.~\ref{fig:Qmin_loglike} and~\ref{fig:jmin_loglike} correspond to the identifiability plane in Fig~\ref{fig:id_bound1}. The log-likelihoods of $Q_{min}$ and $j_{min}$ have first derivative zero over some region as predicted by the identifiability boundaries. The true values of $j_{min}$ and $Q_{min}$ are such that $j_{min}$ is not informed by the simulated experimental data. 

If the experimental data does not inform one of the parameters, then that parameter does not influence the output of our baseline model, and it cannot be accurately inferred. However, as shown by Fig.~\ref{fig:id_bounds}, the identifiability boundaries change with the type of experiment. This behaviour could cause very poor prediction results. Suppose none of the experimental data is informative about $j_{min}$, and inference is performed on the model parameters. Using the model in prediction could result in poor performance especially if predictions are made for an experimental condition in which $j_{min}$ does influence the output of the model. 

It is also necessary to infer robust posteriors in this case. Suppose Gaussian variational inference is selected as the inference method. A Gaussian variational posterior will be computed for each of the parameters to be inferred, providing a unimodal distribution for each. This can be misleading about the Bayesian posterior distribution of the parameter and result in poor uncertainty quantification during prediction. 

We explore two approaches towards the aim of improving model predictions by improving the form of the baseline model. 
First, we update the model based on insight from the shortcomings observed during inference, described in Sec.~\ref{sec:identifiability_updates}. We then pursue a different approach in which model augmentations are learning using machine learning tools to augment the baseline model with previously unmodeled dynamics, discussed in Sec.~\ref{sec:machine_learning_augmentations}.

\subsection{Inference on baseline model}

In this section, we demonstrate inference results using Gaussian variational inference on the baseline model. The purpose of this experiment is to illustrate the shortcomings of the model form itself, not the selected method of inference. Inference is performed using simulated data from the baseline model, and we demonstrate that the posterior predictive results in good prediction performance for some experimental conditions, but poor performance on others. 

Data is generated by simulating the baseline model 10 times up to a final time $T=250s$ for a voltage ramp experiment with $V_R=1.0$, $\log C_v = -8$, $j_{min} = 1.5$, $\sigma = 0.14$, and $Q_{min} = 100$, and add Gaussian random noise $\epsilon \sim \mathcal{N}(0, \eta^2)$, where $\eta = 0.01$, at each time step to simulate measurement noise. Only data from current measurements is considered in the inference process. We then perform Gaussian VI for parameters $\log C_v$, $j_{min}$, and $Q_{min}$ on the baseline model. 

\begin{figure}[h!]
    \centering
    \begin{subfigure}[b]{0.49\textwidth}
        \includegraphics[width=\linewidth]{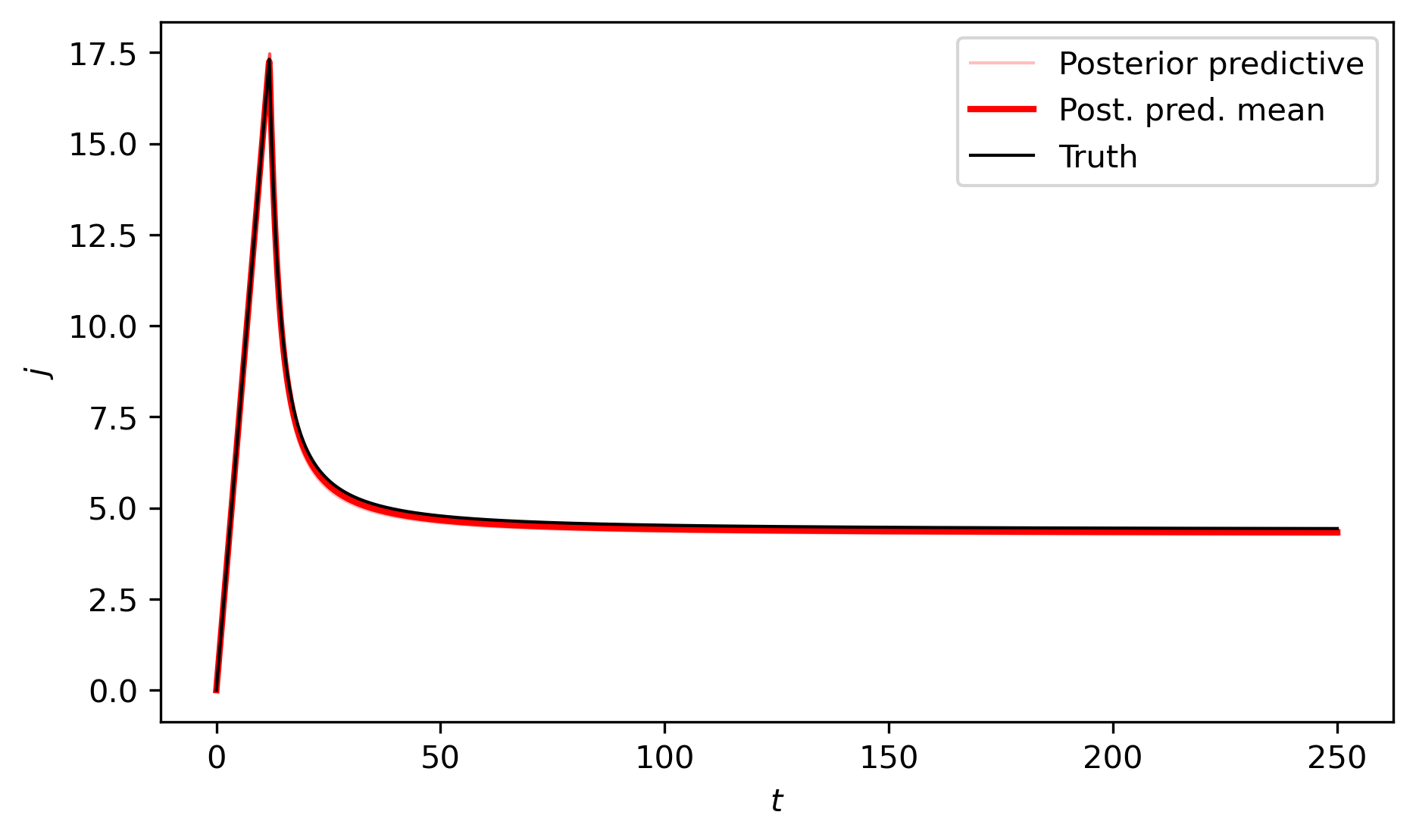}
        \caption{Voltage ramp experiment data, $V_R=1.0$, $\sigma=0.14$, $-\log C_v=8$, $j_{min}=1.5$, $Q_{min}=100$}
        \label{fig:posterior_predictive_example_good}
    \end{subfigure}
    \hfill
    \begin{subfigure}[b]{0.49\textwidth}
        \includegraphics[width=\linewidth]{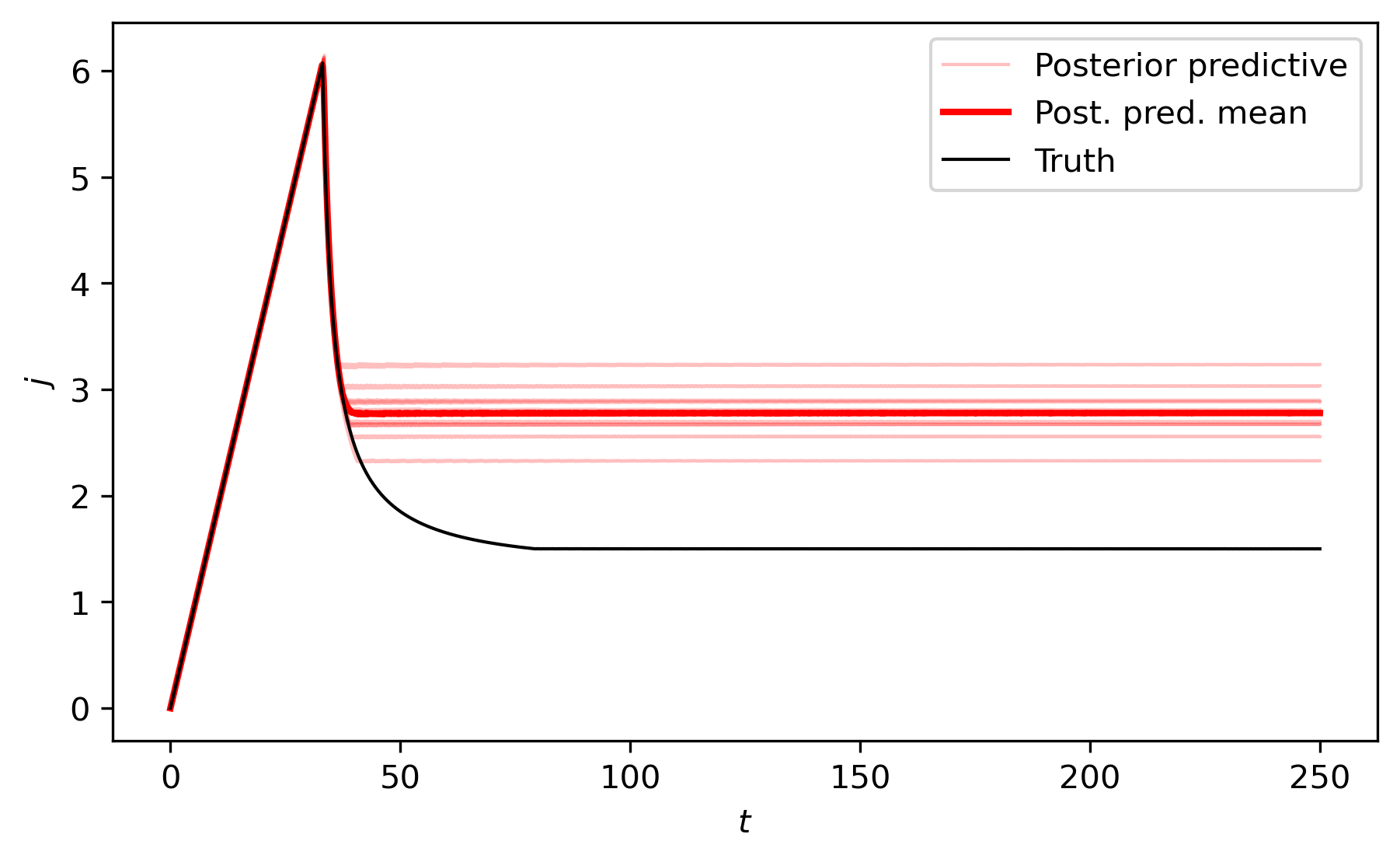}
        \caption{Voltage ramp experiment prediction, $V_R=0.125$, $\sigma=0.14$, $-\log C_v=8$, $j_{min}=1.5$, $Q_{min}=100$}
        \label{fig:posterior_predictive_example_bad}
    \end{subfigure}
    \caption{Posterior predictive results after performing Gaussian VI on data from a simulated voltage ramp experiment. The posterior predictive results in accurate simulations on the data (a), but poor predictions for other experiments (b). This is caused by unidentifiable $j_{min}$ in the data.}
    \label{fig:posterior_predictive_example}
\end{figure}

After learning the variational posterior $q_\phi$, samples are drawn and used to simulate the baseline model to obtain samples from the posterior predictive. Ten of these samples along with the mean are shown in Fig.~\ref{fig:posterior_predictive_example_good}. For the experimental configuration on which the data is generated, accurate and low variance prediction is observed. However, we then use the variational posterior distribution to predict on a voltage ramp experiment in which all parameters are the same except the voltage ramp $V_R = 0.125$. Fig.~\ref{fig:posterior_predictive_example_bad} shows that prediction performance is very poor for this experiment. 

The reason for this stems from the identifiability issues previously discussed in Sec~\ref{sec:identifiability}. In our experiment, we assume that the `true' value of $j_{min}$ is 1.5. However, in the experiment which we gather data from, $j_{min}$ is unidentifiable in the baseline model. Thus the inference results in a variational posterior distribution for $j_{min}$ of $q(j_{min}|\mathcal{D}) = \mathcal{N}(2.66, 0.01)$, which is a low variance but poor estimate of the true parameter value. 
This posterior distribution is then used for prediction in an experiment in which $j_{min}$ \textit{does} have an effect on model output, and because the posterior is not accurate, prediction is also inaccurate.

On top of identifiability issues potentially resulting in poor prediction performance, the parameter $Q_{min}$ is inconsistent accross different experiment types. To illustrate this issue, Gaussian VI is performed on the parameters $C_v$, $j_{min}$, and $Q_{min}$ twice - first using real experimental data from only the voltage ramp experiments and again using only data from the constant current experiments. The maximum a posteriori (MAP) of the variational posterior for each distribution are very different - in particular for $Q_{min}$. Using only voltage ramp experimental data, the MAP of the variational posterior is $Q_{min} \approx 261$; however, using only constant current experimental data, the MAP is located at $Q_{min} \approx 101$. This is indicative of $Q_{min}$ being problematic in allowing the baseline model to accurately predict experimental data with different boundary conditions. Our natural conclusion is that the model is incorrect, and a root cause may lie in a constant $Q_{min}$. Rather, $Q_{min}$ may be a function of the type of experiment being performed rather than a constant to be inferred. In Sec~\ref{sec:identifiability_updates}, physically intuitive model updates are introduced with the aim of alleviating the identifiability concerns of the baseline model and improving the modelling of the minimum charge criterion.

\section{Model updates}\label{sec:model_updates}

The baseline model was shown to exhibit identifiability as well as generalization deficiencies in Sec.~\ref{sec:inference}. In this section, we aim to improve the prediction accuracy and generalizability of the baseline model. First, modifications are made based on the inference results of the baseline model to aid in improving parameter identifiability and minimum charge criterion modeling.

An alternative approach to improving the baseline model is also investigated in which machine-learning augmentations are introduced to model system dynamics which are absent from the baseline model. These augmentations are introduced with an emphasis on interpretability of the augmented model while allowing for greater flexibility in model expressiveness.




\subsection{Inference-informed modifications}\label{sec:identifiability_updates}

Based on the inference experiments of Sec.~\ref{sec:inference}, model updates are proposed to alleviate the observed inadequacies during prediction. The issue of identifiability arises in the baseline model due to the double conditional statement that film deposition begins only if $j > j_{min}$ \textit{and} $Q > Q_{min}$. This conditional statement, in particular $j > j_{min}$, creates non-physical, discontinuous behavior of the model. The film growth rate given by Eq. \ref{eq:baseline_part1} is exactly zero until the conditional statement is true. Assuming that $j_{min} > 0$, the film growth rate will instantaneously increase, and a sharp discontinuity in the film thickness growth occurs.
The model also does not accurately model the cases in which both onset criteria are met, but the minimum current condition is no longer met at a later time. This behavior is one of the reasons that prediction is observed to be inaccurate in Fig.~\ref{fig:posterior_predictive_example_bad}. 
We therefore propose a model update to create a model in which the dynamics are continuous which also allows for film dissolution by replacing the film thickness dynamics in~Eq.~\ref{eq:baseline_part2} with 
\begin{equation}\label{eq:dhdt_updated}
    \frac{dh}{dt} = C_v(j_n - j_{min}) \textrm{ for } Q > Q_{min}, \textrm{ s.t. }h\geqslant 0\; .
\end{equation}
This also gives the benefit of $j_{min}$ being identifiable for all experimental configurations. 

With $j_{min}$ now identifiable, the expressiveness of the parameter is investigated to improve generalizability. Assuming that the minimum charge criterion is related to the concentration of $\ce{OH}^-$ present in the bath,~\ref{app:K} illustrates that $Q_{min}$ is not constant accross experiment types, but depends on a different constant $K$. 
It is shown in~\ref{app:K} that $Q_{min}$ is a function of this constant, and the function differs between the voltage ramp and constant current experiments. For VR (Eq.~\ref{eq:Qmin_VR}) and CC (Eq.~\ref{eq:Qmin_CC}) experiments, this function is given by

\noindent
\begin{tabularx}{\linewidth}{XXX}
\begin{equation}
    Q_{min} = \left ( \frac{81}{128\beta} \right ) ^{1/3} K^{4/3} \label{eq:Qmin_VR}
\end{equation}
&
\begin{equation}
    Q_{min} = \frac{K^2}{j_0}, \label{eq:Qmin_CC}
\end{equation}
\end{tabularx}
where $\beta = \sigma V_R / (\sigma R_{0} + L)$

The updated film thickness dynamics of Eq.~\ref{eq:dhdt_updated} along with introducing the parameter $K$ used to compute the minimum charge criterion constitute an updated model which we dub the `inference-informed' model.

\begin{figure}[h!]
    \centering
    \begin{subfigure}[b]{0.32\textwidth}
        \includegraphics[width=\linewidth]{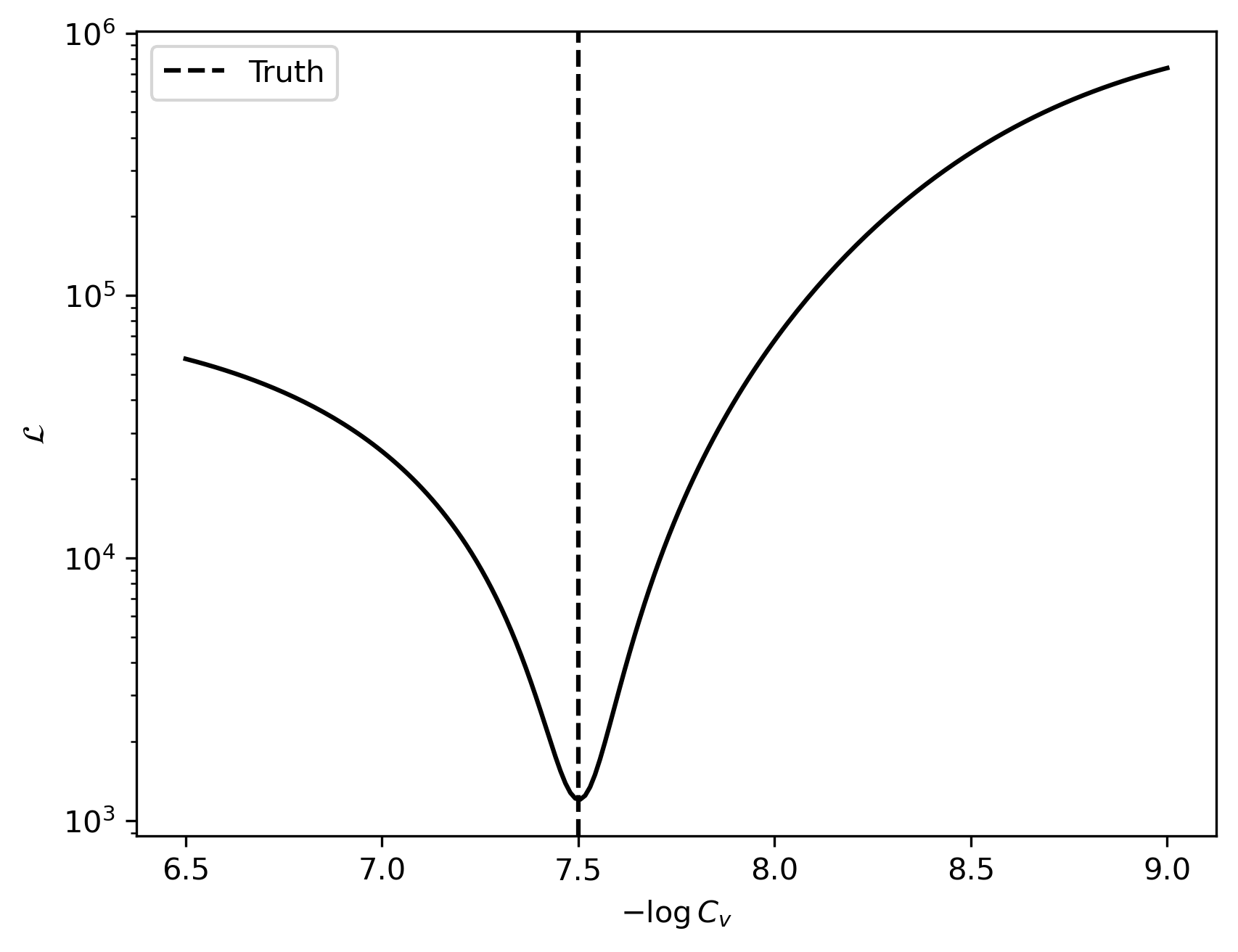}
        \caption{$-\log C_v$}
    \end{subfigure}
    \hfill
    \begin{subfigure}[b]{0.32\textwidth}
        \includegraphics[width=\linewidth]{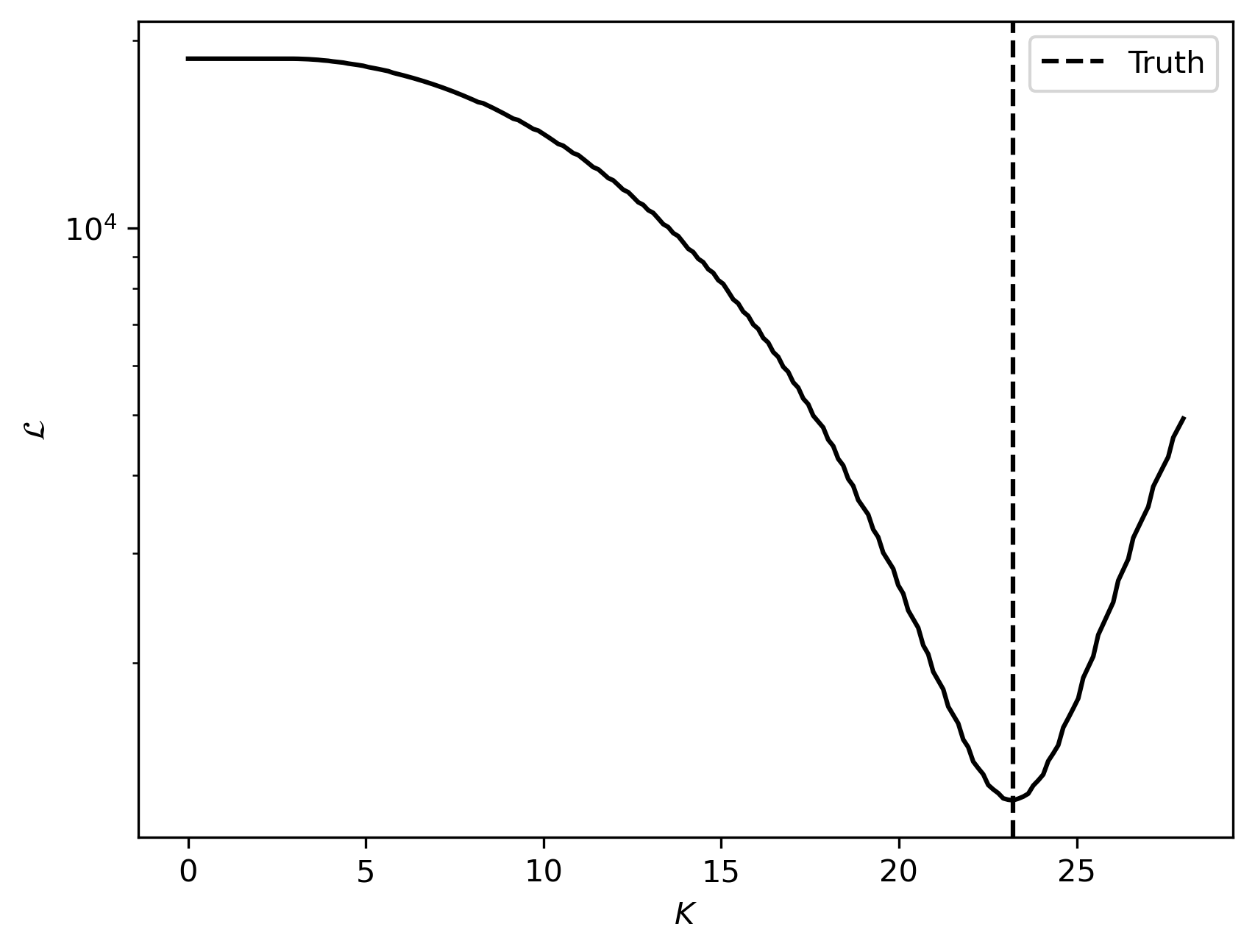}
        \caption{$Q_{min}$}
        \label{fig:K_loglike}
    \end{subfigure}
    \begin{subfigure}[b]{0.32\textwidth}
        \includegraphics[width=\linewidth]{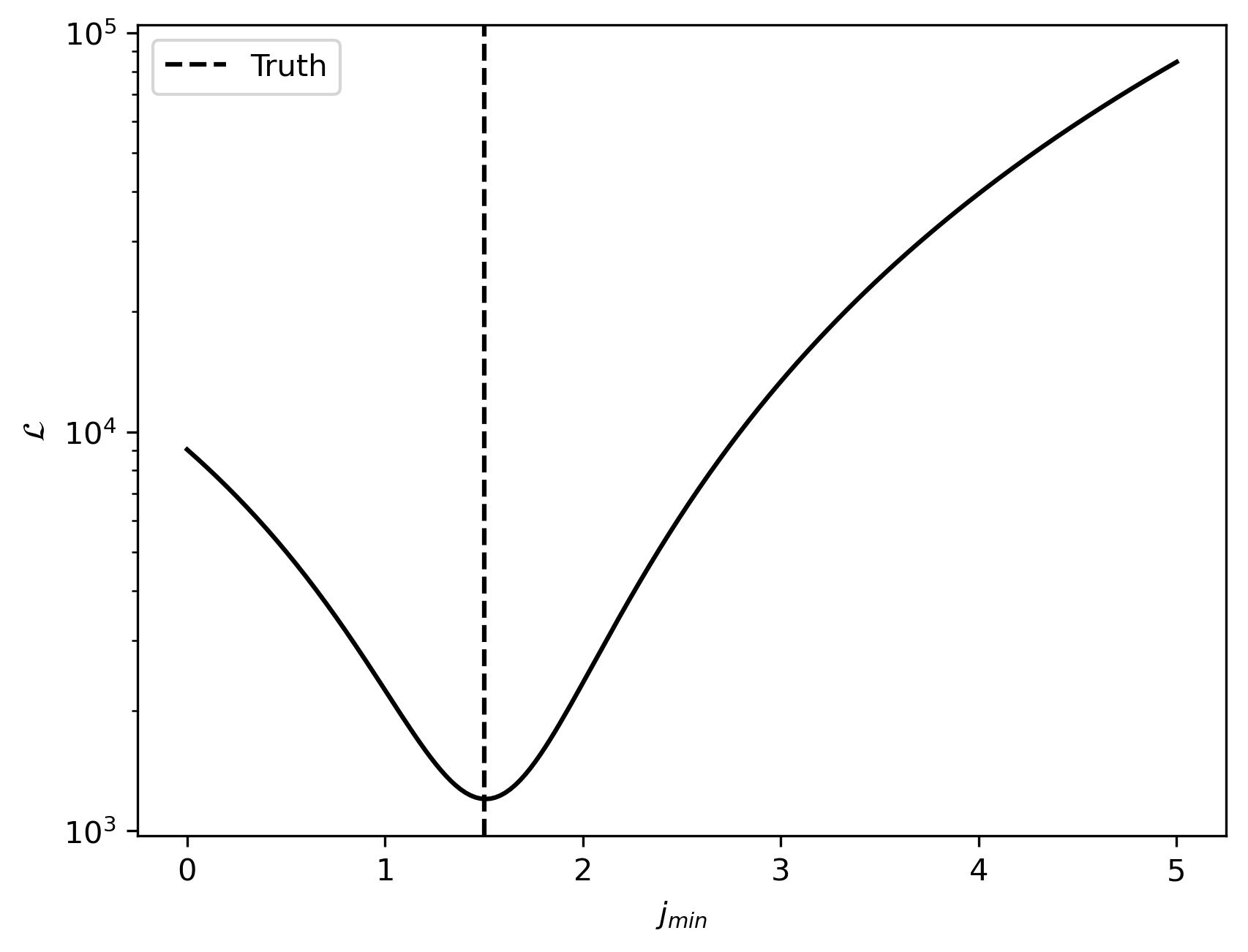}
        \caption{$j_{min}$}
        \label{fig:jmin_loglike}
    \end{subfigure}
    \caption{Negative log-likelihoods computed from simulated data on a voltage ramp experiment using the inference-informed model with experimental conditions $V_R = 0.125$, $\sigma=0.14$, $-\log C_v=8.5$, $Q_{min} = 100.0$ ($K=23.2$), and $j_{min} = 1.5$.}
    \label{fig:id_loglike}
\end{figure}

Performing the same exercise to visualize the negative log-likelihood as in Fig.~\ref{fig:unid_loglike}, we visualize the negative log-likelihood of the inference-informed model on artificial data to illustrate that all parameters are now identifiable. In this case, we simulate the data from the same experimental configuration as Fig.~\ref{fig:unid_loglike}, which is a VR experiment with $V_R = 0.125$, $\sigma=0.14$, $-\log C_v=8.5$, $Q_{min} = 100.0$, and $j_{min} = 1.0$. However, the parameter $K$ is used instead of $Q_{min}$; thus the value of $K$ is found using the relationship in Eq.~\ref{eq:Qmin_VR}, obtaining $K=23.2$. The negative log-likelihoods of $-\log C_v$, $K$, and $j_{min}$ are illustrated in Fig.~\ref{fig:id_loglike}, and all parameters exhibit a global minimum at the true parameter values, indicating that all are now identifiable. 

\subsubsection{Model Comparisons}
The baseline model and the inference-informed model are directly compared by performing Bayesian inference using the gridding approach discussed in Sec.~\ref{sec:gridding}. We then predict for each model based on the maximum a posteriori and compute the negative log-likelihood of the data. Lower values of the negative log-likelihood at the MAP correspond to a model which better fits the experimental data. 

The MAP of the approximated posterior using the gridding approach assuming the baseline model is located at $\theta_{MAP} = [-\log C_v, Q_{min}, j_{min}] = [7.58, 173.5, 0.0]$ with a negative log-likelihood at the MAP of $3.35 \times 10^7$. Assuming the updated model, the MAP is located at $\theta_{MAP} = [-\log C_v, K, j_{min}] = [7.32, 44.2, 0.63]$ while the negative log-likelihood is $2.65 \times 10^7$. 
This suggests that the inference-informed model performs better than the baseline model by having the potential to represent the experimental data more accurately. 

\begin{figure}[h!]
    \centering
    \begin{subfigure}[b]{0.4\textwidth}
        \includegraphics[width=\linewidth]{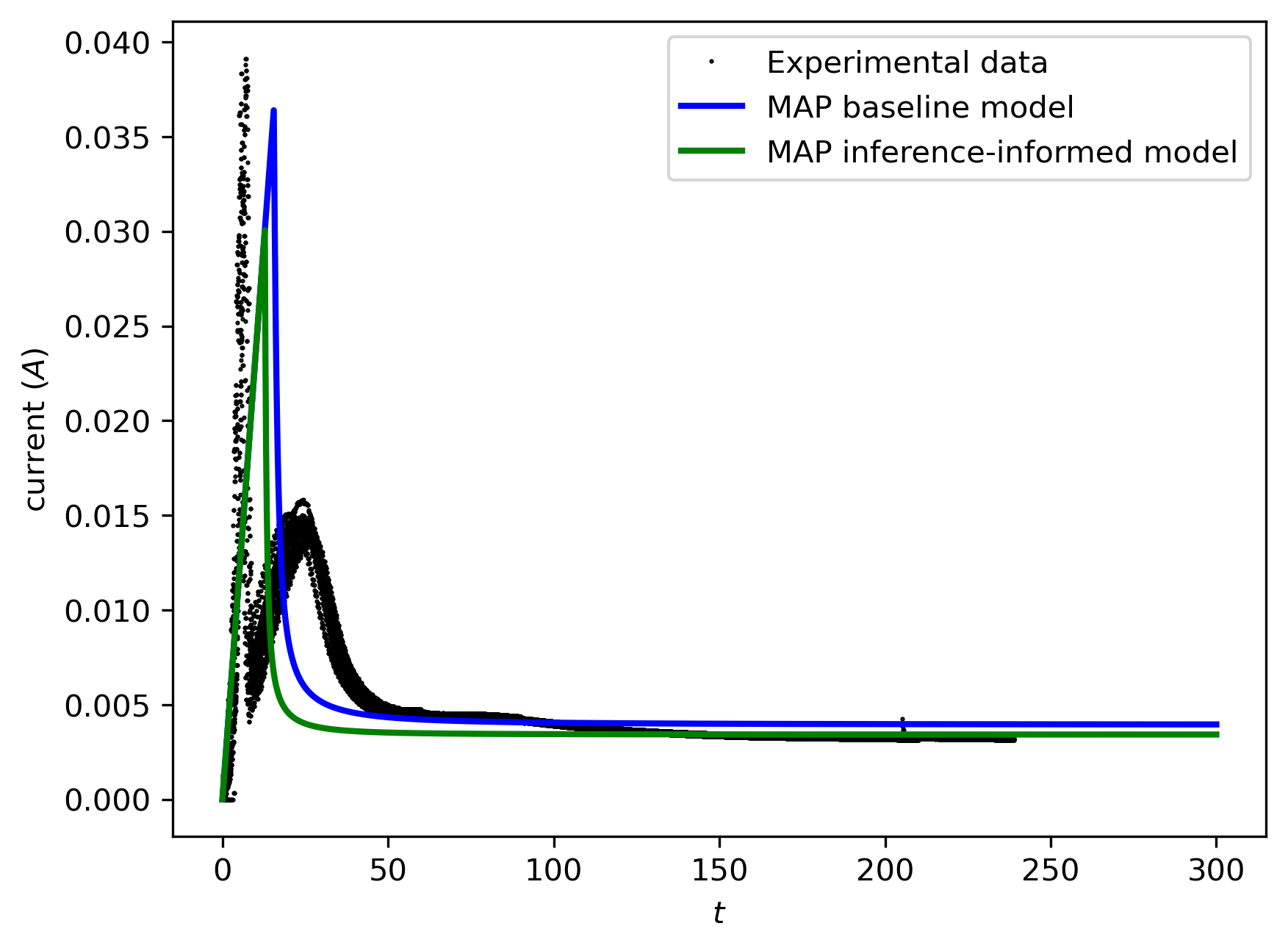}
        \caption{Voltage ramp experiment prediction, $V_R=1.0 V/s$}
        \label{fig:posterior_predictive_example_good}
    \end{subfigure}
    \hfill
    \begin{subfigure}[b]{0.4\textwidth}
        \includegraphics[width=\linewidth]{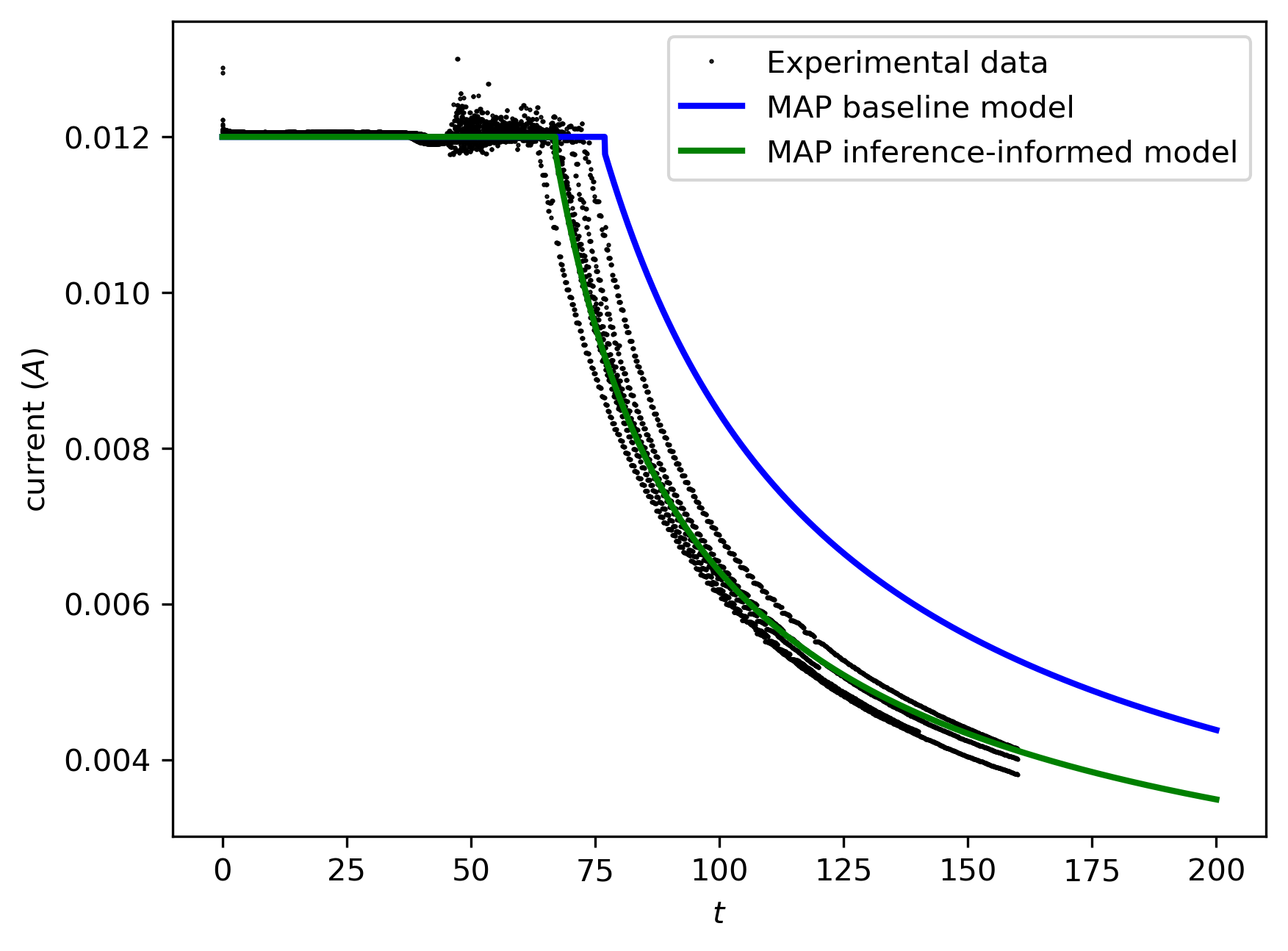}
        \caption{Constant current experiment prediction, $j_0=7.5mA$}
    \end{subfigure}
    \caption{Comparisons between current prediction on the baseline model and inference-informed model at the MAP for each on (a) voltage ramp experiment with $V_R = 1.0 V/s$ and (b) constant current experiment with $j_0 = 7.5 mA$.}
    \label{fig:model_map_comparisons}
\end{figure}
\begin{figure}[h!]
    \centering
    \begin{subfigure}[b]{0.4\textwidth}
        \includegraphics[width=\linewidth]{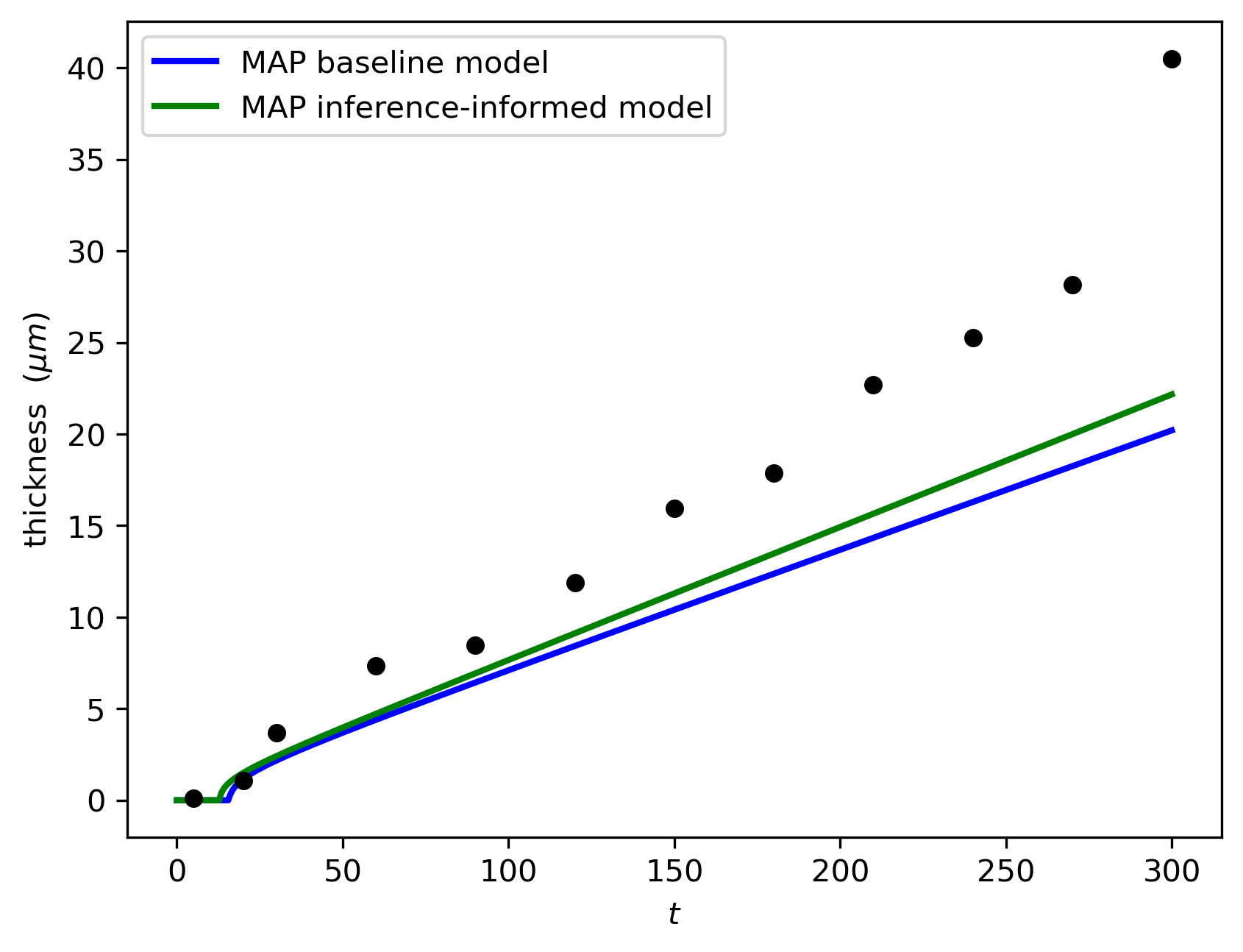}
        \caption{Voltage ramp experiment prediction, $V_R=1.0$ (film thickness)}
    \end{subfigure}
    \hfill
    \begin{subfigure}[b]{0.4\textwidth}
        \includegraphics[width=\linewidth]{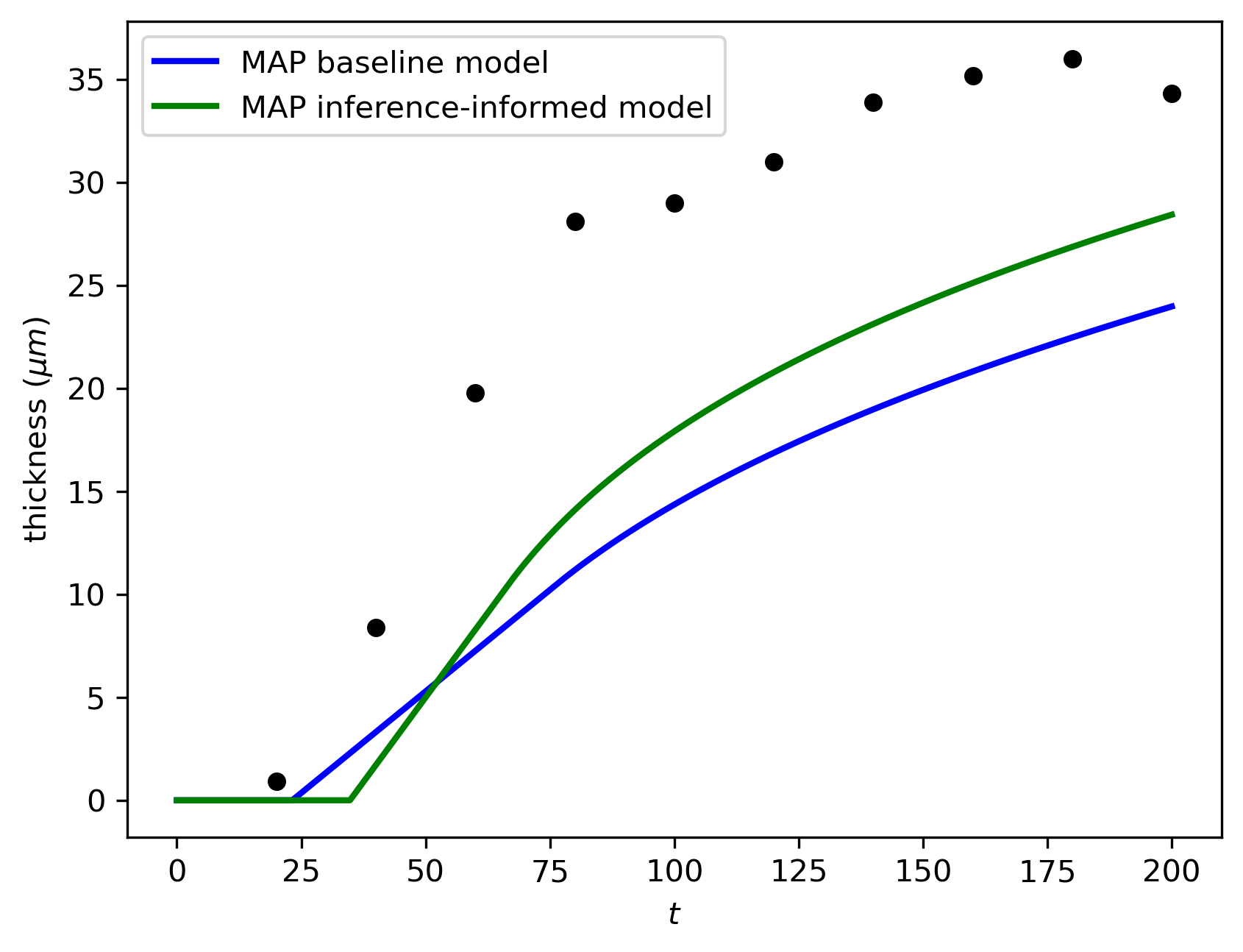}
        \caption{Constant current experiment prediction, $j_0=7.5mA$ (film thickness)}
    \end{subfigure}
    \caption{Comparisons between film thickness prediction on the baseline model and inference-informed model at the MAP for each.}
    \label{fig:model_map_thickness}
\end{figure}

Each model prediction at the map is compared to the experimental data in Fig~\ref{fig:model_map_comparisons}. The current predictions are shown for 2 experiments: VR with $V_R = 1.0 V/s$ and CC with $j_0 = 7.5 mA$. Predictions for all experiments on current, resistance, and thickness data are included in~\ref{app:MAP}. Prediction of the inference-informed model for constant current experiments exhibits significant improvement over the baseline model, likely due to the improved parameterization of the minimum charge criterion $Q_{min}$. However, there is clearly some physical behavior which exists in the voltage ramp experiments which is not present in our model. For instance, there are two current `peaks' in experimental data for voltage ramp experiments, but only one is possible in model predictions. Additionally, although the thickness prediction of the inference-informed model is closer to the experimental data than the baseline model, Fig.~\ref{fig:model_map_thickness} illustrates that neither are particularly accurate. Ultimately, thickness prediction is the most important quantity of interest. We thus turn towards improving the model through augmenting the model forming, and learning the augmentations via a machine-learning based framework.

\subsection{Machine-learning augmentations}\label{sec:machine_learning_augmentations}
Modifying the model using insights from the inference process aid in understanding some shortcomings due to model form, but results in limited improvement to predictive performance. Additional physical behavior is missing from the model itself, which results in poor performance even when the optimal model parameters are identified. There are two main features absent from the baseline model: the presence of two peaks in the current dynamics for the voltage ramp experiments, and smooth behavior of those peaks. Unstable behavior in data for constant current experiments observed just before the drop in current is due to the current controller in the experiments, which we do not account for in our model. 

To incorporate the two missing physical features in our model, we turn to machine learning with an emphasis on interpretability. In particular, we augment the right hand side of the dynamics model with parameterized neural networks and train the augmentations using NeuralODE~\cite{NeuralODE}. This can be used effectively to utilize adjoint-based gradient computation alongside standard machine learning pipelines to learn flexible augmentations in the right hand side of a dynamical system. 

We first attempt to characterize the underlying physics of each of the two peaks present in the experimental current data for voltage ramp experiments. The second peak corresponds to the onset of deposition, which is previously modeled in the baseline and inference-informed models. As film deposition begins, resistance increases and current decreases. However, the baseline model assumes an instantaneous onset of film deposition which results in a discontinuous and physically impossible film growth rate. Experimental data illustrates smooth transitions to film growth in current data, further supporting the need for additional modeling. We propose an augmentation to this model which has the additional benefit of modeling the deposition onset time without the need of threshold parameters. We propose multiplying the right hand side of Eq.~\ref{eq:baseline_part2} by a learnable term $g_\phi(V, Q)$ which varies smoothly between zero and one. This term is dubbed the `coverage fraction model', and it is a function of voltage and charge only. It is found empirically that using both voltage and charge as inputs to the model results in the best predictive performance. It is a function of charge due to the dependence of the deposition onset time of the baseline model on charge, and it is a function of voltage due to a change in the time derivative of voltage across experiment types. 

The first peak present in the current data for voltage ramp experiments is caused by a phenomenon ignored by the baseline model altogether. This peak may be caused by an oxygen-reduction reaction~\cite{HisashiNAGAI2012} which occurs at the anode as the voltage increases. The relationship between electric potential and current in redox reactions is described by the Butler-Volmer equation~\cite{DICKINSON2020114145} of the form 
\begin{equation}\label{eq:butler-volmer}
    j = j_0 \left [ \exp \left (aV \right ) - \exp(bV)\right ] \; ,
\end{equation}
where the parameters $a$ and $b$ depend on many physically relevant parameters. However, the particular form is not relevant to the discussion here as we aim to learn this component of the model while keeping the general form to ensure that the model is physically-interpretable. We assume that the combined effect of this redox reaction results in a resistance `source' such that $j = c_1 \exp(c_2V)$, based on the form of Eq.~\ref{eq:butler-volmer}.

After some critical point is reached (which we do not explicitly model, but likely corresponds to some minimum charge criterion), we assume that the $\ce{OH^-}$ starts being diffused at some rate~\cite{HisashiNAGAI2012} which decreases the $\ce{OH^-}$ concentration. As the redox reaction continues, we assume that the combined effect of these two reactions results in a different exponent such that $j = c_3\exp(c_4V)$, where $c_4 < 0$. 

To augment the voltage ramp experiment model with this behavior of switching between two different exponents resulting from a combination of reactions, we use an exponential function in which the argument includes a hyperbolic tangent function which can vary smoothly between two different values. We thus assume an augmented model of the form 
\begin{equation}\label{eq:j_aug}
j = \frac{\sigma V}{\sigma R_{film} + L} + c_1 \exp(c_2 V f_\theta(V,Q)), \quad  f_\theta(V,Q) = \tanh (-\hat{f}_\theta (V,Q)) \; ,
\end{equation}
and $\hat{f}_\theta$ is a parameterized function such as a feed-forward neural network (FNN). The first term in Eq.~\ref{eq:j_aug} corresponds to the baseline model current in Eq.~\ref{eq:bc_vr}, but the second term 
is informed by the aforementioned physical processes while allowing flexibility in learning the particular dynamics. Using the $\tanh$ function allows a smooth transition between two different exponents, effectively switching between the two dominant reactions at the beginning of voltage ramp experiments. We have added additional parameters to the model to allow for more flexibility. This model augmentation is then added to the current in voltage ramp experiments to capture the behavior of the first peak.

\begin{figure}[h!]
    \centering
    \begin{subfigure}[b]{0.42\textwidth}
        \includegraphics[width=\linewidth]{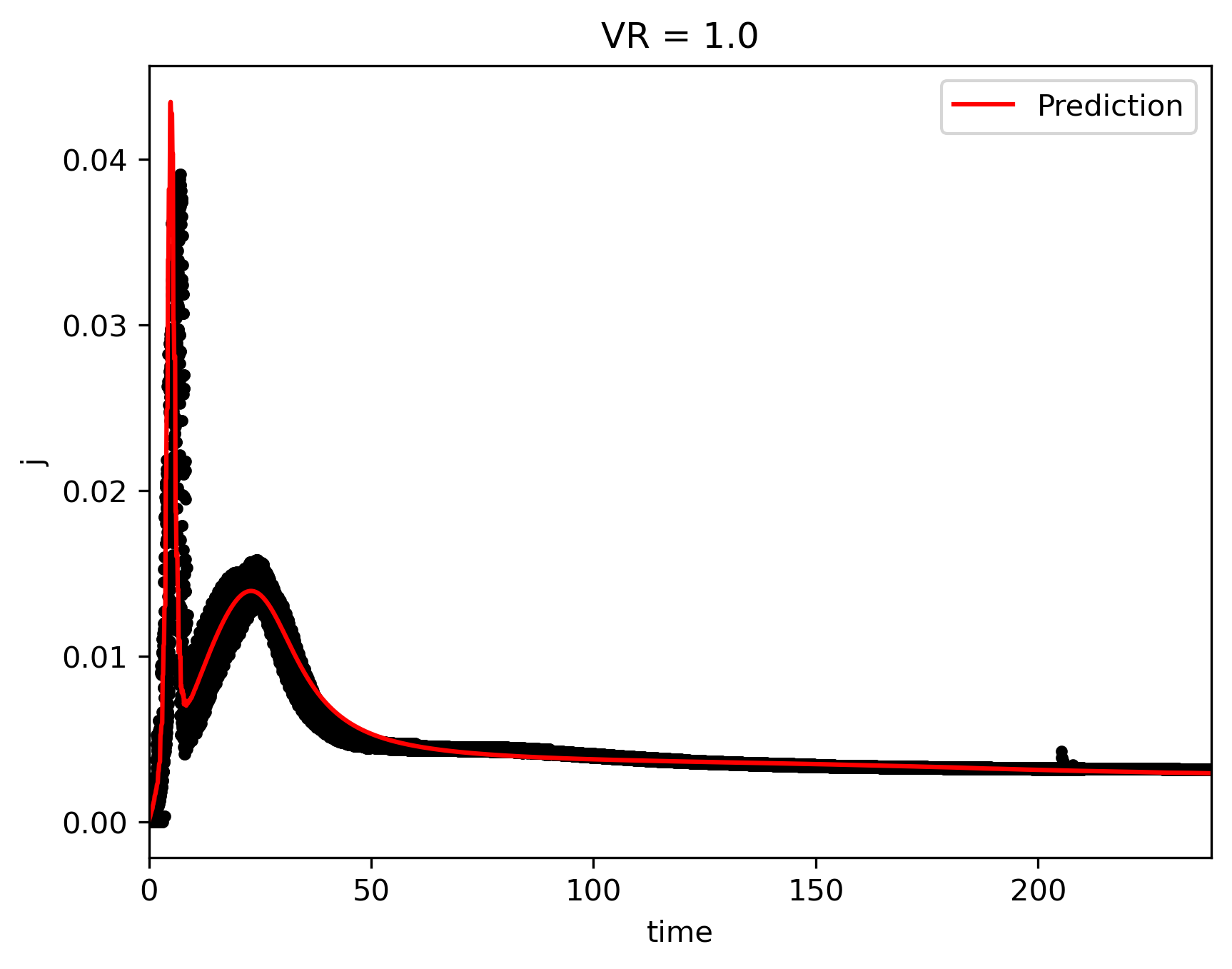}
        \caption{Voltage ramp experiment prediction, $V_R=1.0 V/s$}
        \label{fig:posterior_predictive_example_good}
    \end{subfigure}
    \hfill
    \begin{subfigure}[b]{0.42\textwidth}
        \includegraphics[width=\linewidth]{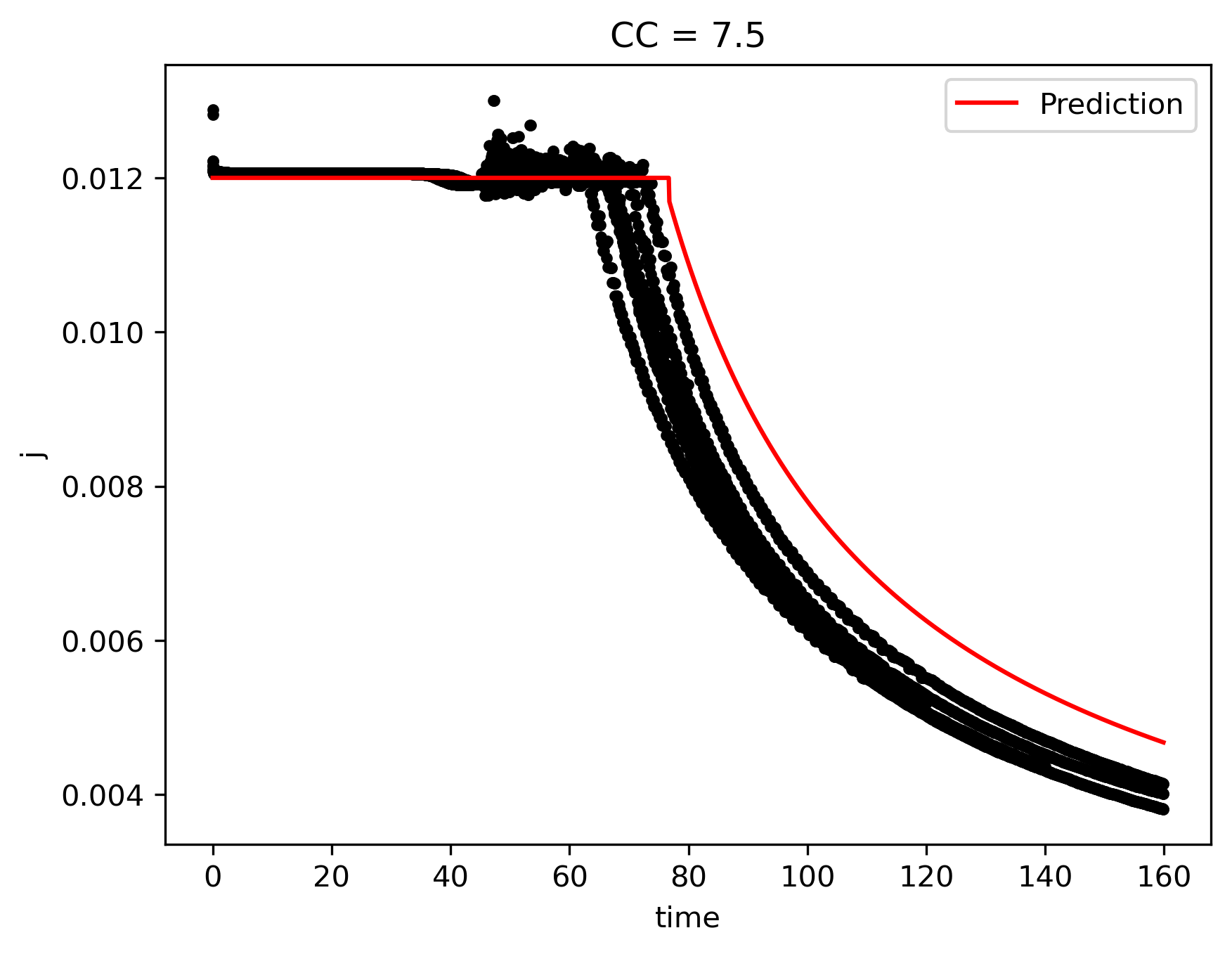}
        \caption{Constant current experiment prediction, $j_0=7.5mA$}
    \end{subfigure}
    \caption{Current prediction on the machine-learning augmented model trained with the first peak model.
    }
    \label{fig:first_peak}
\end{figure}

The augmented model is finally expressed by
\begin{align}\label{eq:augmented_VR}
    f_\theta(V,Q) = \tanh{(-\hat{f}_\theta(V,Q))} & , \quad g_\phi(V,Q) = \frac{1}{1+c_3\exp{(-c_4\hat{g}_\phi(V,Q))}} \\
    \frac{dR_{film}}{dt} = g_\phi(V)\rho(j)j C_v & , \quad j = c_1( \exp(c_2Vf_\theta(V))) +  \frac{\sigma V}{\sigma R_{film} + L} \; ,
\end{align}
for VR experiments and 
\begin{equation}\label{eq:augmented_VR}
    g_\phi(V,Q) = \frac{1}{1+c_3\exp{(-c_4\hat{g}_\phi(V,Q))}}, \quad \frac{dR_{film}}{dt} = g_\phi(V)\rho(j)j C_v \\,
\end{equation}
for CC experiments. Note that the relationship between voltage, current, and resistance are unchanged from the baseline model in the CC experiments even with the machine learning augmentations introduced to the model. The model augmentation functions $\hat{g}_\phi$ and $\hat{f}_\theta$ are parameterized by FNN's with 4 layers, 8 nodes per layer, and ReLU activation functions. These are trained using the NeuralODE framework using only current data from the 3 VR experiments. Thus, applying the model on the CC experiments is purely prediction as none of the data is seen during training. We learn the parameters $\theta, \phi, C_v$, and $\sigma$ in our experiments and show prediction results. Figure~\ref{fig:first_peak} illustrates the results of the final trained model, better capturing the `double-peaked' nature of the current in experimental data for voltage ramp experiments. Here we show current prediction for only two experimental configurations, but current and resistance predictions for all other experimental configurations are provided in~\ref{app:first_peak}. 

\begin{figure}[h!]
    \centering
    \begin{subfigure}[b]{0.32\textwidth}
        \includegraphics[width=\linewidth]{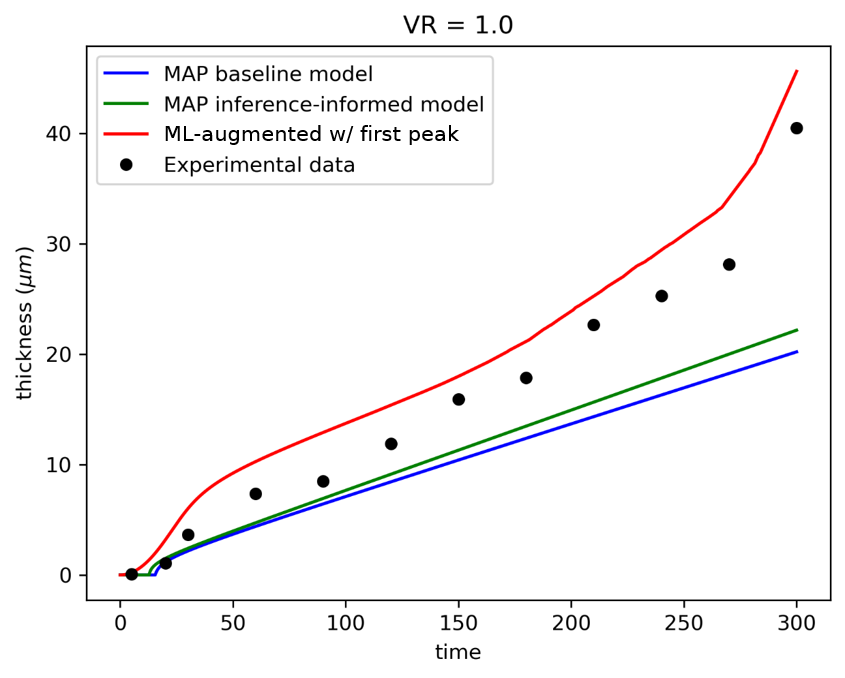}
        \caption{Voltage ramp experiment prediction, $V_R=1.0$ (film thickness)}
    \end{subfigure}
    \hfill    
    \begin{subfigure}[b]{0.32\textwidth}
        \includegraphics[width=\linewidth]{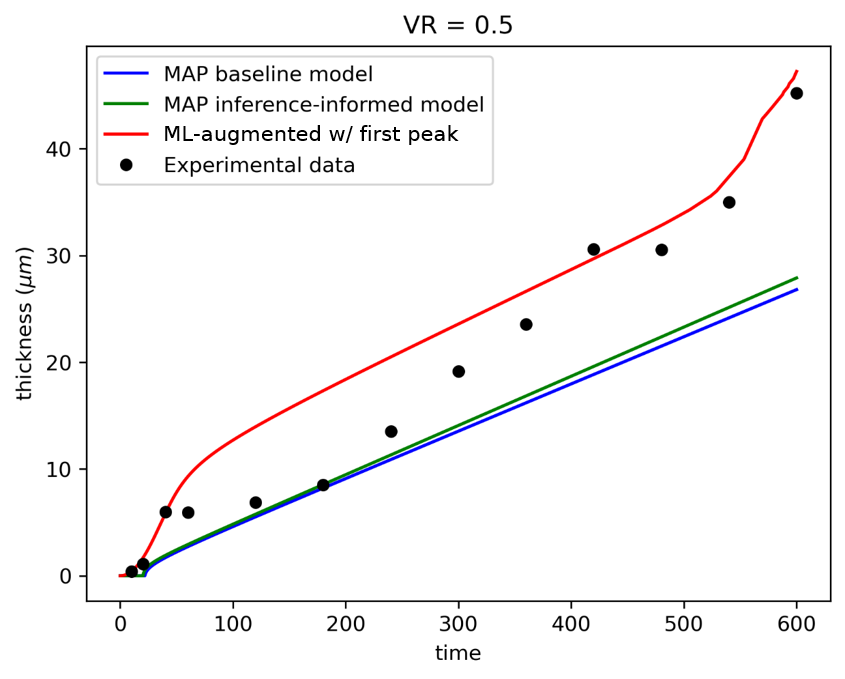}
        \caption{Voltage ramp experiment prediction, $V_R=0.5 V/s$ (film thickness)}
    \end{subfigure}
    \hfill
    \begin{subfigure}[b]{0.32\textwidth}
        \includegraphics[width=\linewidth]{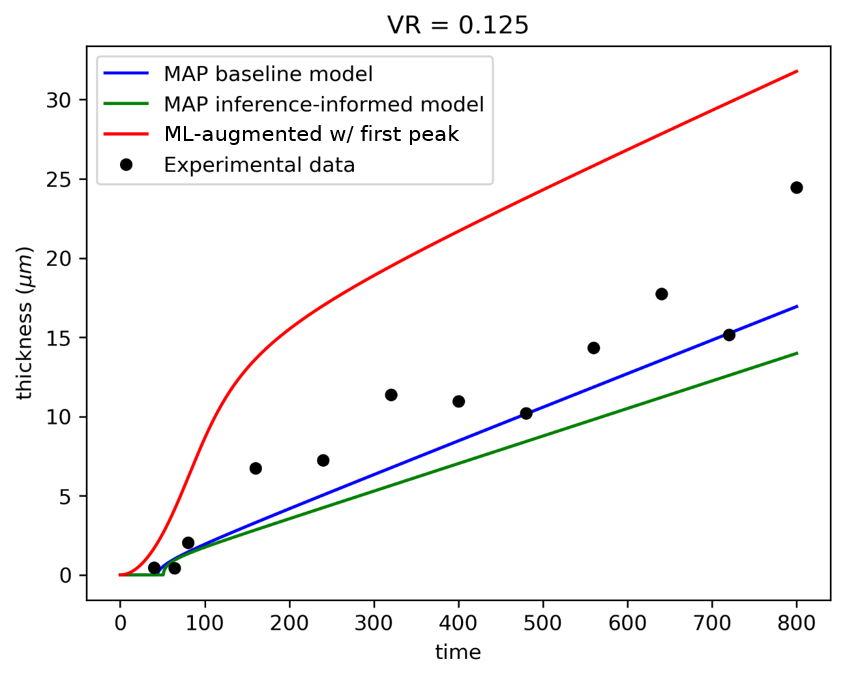}
        \caption{Voltage ramp experiment prediction, $V_R=0.125 V/s$ (film thickness)}
    \end{subfigure}
    \hfill
    \begin{subfigure}[b]{0.32\textwidth}
        \includegraphics[width=\linewidth]{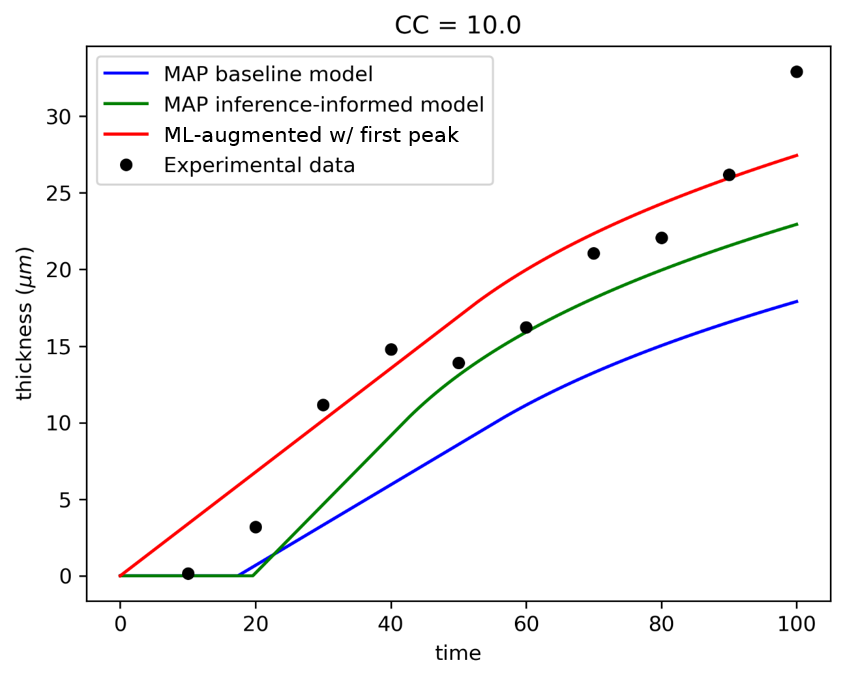}
        \caption{Constant resistance experiment prediction, $j_0=10.0mA$ (film thickness)}
    \end{subfigure}
    \hfill
    \begin{subfigure}[b]{0.32\textwidth}
        \includegraphics[width=\linewidth]{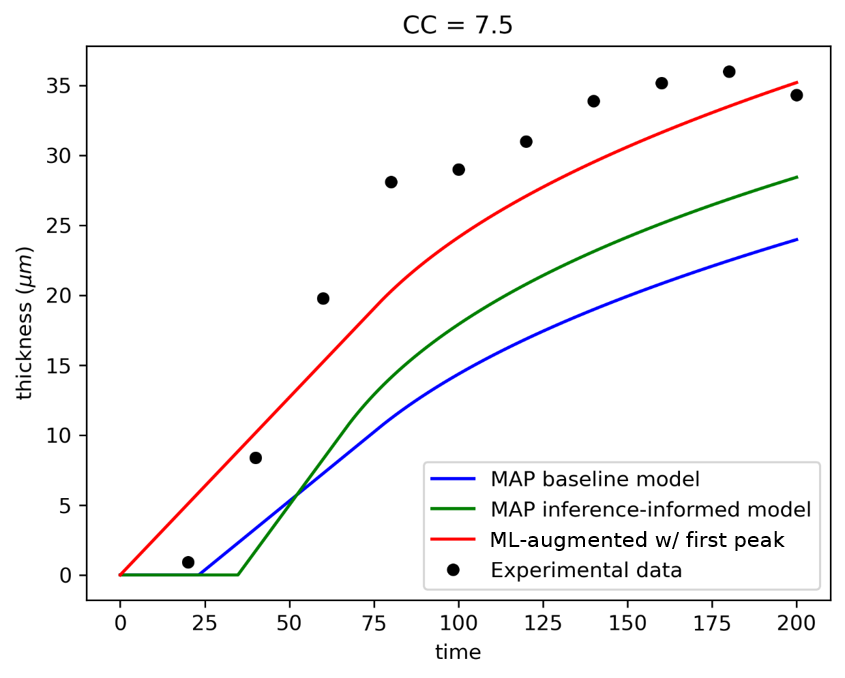}
        \caption{Constant current experiment prediction, $j_0=7.5mA$ (film thickness)}
    \end{subfigure}
    \hfill
    \begin{subfigure}[b]{0.32\textwidth}
        \includegraphics[width=\linewidth]{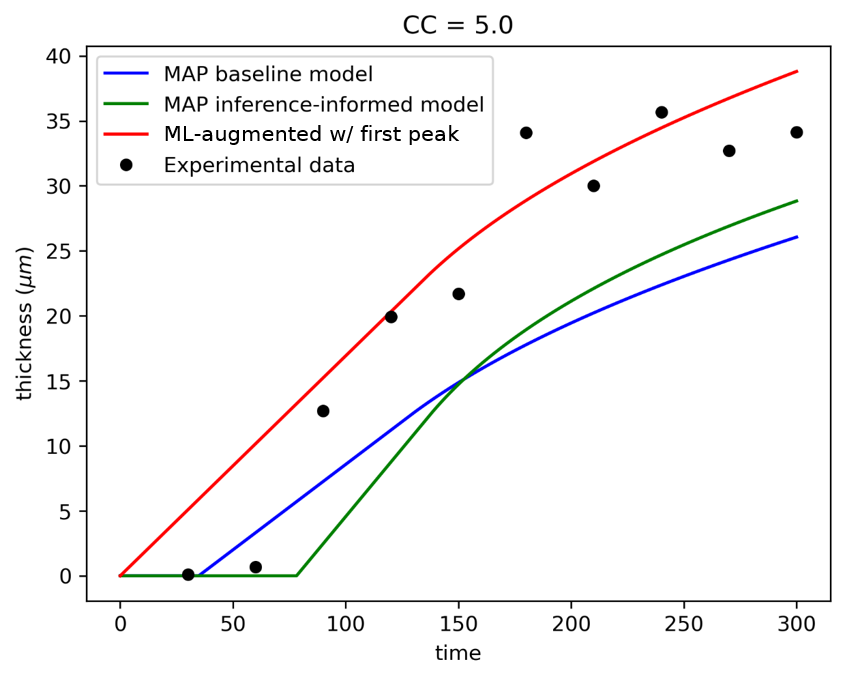}
        \caption{Constant current experiment prediction, $j_0=5.0mA$ (film thickness)}
    \end{subfigure}
    \caption{Comparisons between film thickness prediction on the baseline model, inference-informed model, and ML-augmented model with first peak.
    }
    \label{fig:thickness_first_peak}
\end{figure}

Further, thickness predictions for all experimental configurations are illustrated in Fig.~\ref{fig:thickness_first_peak} along with predictions from the baseline and inference-informed models for comparison. Film thickness predictions show an improvement over the baseline model in all cases except for a single experimental configuration: voltage ramp experiments with $V_R = 0.125$. We expect that this is due to some unmodeled behavior in the low voltage regime which is difficult to capture. However, in constant current experiments and larger voltage values in voltage ramp experiments, our augmentations provide significant improvement to model predictions. 


\begin{figure}[h!]
    \centering
    \begin{subfigure}[b]{0.42\textwidth}
        \includegraphics[width=\linewidth]{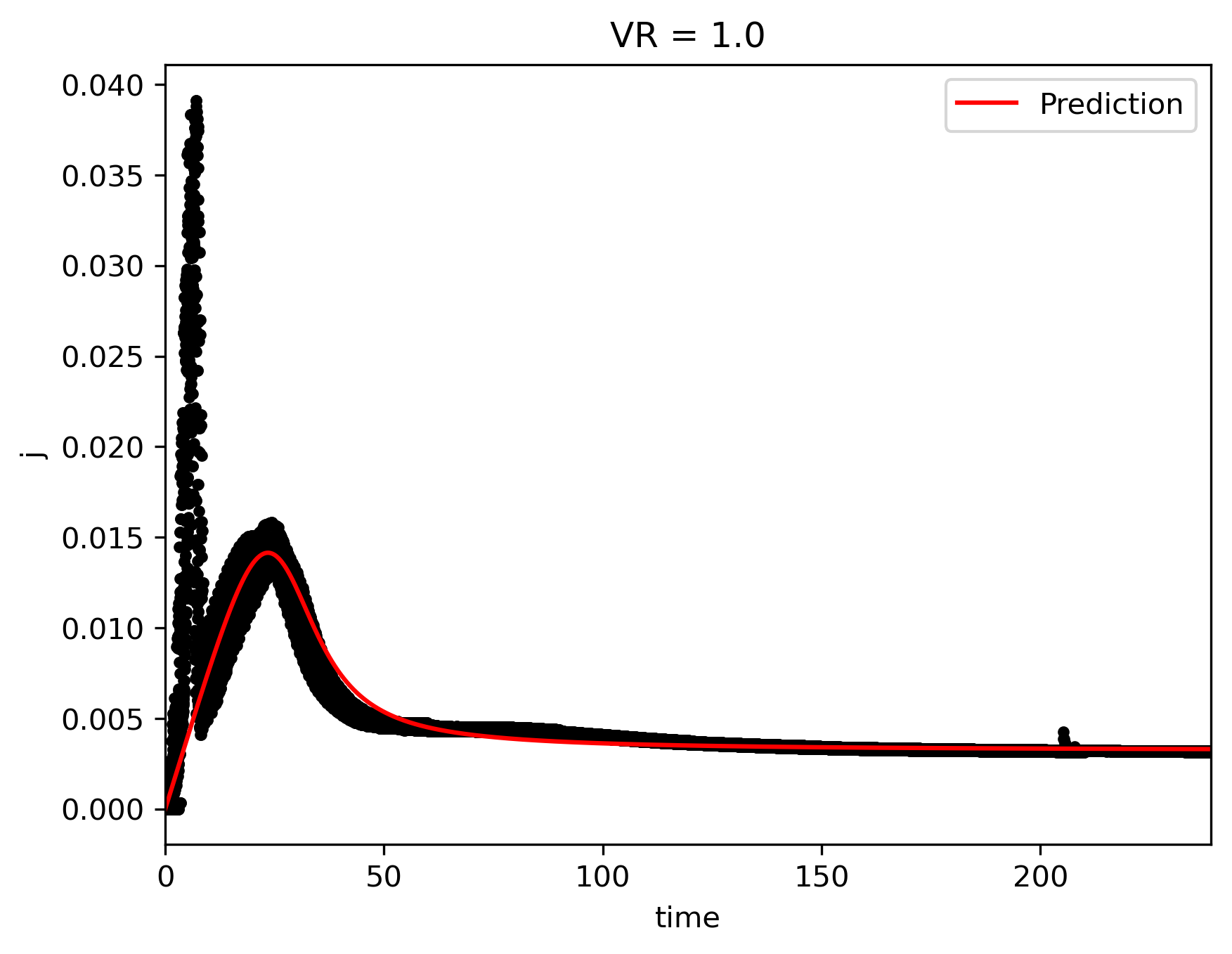}
        \caption{Voltage ramp experiment prediction, $V_R=1.0 V/s$}
        \label{fig:posterior_predictive_example_good}
    \end{subfigure}
    \hfill
    \begin{subfigure}[b]{0.42\textwidth}
        \includegraphics[width=\linewidth]{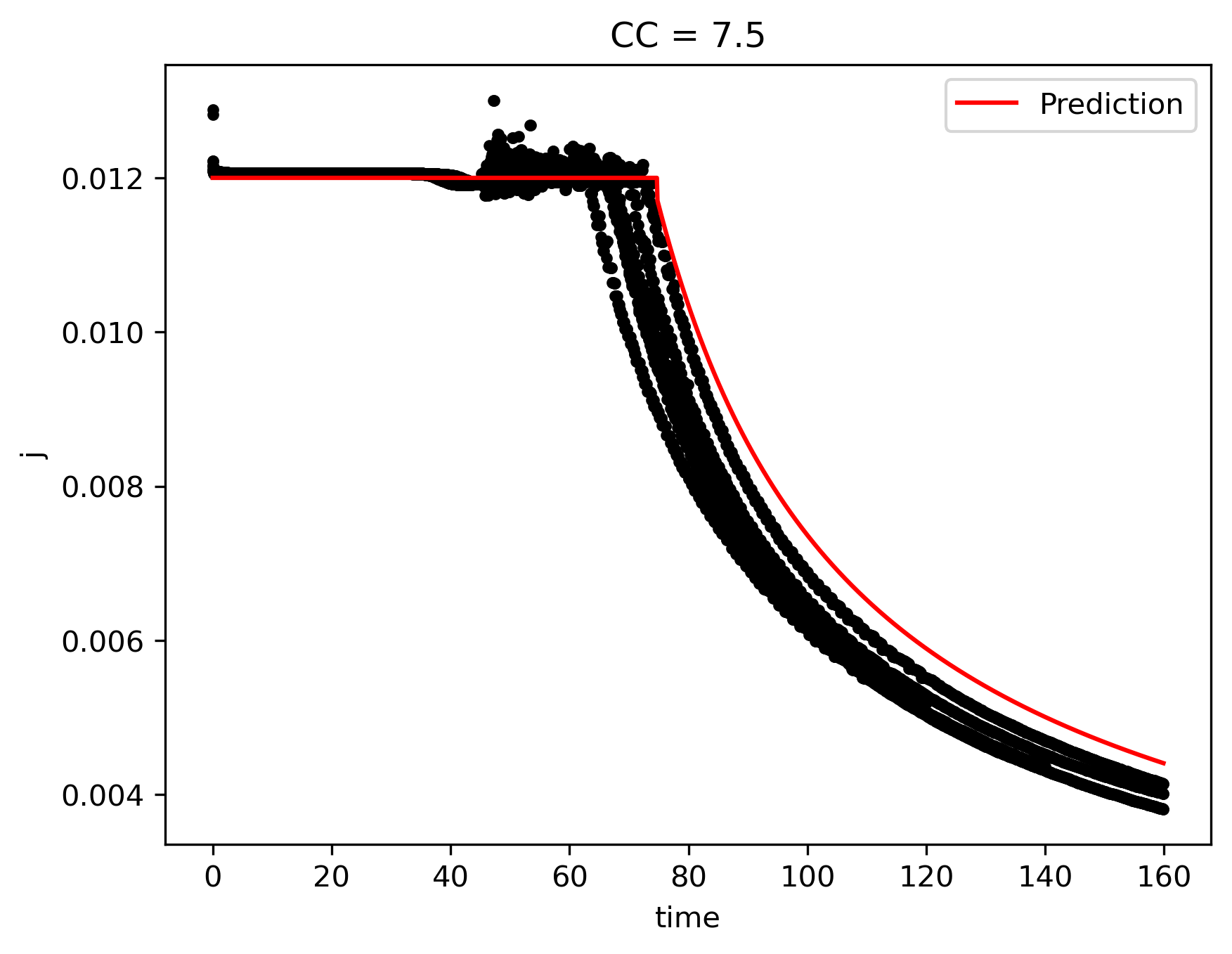}
        \caption{Constant current experiment prediction, $j_0=7.5mA$}
    \end{subfigure}
    \caption{Current prediction on the machine-learning augmented model trained without the first peak model, shown for (a) voltage ramp experiment with $V_R = 1.0 V/s$ and (b) constant current experiment with $j_0 = 7.5 mA$.}
    \label{fig:no_first_peak}
\end{figure}

\begin{figure}[h!]
    \centering
    \begin{subfigure}[b]{0.32\textwidth}
        \includegraphics[width=\linewidth]{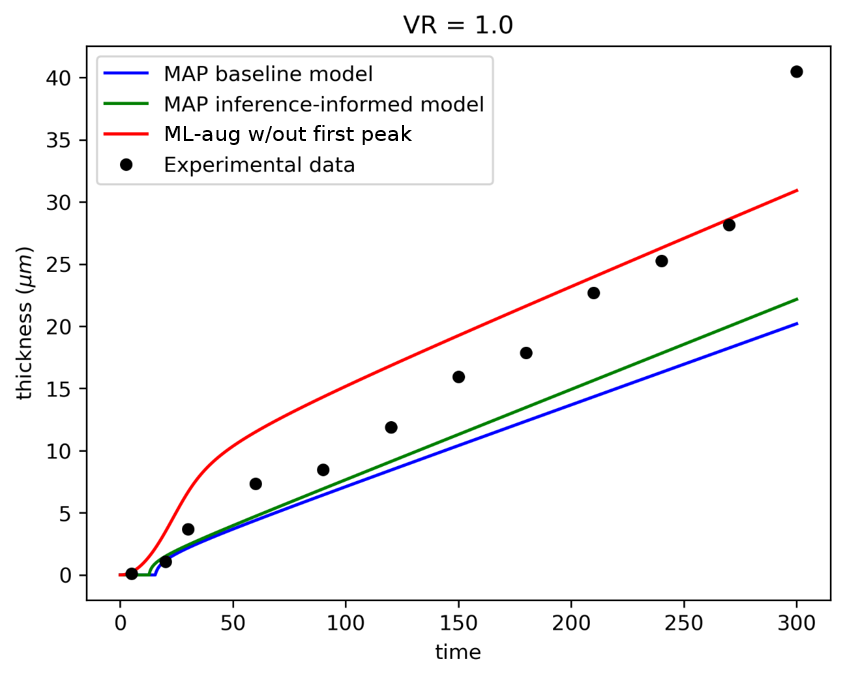}
        \caption{Voltage ramp experiment prediction, $V_R=1.0$ (film thickness)}
    \end{subfigure}
    \hfill    
    \begin{subfigure}[b]{0.32\textwidth}
        \includegraphics[width=\linewidth]{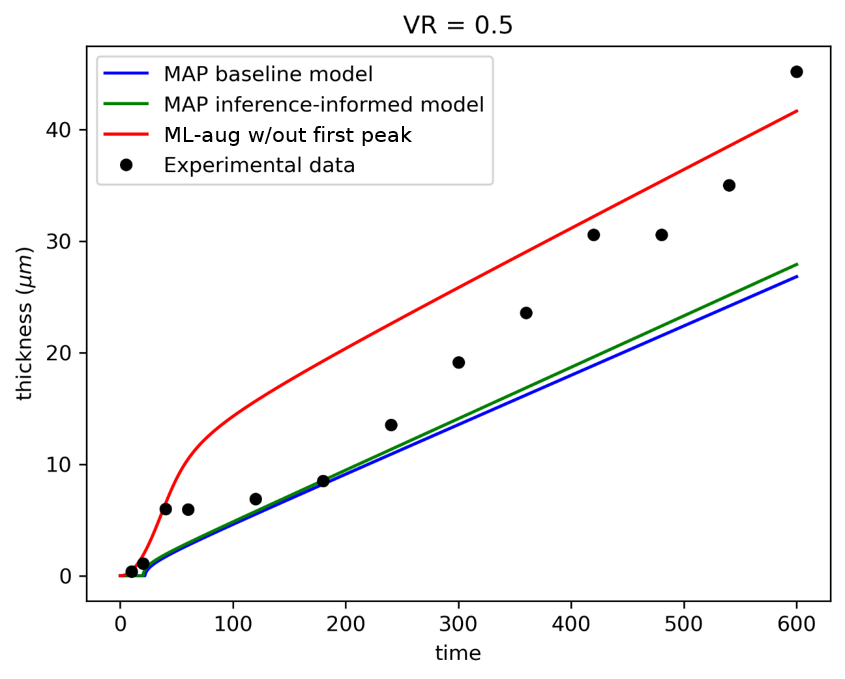}
        \caption{Voltage ramp experiment prediction, $V_R=0.5 V/s$ (film thickness)}
    \end{subfigure}
    \hfill
    \begin{subfigure}[b]{0.32\textwidth}
        \includegraphics[width=\linewidth]{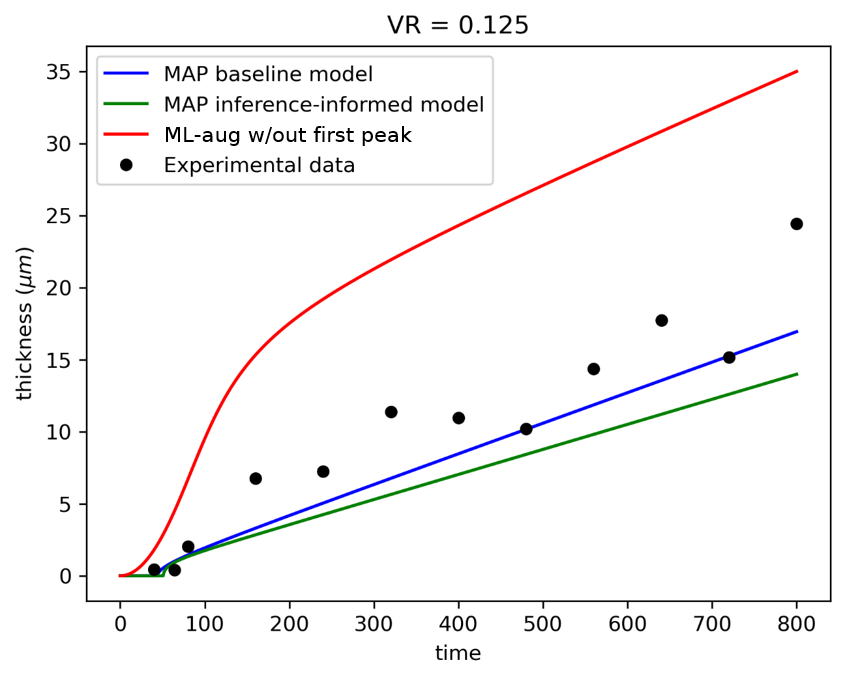}
        \caption{Voltage ramp experiment prediction, $V_R=0.125 V/s$ (film thickness)}
    \end{subfigure}
    \hfill
    \begin{subfigure}[b]{0.32\textwidth}
        \includegraphics[width=\linewidth]{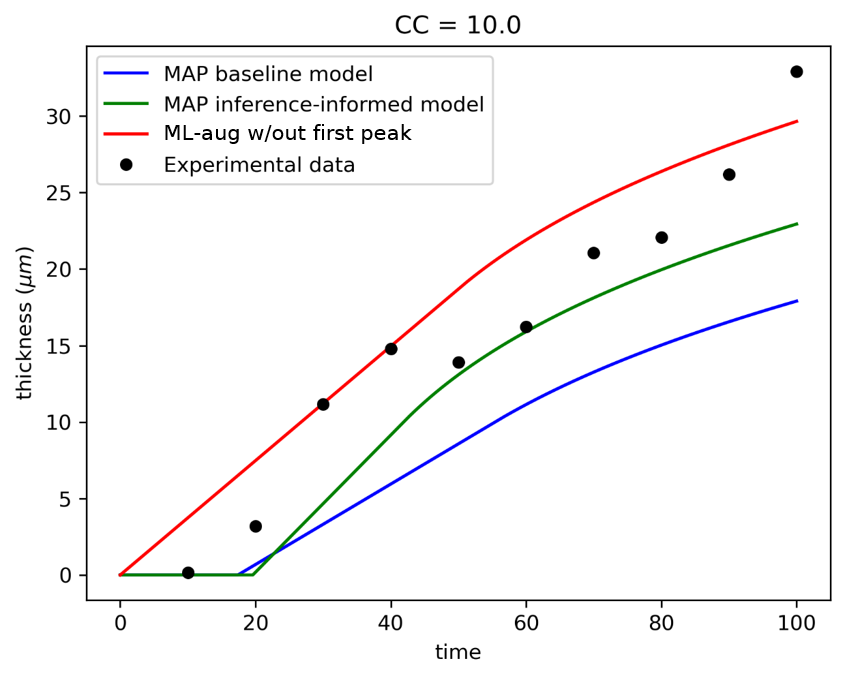}
        \caption{Constant resistance experiment prediction, $j_0=10.0mA$ (film thickness)}
    \end{subfigure}
    \hfill
    \begin{subfigure}[b]{0.32\textwidth}
        \includegraphics[width=\linewidth]{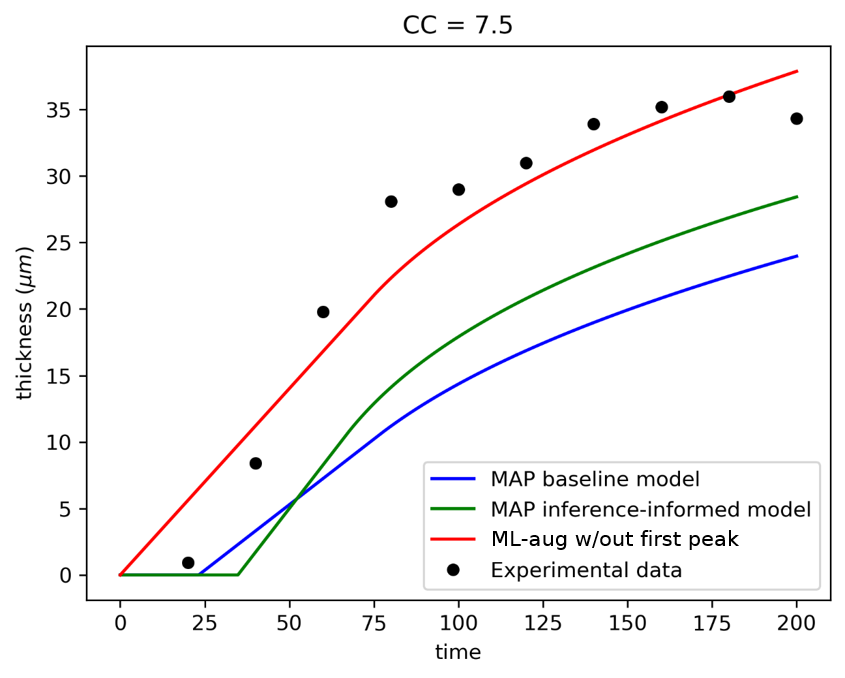}
        \caption{Constant current experiment prediction, $j_0=7.5mA$ (film thickness)}
    \end{subfigure}
    \hfill
    \begin{subfigure}[b]{0.32\textwidth}
        \includegraphics[width=\linewidth]{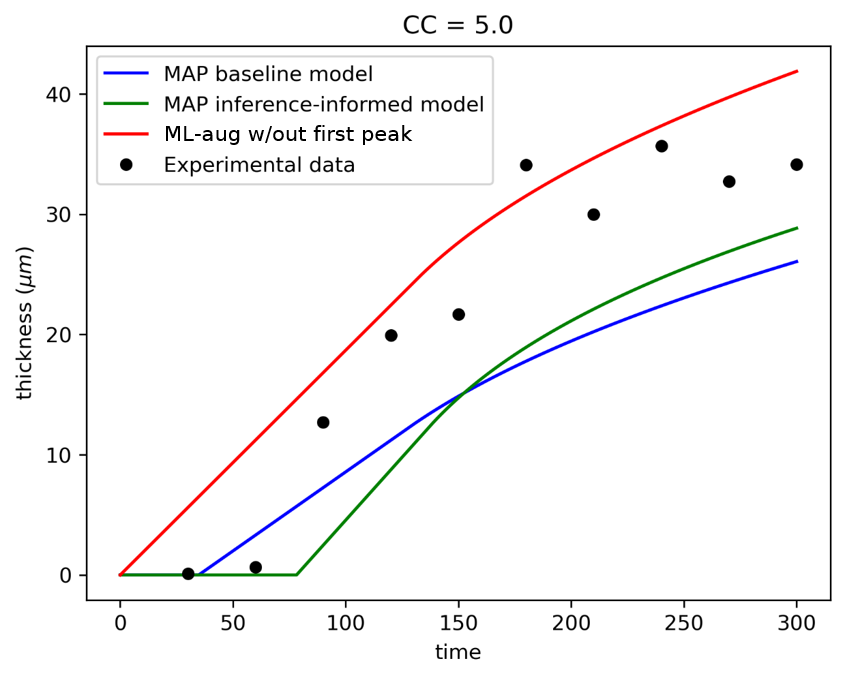}
        \caption{Constant current experiment prediction, $j_0=5.0mA$ (film thickness)}
    \end{subfigure}
    \caption{Comparisons between film thickness prediction on the baseline model, inference-informed model, and ML-augmented model without first peak.}
    \label{fig:thickness_no_first_peak}
\end{figure}

Learning the first peak model function $f_\theta$ in the NeuralODE framework is significantly more expensive computationally than training only the coverage fraction model $g_\phi$ due to the large gradient change requiring a finely-discretized time step to adequately resolve. Additionally, the first peak model performs notably poorly for the voltage ramp experiment in the low voltage ramp regime. We thus remove the `first peak' model and retrain only the augmentation function $g_\phi(V, Q)$ by masking out the experimental data corresponding to the first peak in current for voltage ramp experiments. This results in more efficient training and thickness prediction while providing quite similar predictions when the first peak model is included. These results show that modelling the dynamics of the first peak may unnecessarily decrease computational efficiency while still providing more accurate thickness prediction, which is ultimately the goal. Current predictions for two experiments (only voltage ramp and one constant current) are shown in Fig.~\ref{fig:no_first_peak}, and thickness predictions are shown and compared to all other models presented in Fig.~\ref{fig:thickness_no_first_peak}. These results again illustrate a significant improvement in thickness prediction without the need for including the more expensive first peak model, indicating that modeling such behavior may be largely unnecessary for predicting thickness. However, prediction in the low voltage ramp regime is actually worse than with the first peak model included, further supporting our hypothesis that there exist additional unmodeled dynamics in such cases. Again, all other predictions for current and resistance using our model augmentations without the first peak model included are provided in~\ref{app:no_first_peak}.

\section{Conclusions}\label{sec:conclusion} 
We presented a comprehensive framework to introduce adaptable yet interpretable machine learning-based model augmentations. Our initial objective of parameter inference unveiled significant limitations in the existing baseline model, necessitating the integration of diverse model refinements aimed at rectifying these deficiencies.

Even with such enhancements, however, certain physical behaviors inherent in experimental data were not fully replicated by the models. Notably, instances such as the emergence of two peaks in current data during voltage ramp experiments and the smooth evolution of current data remained elusive. To address these disparities, we devised physically meaningful augmentations, seamlessly integrated them into the model, and harnessed a NeuralODE framework for implementing learnability. This approach to augmentation construction maintains interpretability while harnessing the expressive power of neural networks to allow enhanced adaptability.

The integration of these augmentations yields models which demonstrate improved predictive accuracy when compared against empirical data. Nonetheless, we recognize that modeling the 'first peak' entails considerable training costs, and our findings reveal that, despite its omission, accurate film thickness predictions can still be achieved with a notable reduction in offline computational burdens.

Integral to our methodology is an in-depth procedure of evaluating the model's behavior prior to augmentation. This diagnostic analysis is pivotal, as it grants insights into the limitations of the model. By first grasping these limitations, we gather information on how to design augmentations that retain physical interpretability, while simultaneously fostering flexibility and removing the necessity for manual crafting and fine-tuning of model structures. This practice holds immense significance not only for machine learning-driven dynamical system understanding but also for broader applications in physics-based domains. It facilitates augmentation strategies that preserve interpretability, thereby opening avenues for model improvements.

 We acknowledge that other problem settings may introduce different intricacies; nevertheless, the systematic process outlined in this study remains invaluable for guiding future applications in different problems. This study offers a variety of insights, equipping researchers with tools to assess and elevate the forms of model representations for diverse dynamical systems. Ultimately, our work fosters a bridge between machine learning and physics, ensuring not only advancements in the modeling process but also revealing the underlying system dynamics.

\section*{Acknowledgements}
This work was supported at the University of Michigan by Ford Motor Company under the grant "Hybrid Physics-Machine Learning models for Electro-deposition"

\clearpage

\appendix
\section{Voltage ramp}
For the electric field, we solve a Poisson equation with Robin boundary condition on the interface bath/film and Dirichlet condition at the anode.
\begin{align}
\sigma_{\rm{bath}}\frac{\partial^2\phi}{\partial x^2}&=0 &\textrm{ in the bath}\\
\phi-R_{\textrm{film}} \sigma_{\rm{bath}}\frac{\partial \phi}{\partial x}&=0 &\textrm{ at the interface film-bath}\\
\phi_{\rm{anode}}(t) &= \phi_{t=0} + \phi_{\rm{ramp}}(t) & \textrm{ at the anode}
\end{align}
where $\sigma_{\rm{bath}}$ is the bath conductivity, $j$ is the normal component of the current density, $\phi$ is the electrical potential, $R_{\textrm{film}}$ is the film resistance and $h$ is the film thickness.

\section{Constant current}
For the electric field, we solve a Poisson equation with Robin boundary condition on the interface bath/film and Neumann condition at the anode.

\begin{align}
\sigma_{\rm{bath}}\frac{\partial^2\phi}{\partial x^2}&=0 &\textrm{ in the bath}\\
\phi-R_{\textrm{film}} \sigma_{\rm{bath}}\frac{\partial \phi}{\partial x}&=0 &\textrm{ at the interface film-bath}\\
\sigma_{\rm{bath}}\frac{\partial \phi}{\partial x}&= j_0& \textrm{ at the anode}
\end{align}

\section{Evolution of the concentration of OH$^-$}\label{app:K}

\subsection{Solution of the diffusion equation}
We consider the diffusion equation in the domain [0;L]
\begin{equation}
\frac{\partial u}{\partial t}=D \frac{\partial^2 u}{\partial x}
\end{equation}
with the following boundary conditions:
\begin{equation}
\left.\frac{\partial u}{\partial x}\right|_{x=0}=-\frac{j(t)}{DF}=g(t) \quad \textrm{ and } \quad u(x=L,t)=h(t)
\end{equation}
where u is the OH$^-$ concentration, $F$ is the Faraday constant, $D$ is the diffusion coefficient, $h$ and $g$ are time-dependent functions.
The initial condition is defined by $u(x, 0)= u_0$, $u_0$ being the initial concentration.

Applying the Laplace transform
\begin{align}
\mathcal{L}\left\{\frac{\partial u}{\partial t}-D \frac{\partial^2 u}{\partial x}\right\}
= \mathcal{L}\left\{\frac{\partial u}{\partial t}\right\}-D \left\{\frac{\partial^2 u}{\partial x}\right\}
=& s \ub-u_0 - D\ub_{xx}=0
\end{align}

The solution of the homogeneous equation $s \ub - D\ub_{xx}=0$ is given by
\begin{equation}
\ub_h(x,s)=C_1 \exp(r_1 x)+C_2 \exp(r_2 x)
\end{equation}
where $r_1=\sqrt{s/D}$ and $r_2=-\sqrt{s/D}$

The specific solution, assuming a constant $\ub_{sp.}=A$ is given by $\ub_{sp.}=u_0/s$. Therefore, the solution has the following form
\begin{equation}
\ub(x,s) = C_1 \exp(r_1 x)+C_2 \exp(r_2 x) +\frac{u_0}{s}
\end{equation}
where $C_1$ and $C_2$ are functions derived using the boundary and initial conditions.


Using the Neumann boundary condition at x=0 and the definition $\mathcal{L}\left\{g(t)\right\}= G(s)$, then
\begin{equation}
\ub_x(x=0,s)=G(s)= C_1 r_1 + C_2 r_2.
\end{equation}

Therefore $C_1=\dfrac{G(s)-C_2 r_2}{r_1}=\dfrac{G(s)}{r_1}+C_2$.

Hence
\begin{equation}
\ub(x,s) = \left(\frac{G(s)}{r_1}+C_2\right) \exp(r_1 x)+C_2 \exp(r_2 x) +\frac{u_0}{s}
\end{equation}

Using the Dirichlet boundary condition at x=L:
\begin{equation}
\ub(x=L,s)=H(s)=\left(\dfrac{G(s)}{r_1}+C_2\right) \exp(r_1 L)+C_2 \exp(r_2 L) +\dfrac{u_0}{s},
\end{equation}
the solution is given by
\begin{equation}
\ub(x,s)=\left[\dfrac{G(s)}{r_1}+\dfrac{H(s)-\dfrac{u_0}{s}-\dfrac{G(s)}{r_1}\exp(r_1L)}{\exp(r_1L)+\exp(r_2L)}\right]\exp(r_1x)
+\left[\dfrac{H(s)-\dfrac{u_0}{s}-\dfrac{G(s)}{r_1}\exp(r_1L)}{\exp(r_1L)+\exp(r_2L)}\right]\exp(r_2x)
+\frac{u_0}{s}
\end{equation}

If $L\rightarrow\infty$ and $h(t)=u_0$ (i.e. $H(s)=u_0/s$), the solution simplifies to
\begin{align}
\ub(x,s)
=&\left[\dfrac{G(s)}{r_1}-\dfrac{G(s)}{r_1}\right]\exp(r_1x)
+\left[-\dfrac{G(s)}{r_1}\right]\exp(r_2x)
+\frac{u_0}{s}\\
=&-\dfrac{G(s)}{r_1}\exp(r_2x)+\frac{u_0}{s}
\end{align}

Using the inverse Laplace transform yields
\begin{align}
\mathcal{L}^{-1}\left\{\ub(x,s)\right\}
=&\mathcal{L}^{-1}\left\{-G(s)\sqrt{\frac{D}{s}}\exp\left(-\sqrt{\frac{s}{D}}x\right)+\frac{u_0}{s}\right\}\\
=&-\sqrt{D}\mathcal{L}^{-1}\left\{G(s)\frac{1}{\sqrt{s}}\exp\left(-\frac{x}{\sqrt{D}}\sqrt{s}\right)\right\}+u_0.
\end{align}

Knowing that
\begin{align}
\mathcal{L}^{-1}\left\{\frac{1}{\sqrt{s}}\exp\left(-\frac{x}{\sqrt{D}}\sqrt{s}\right)\right\}
=&\frac{1}{\sqrt{\pi t}}\exp\left(-\frac{x^2}{4Dt}\right)
\end{align}
and
\begin{align}
\mathcal{L}^{-1}\left\{G(s)K(s)\right\}=\int_{0}^{t}g(\tau) k(t-\tau)d\tau,
\end{align}
%
%
the concentration as a function of space and time is given by
\begin{align}
u(x,t) =u_0+\frac{1}{F\sqrt{\pi D}}\int_{0}^{t}j(\tau)\frac{1}{\sqrt{t-\tau}}\exp\left(-\frac{x^2}{4D(t-\tau)}\right)d\tau.
\end{align}

The concentration at the cathode (x=0) is given by
\begin{equation}
u(0,t)=u_0+\frac{1}{F\sqrt{\pi D}}\int_{0}^{t}j(\tau)\frac{1}{\sqrt{t-\tau}}d\tau.	
\end{equation}

Let $u_{\min}$ be the minimum concentration required to start deposition and $\xi$ be the time when the minimum concentration is reached. Therefore
\begin{align}
u_{\min}=u_0 + \frac{1}{F\sqrt{\pi D}}\int_{0}^{\xi}j(\tau)\frac{1}{\sqrt{\xi-\tau}}d\tau.
\end{align}

Let $K=\frac{1}{2}\left( u_{\min}-u_0 \right) F\sqrt{\pi D}$ be a constant characterizing the deposition onset.

%
%

\subsubsection{Constant current density}
For a constant current density $j$, the concentration is defined as
\begin{equation}
u(0,t)=u_0+\frac{2 j_0}{F\sqrt{\pi D}} \sqrt{t}
\end{equation}
and the electric charge is defined as
\begin{equation}
Q(t)=\int_{0}^{t}jdt=j_0t.
\end{equation}

At the deposition onset, this yields the well-known Sand's equation \cite{SandEquation}:
\begin{equation}
\frac{1}{2}\left( u_{\min}-u_0 \right) F\sqrt{\pi D}=j_0 \sqrt{\xi}.
\end{equation}

The minimum charge can be derived accordingly:
\begin{equation}
Q_{\min}=j_0 \xi=\frac{K^2}{j_0}
\end{equation}

\subsubsection{Linear voltage ramp}
If the voltage ramp is a linear function of time. Before the deposition, due to the Ohm's law, the current density is a linear function of time $j = \beta t+j_0$ and the concentration is expressed as
\begin{align}
u(0,t)
=&u_0 + \frac{1}{F\sqrt{\pi D}}\int_{0}^{t}j(\tau)\frac{1}{\sqrt{t-\tau}}d\tau\\
=&u_0 + \frac{1}{F\sqrt{\pi D}}\left[2j_0\sqrt{t}+2\int_{0}^{t}\frac{\partial j}{\partial \tau}\sqrt{t-\tau}d\tau\right]\\
=&u_0 + \frac{1}{F\sqrt{\pi D}}\left[2j_0\sqrt{t}+\frac{4}{3}\beta t^{3/2}\right].
\end{align}

At the deposition onset, the induction time is the solution of the following equation
%
\begin{equation}
\beta \xi^{3/2}  + \frac{3}{2}j_0\xi^{1/2}=\frac{3}{2}K.
\end{equation}


Assuming that the voltage at the anode is initially 0, we can derive the following expression for the induction time
\begin{equation}
\xi = \beta^{-2/3}\left[\frac{3}{2} K \right]^{2/3} 
\end{equation}


Finally, the minimum electric charge for the linear voltage ramp is
\begin{align}
Q_{\min}=&\left(\frac{81}{128\beta}\right)^{1/3} K^{4/3}.
\end{align}

\section{Baseline / Inference-informed model comparisons}\label{app:MAP}
\begin{figure}[h!]
    \centering
    \begin{subfigure}[b]{0.32\textwidth}
        \includegraphics[width=\linewidth]{TikzPictures/MAP_model_comparisons_current_VR1.0.png}
        \caption{Voltage ramp experiment prediction, $V_R=1.0 V/s$ (current)}
    \end{subfigure}
    \hfill
    \begin{subfigure}[b]{0.32\textwidth}
        \includegraphics[width=\linewidth]{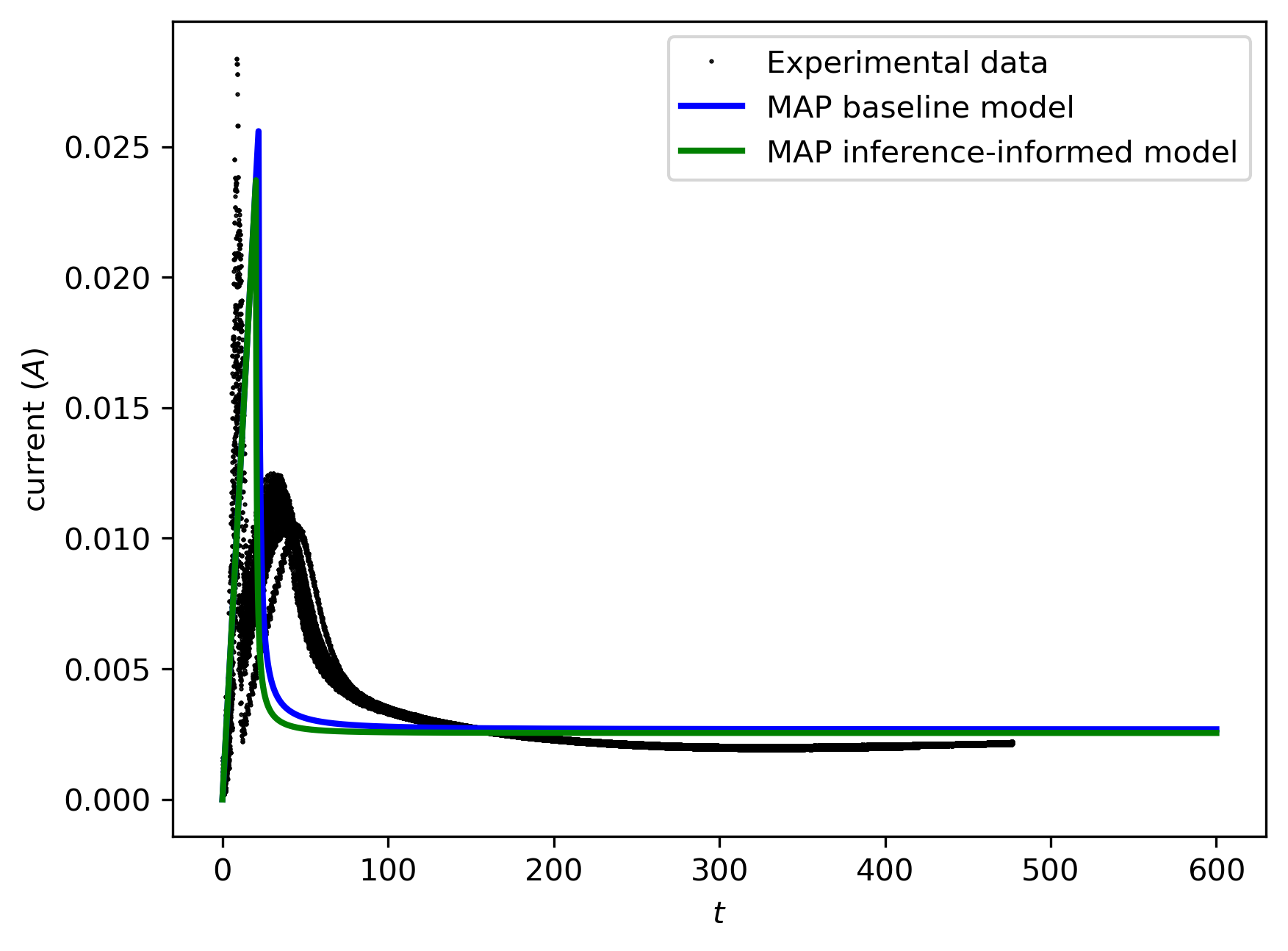}
        \caption{Voltage ramp experiment prediction, $V_R=0.5 V/s$ (current)}
    \end{subfigure}
    \hfill
    \begin{subfigure}[b]{0.32\textwidth}
        \includegraphics[width=\linewidth]{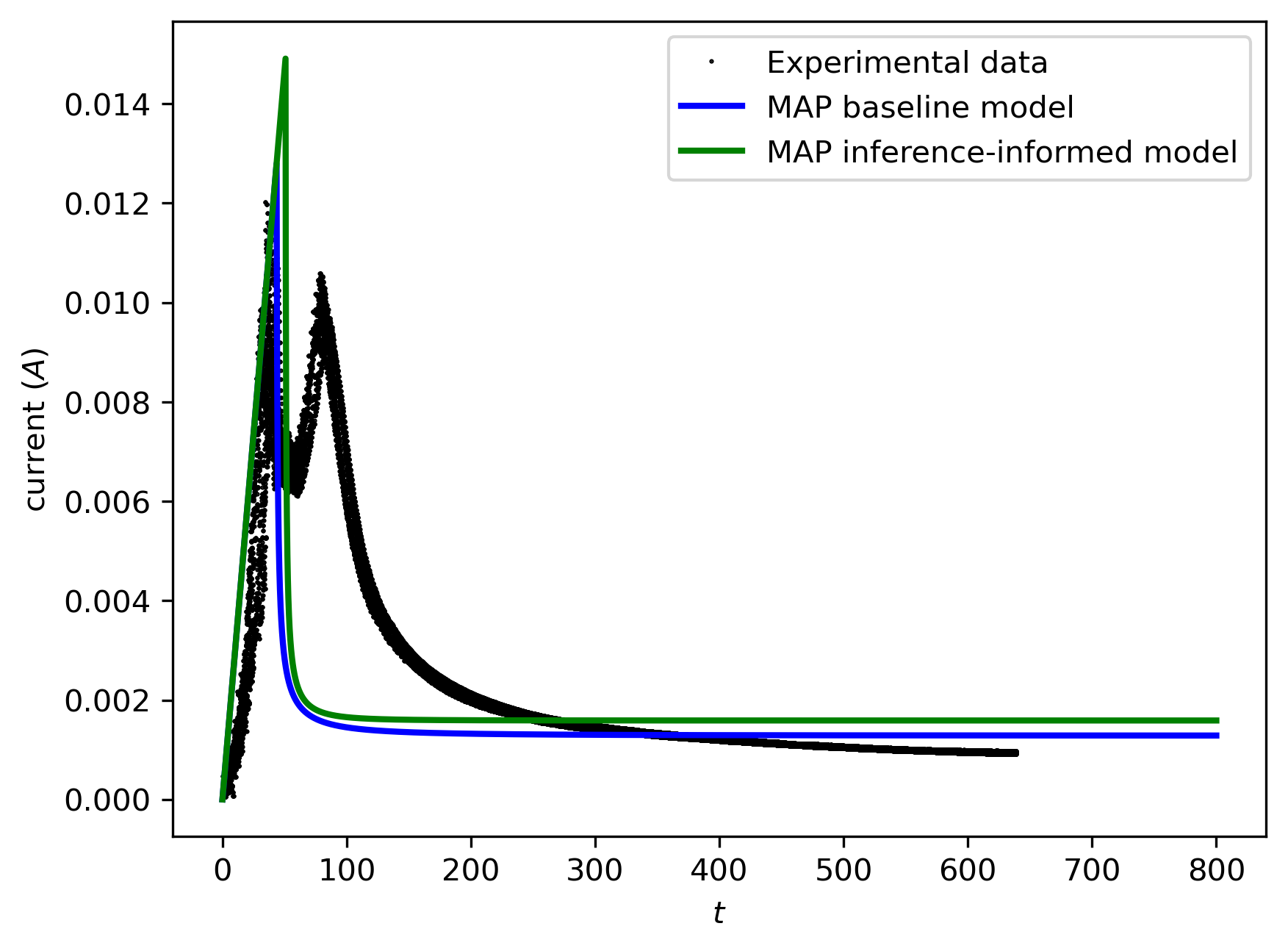}
        \caption{Voltage ramp experiment prediction, $V_R=0.125 V/s$ (current)}
    \end{subfigure}
    \hfill
    \begin{subfigure}[b]{0.32\textwidth}
        \includegraphics[width=\linewidth]{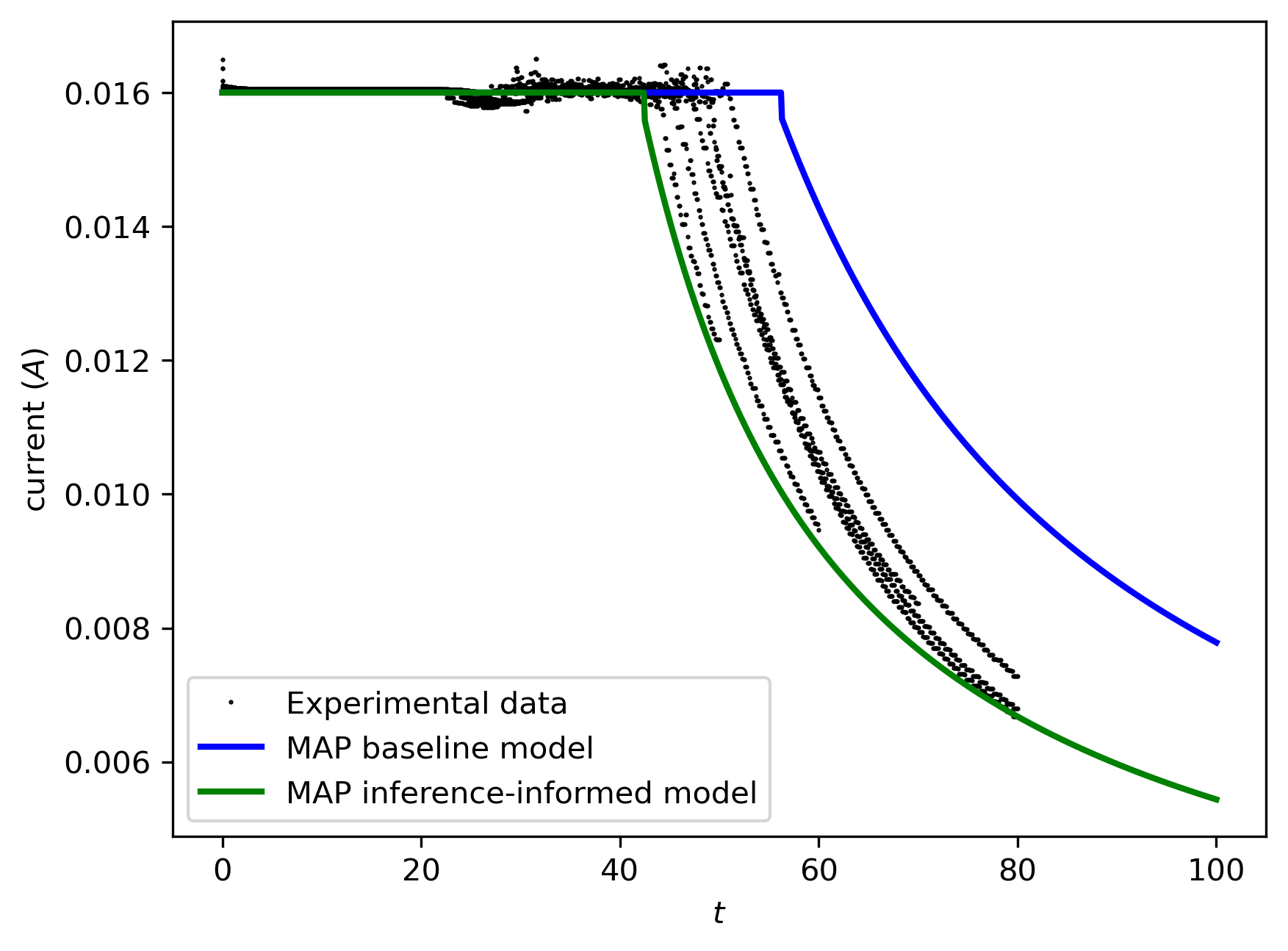}
        \caption{Constant current experiment prediction, $j_0=10.0mA$ (current)}
    \end{subfigure}
    \hfill
    \begin{subfigure}[b]{0.32\textwidth}
        \includegraphics[width=\linewidth]{TikzPictures/MAP_model_comparisons_current_CC7.5.png}
        \caption{Constant current experiment prediction, $j_0=7.5mA$ (current)}
    \end{subfigure}
    \hfill
    \begin{subfigure}[b]{0.32\textwidth}
        \includegraphics[width=\linewidth]{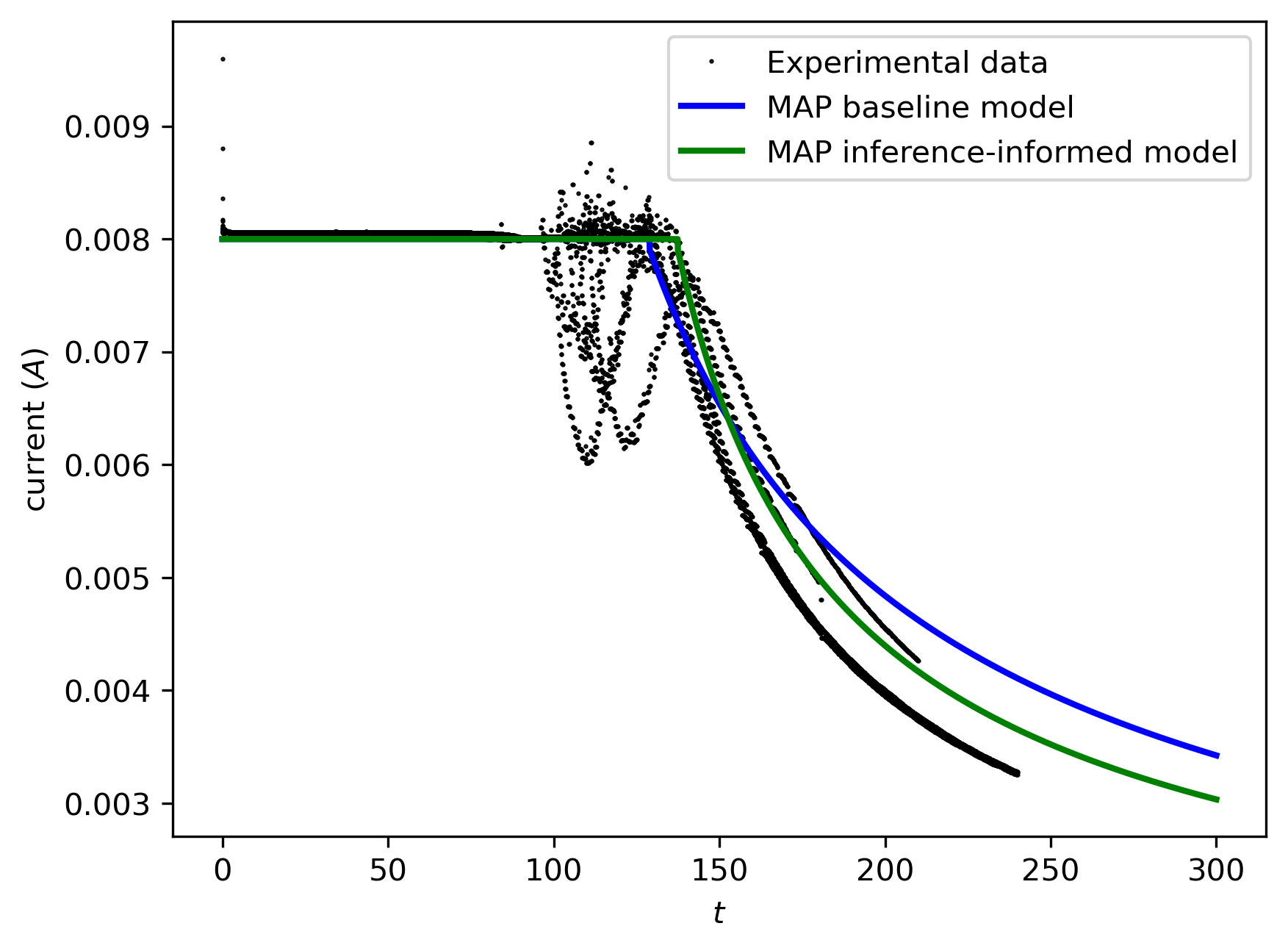}
        \caption{Constant current experiment prediction, $j_0=5.0mA$ (current)}
    \end{subfigure}
    \caption{Comparisons between current prediction on the baseline model and inference-informed model at the MAP for each.}
    \label{fig:app_map_current}
\end{figure}

\begin{figure}[h!]
    \centering
    \begin{subfigure}[b]{0.32\textwidth}
        \includegraphics[width=\linewidth]{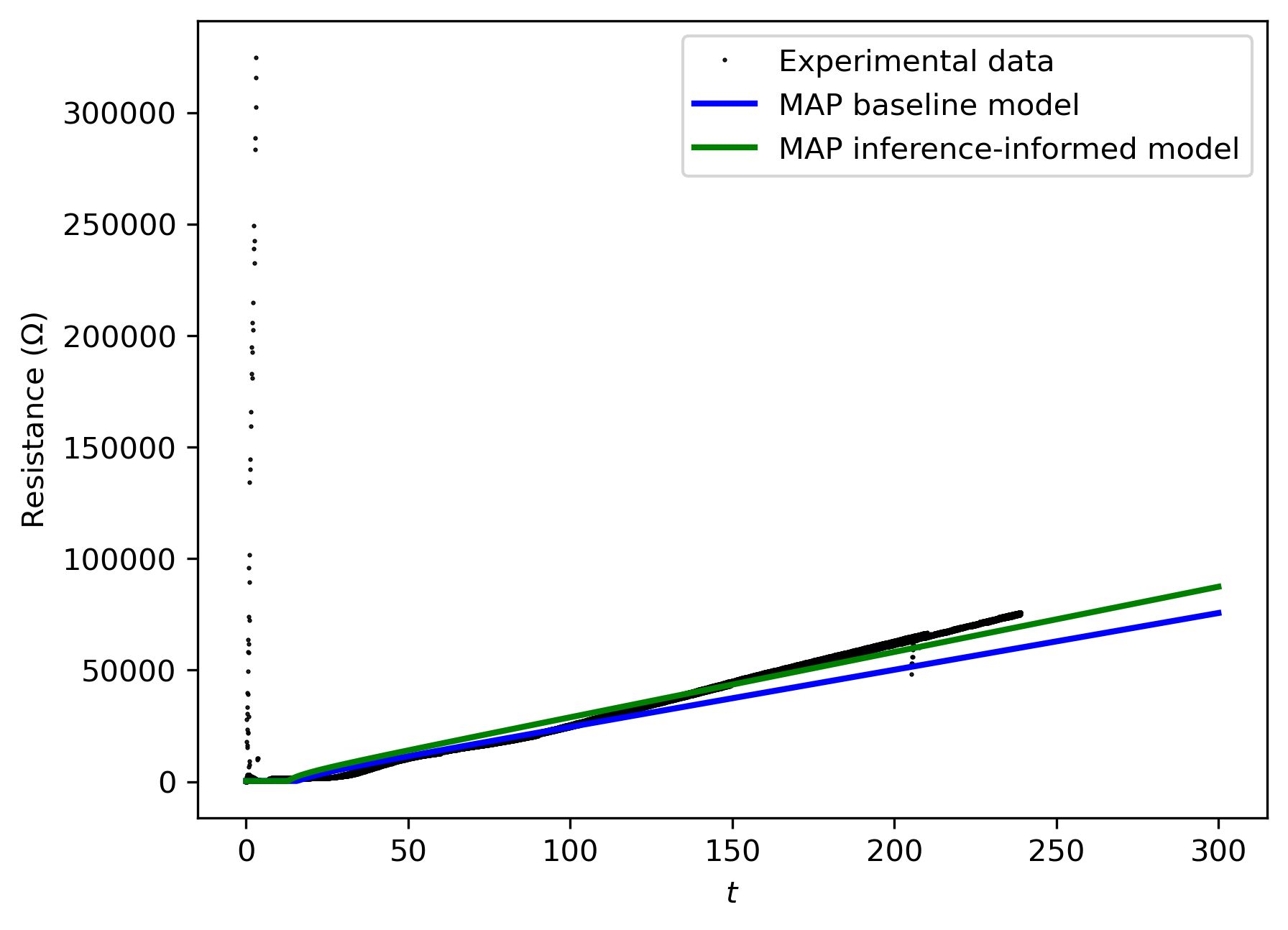}
        \caption{Voltage ramp experiment prediction, $V_R=1.0$ (resistance)}
    \end{subfigure}
    \hfill
    \begin{subfigure}[b]{0.32\textwidth}
        \includegraphics[width=\linewidth]{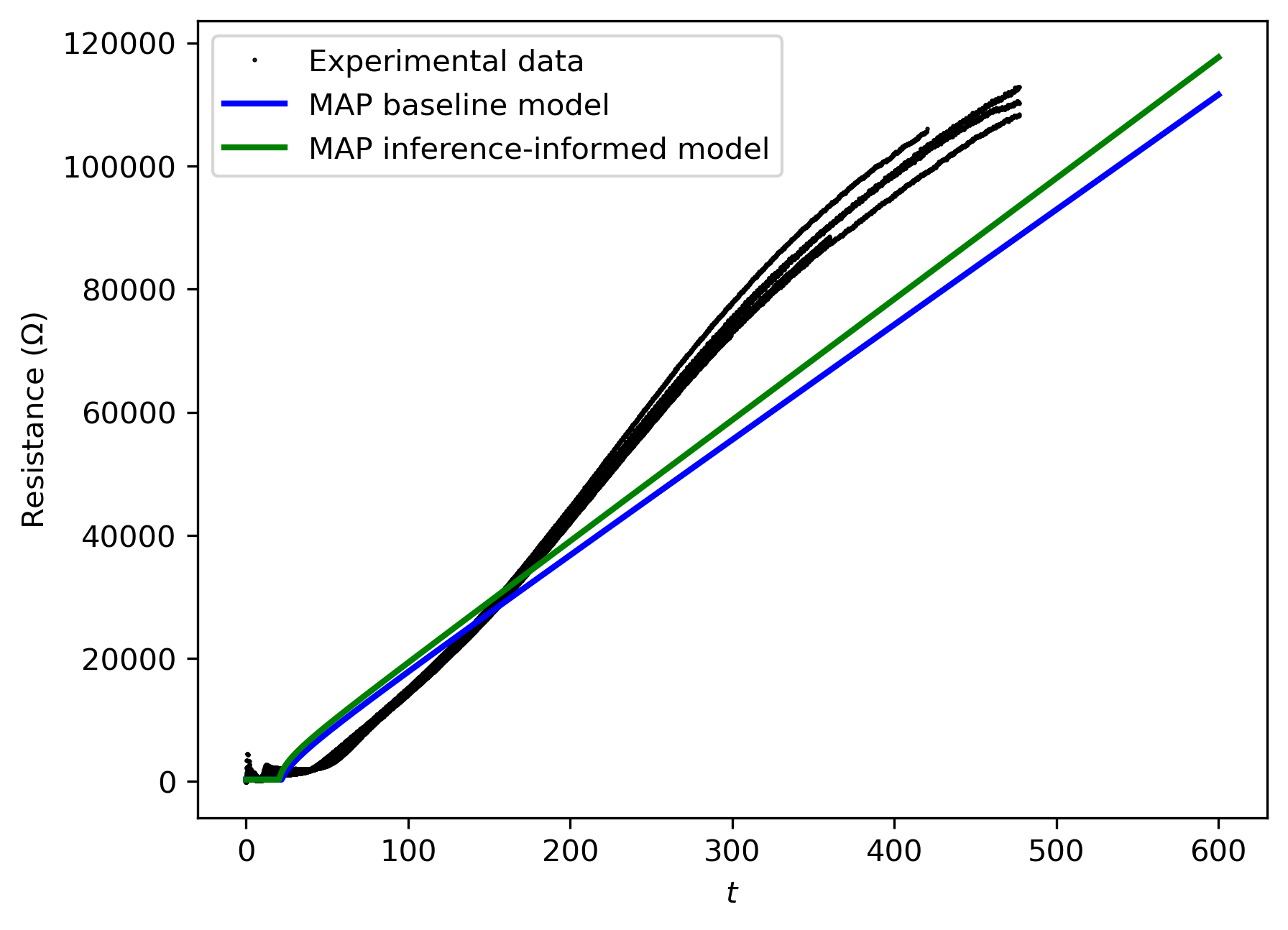}
        \caption{Voltage ramp experiment prediction, $V_R=0.5 V/s$ (resistance)}
    \end{subfigure}
    \hfill
    \begin{subfigure}[b]{0.32\textwidth}
        \includegraphics[width=\linewidth]{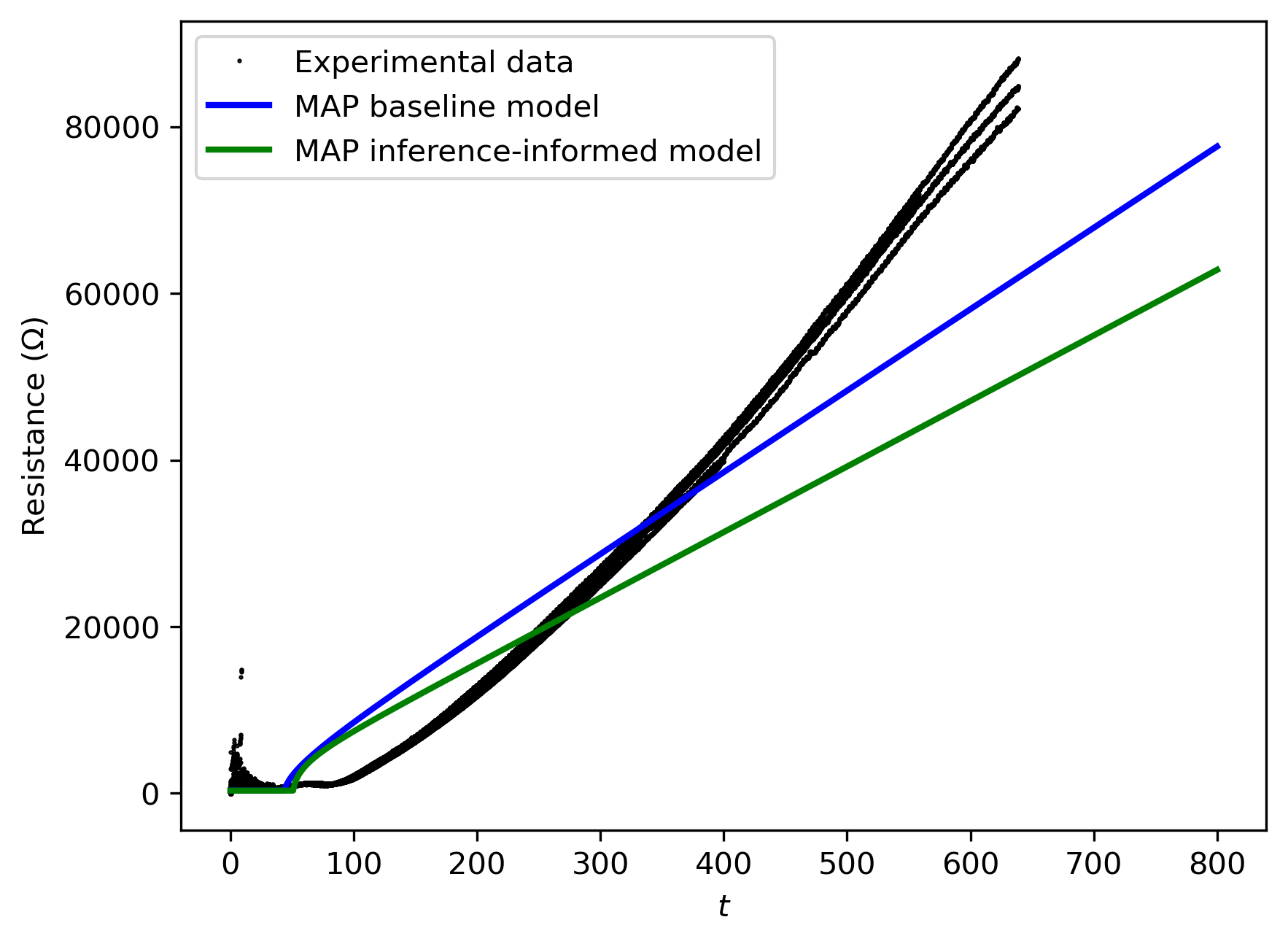}
        \caption{Voltage ramp experiment prediction, $V_R=0.125 V/s$ (resistance)}
    \end{subfigure}
    \hfill
    \begin{subfigure}[b]{0.32\textwidth}
        \includegraphics[width=\linewidth]{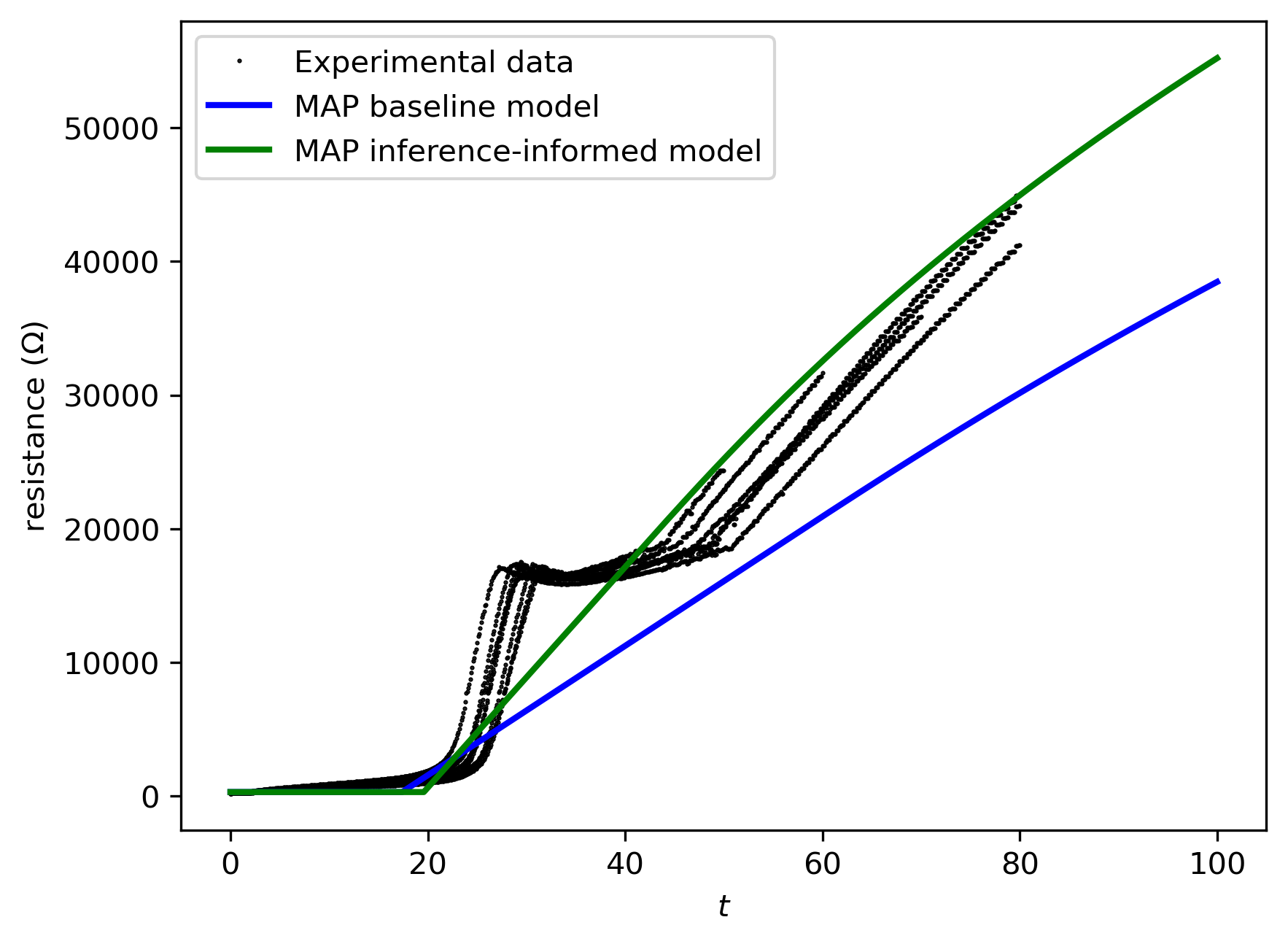}
        \caption{Constant resistance experiment prediction, $j_0=10.0mA$ (resistance)}
    \end{subfigure}
    \hfill
    \begin{subfigure}[b]{0.32\textwidth}
        \includegraphics[width=\linewidth]{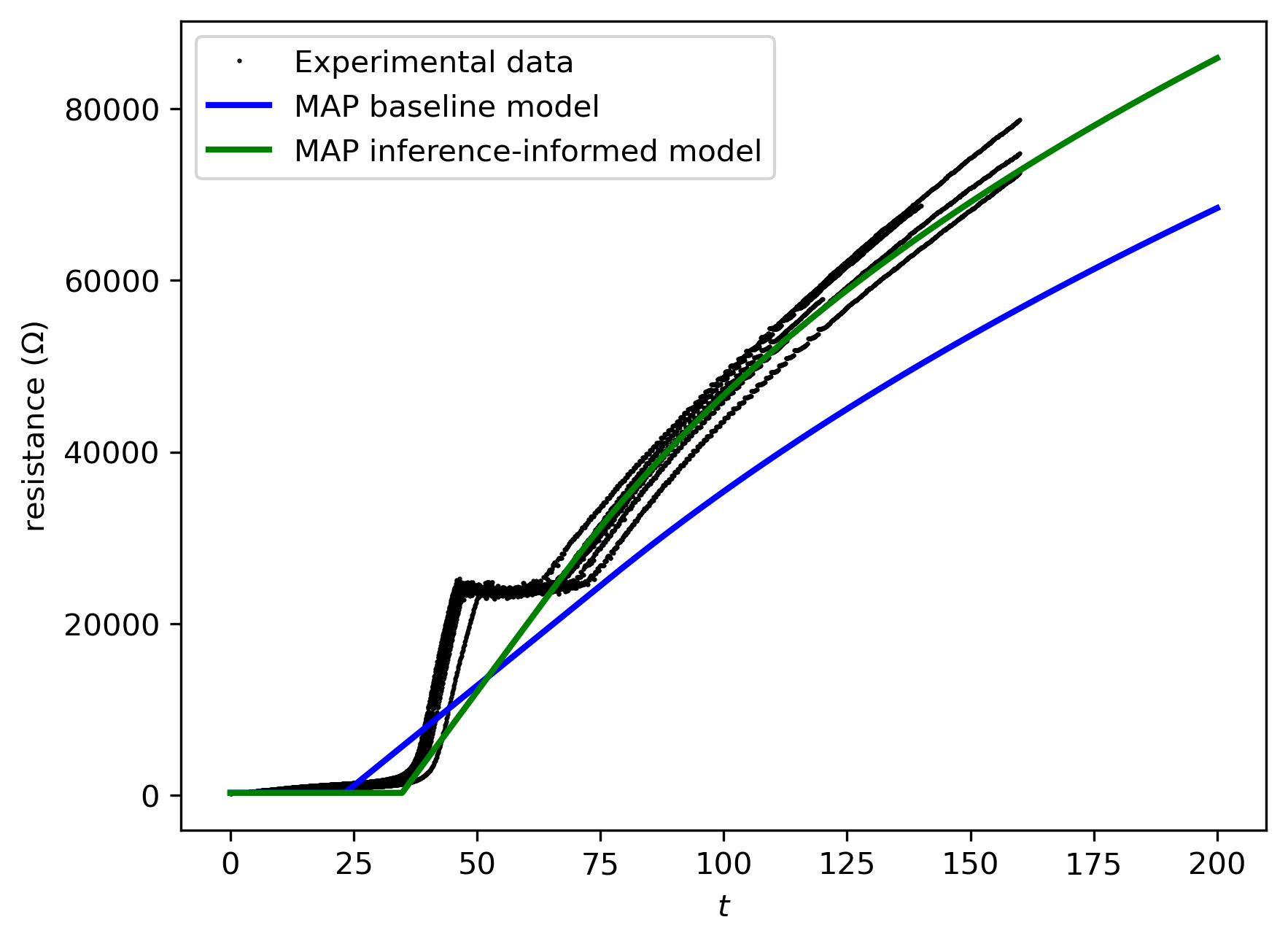}
        \caption{Constant current experiment prediction, $j_0=7.5mA$ (resistance)}
    \end{subfigure}
    \hfill
    \begin{subfigure}[b]{0.32\textwidth}
        \includegraphics[width=\linewidth]{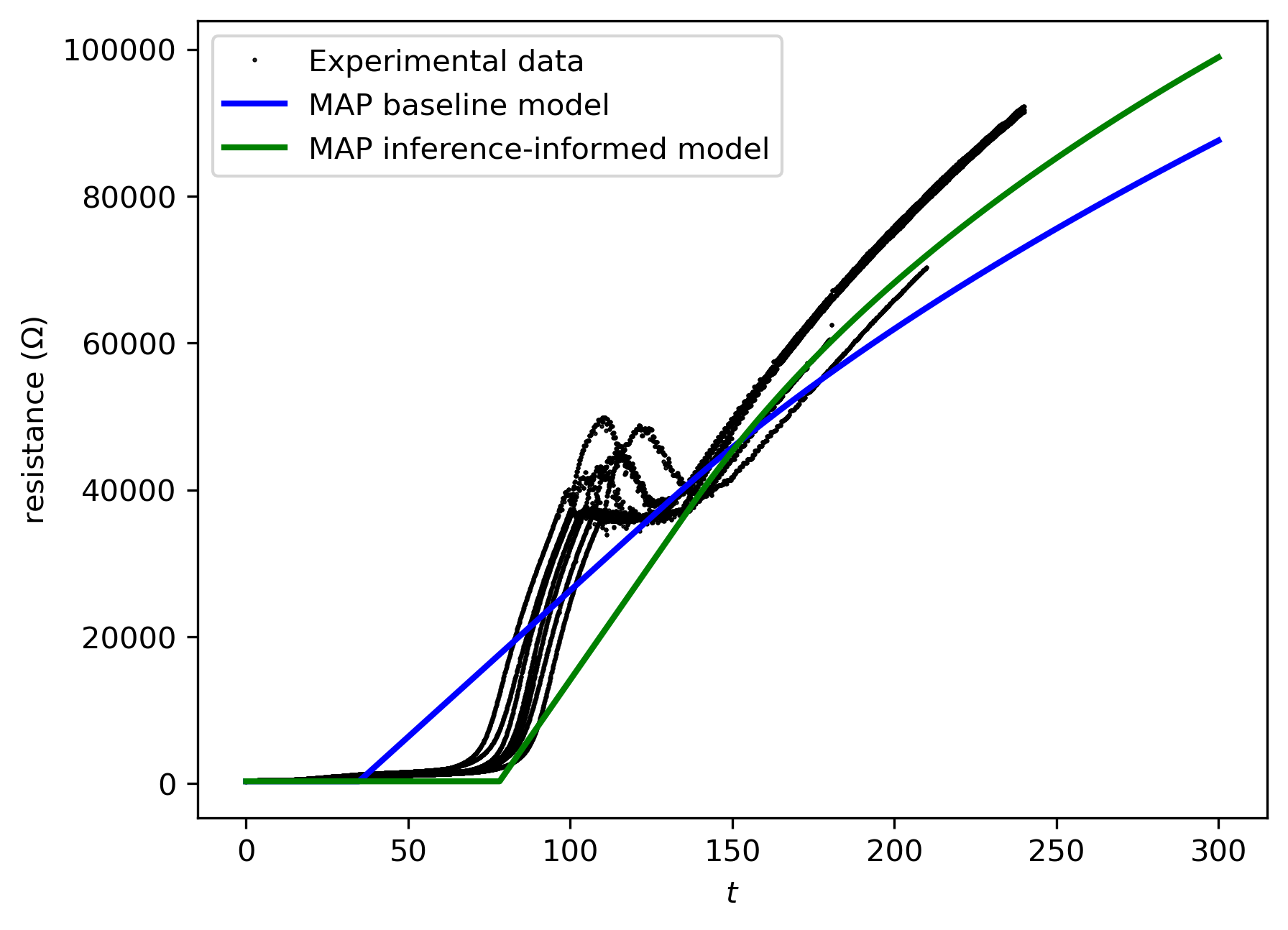}
        \caption{Constant current experiment prediction, $j_0=5.0mA$ (resistance)}
    \end{subfigure}
    \caption{Comparisons between resistance prediction on the baseline model and inference-informed model at the MAP for each.}
    \label{fig:app_map_resistance}
\end{figure}

We include here in Figs~\ref{fig:app_map_current}-~\ref{fig:app_map_resistance} all comparisons between the baseline and inference-informed models. This includes current and resistance comparisons for all 6 experiments. The inference-informed model performs better than the baseline model in all experiments, but the notably greater improvement on predictions for CC experiments. However, thickness prediction is still inaccurate in many cases.

\section{ML-augmented model with first-peak}\label{app:first_peak}
\begin{figure}[h!]
    \centering
    \begin{subfigure}[b]{0.32\textwidth}
        \includegraphics[width=\linewidth]{TikzPictures/ecoat_neode_cur_pred_VR1.0_first_peak.png}
        \caption{Voltage ramp experiment prediction, $V_R=1.0 V/s$ (current)}
    \end{subfigure}
    \hfill
    \begin{subfigure}[b]{0.32\textwidth}
        \includegraphics[width=\linewidth]{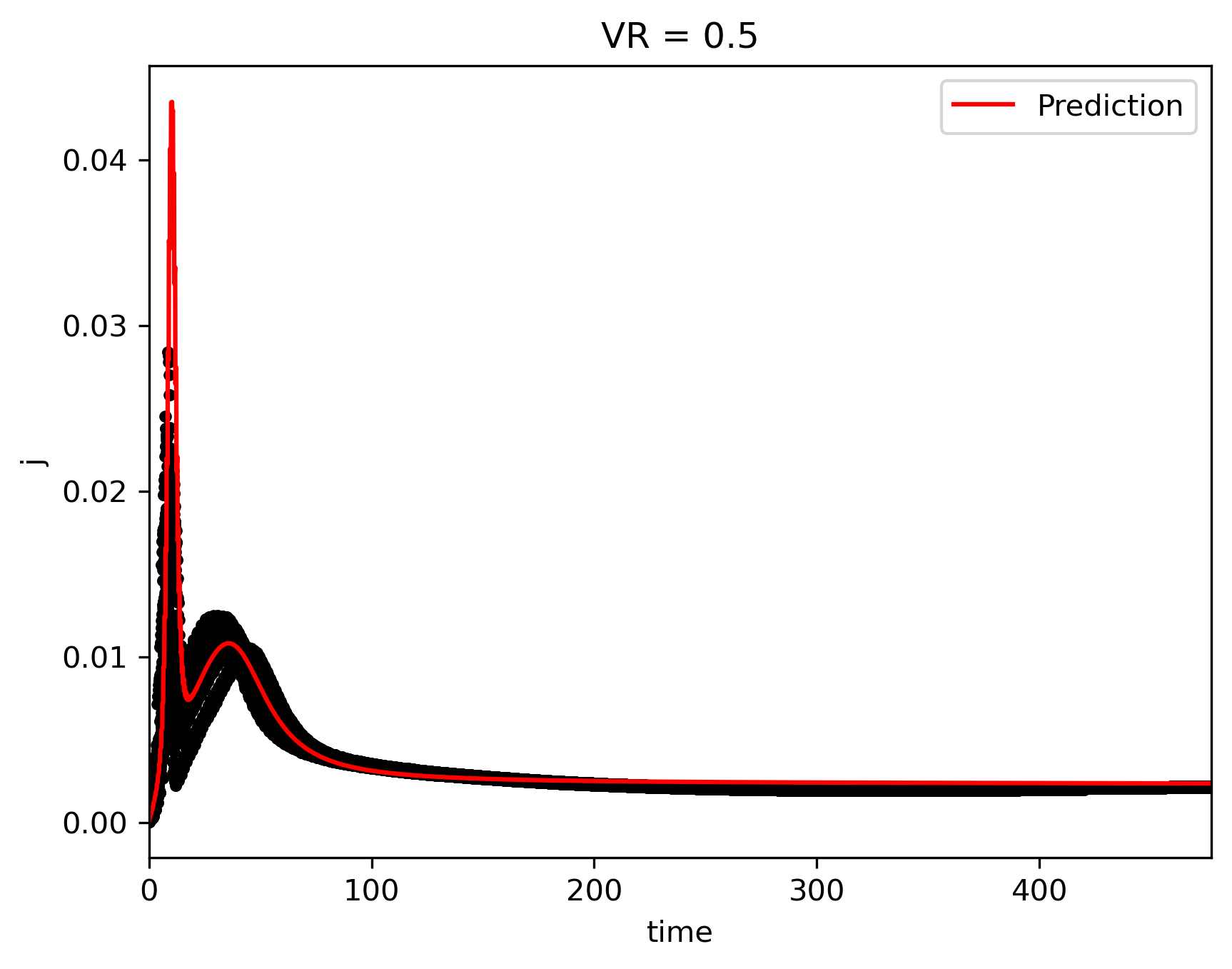}
        \caption{Voltage ramp experiment prediction, $V_R=0.5 V/s$ (current)}
    \end{subfigure}
    \hfill
    \begin{subfigure}[b]{0.32\textwidth}
        \includegraphics[width=\linewidth]{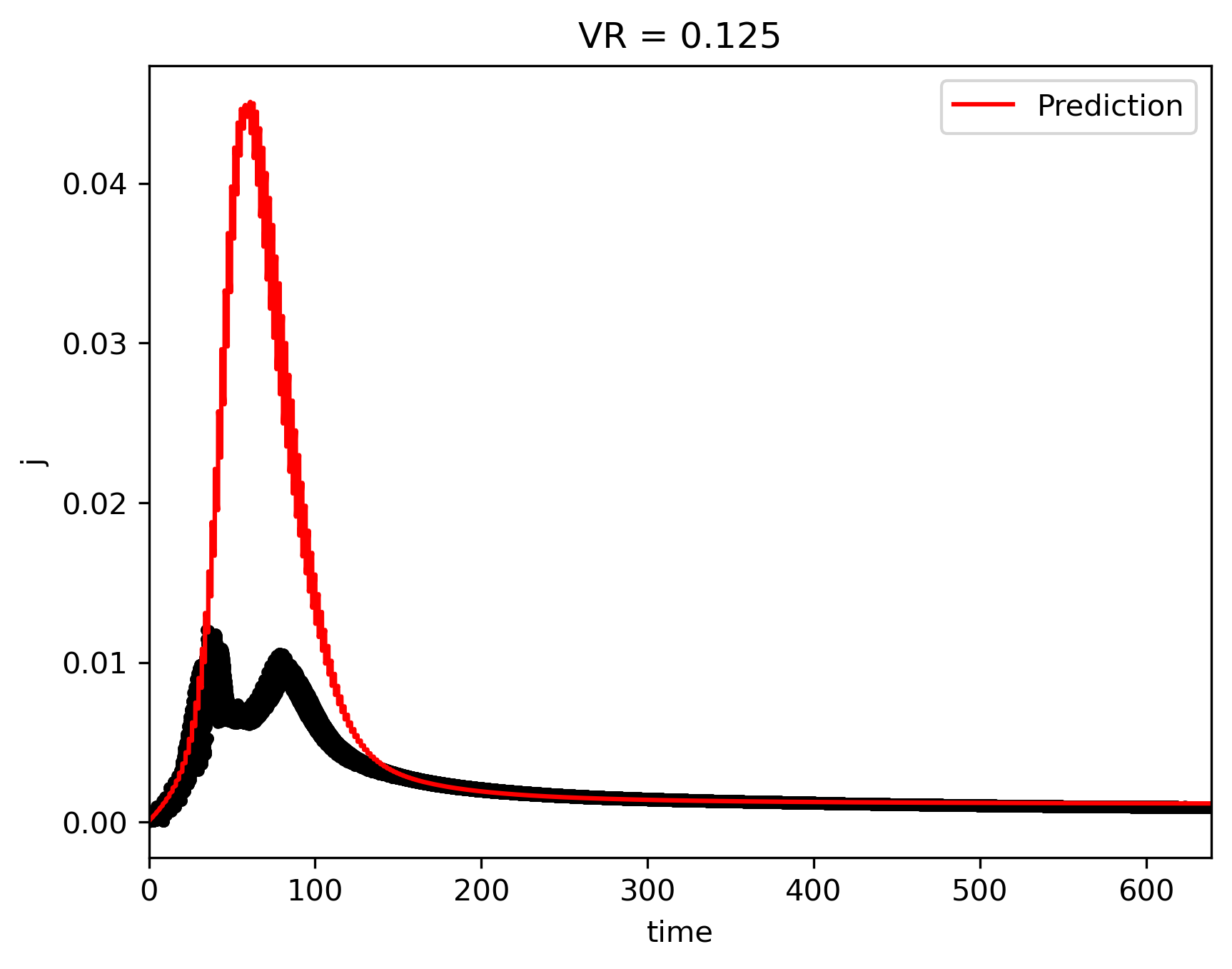}
        \caption{Voltage ramp experiment prediction, $V_R=0.125 V/s$ (current)}
    \end{subfigure}
    \hfill
    \begin{subfigure}[b]{0.32\textwidth}
        \includegraphics[width=\linewidth]{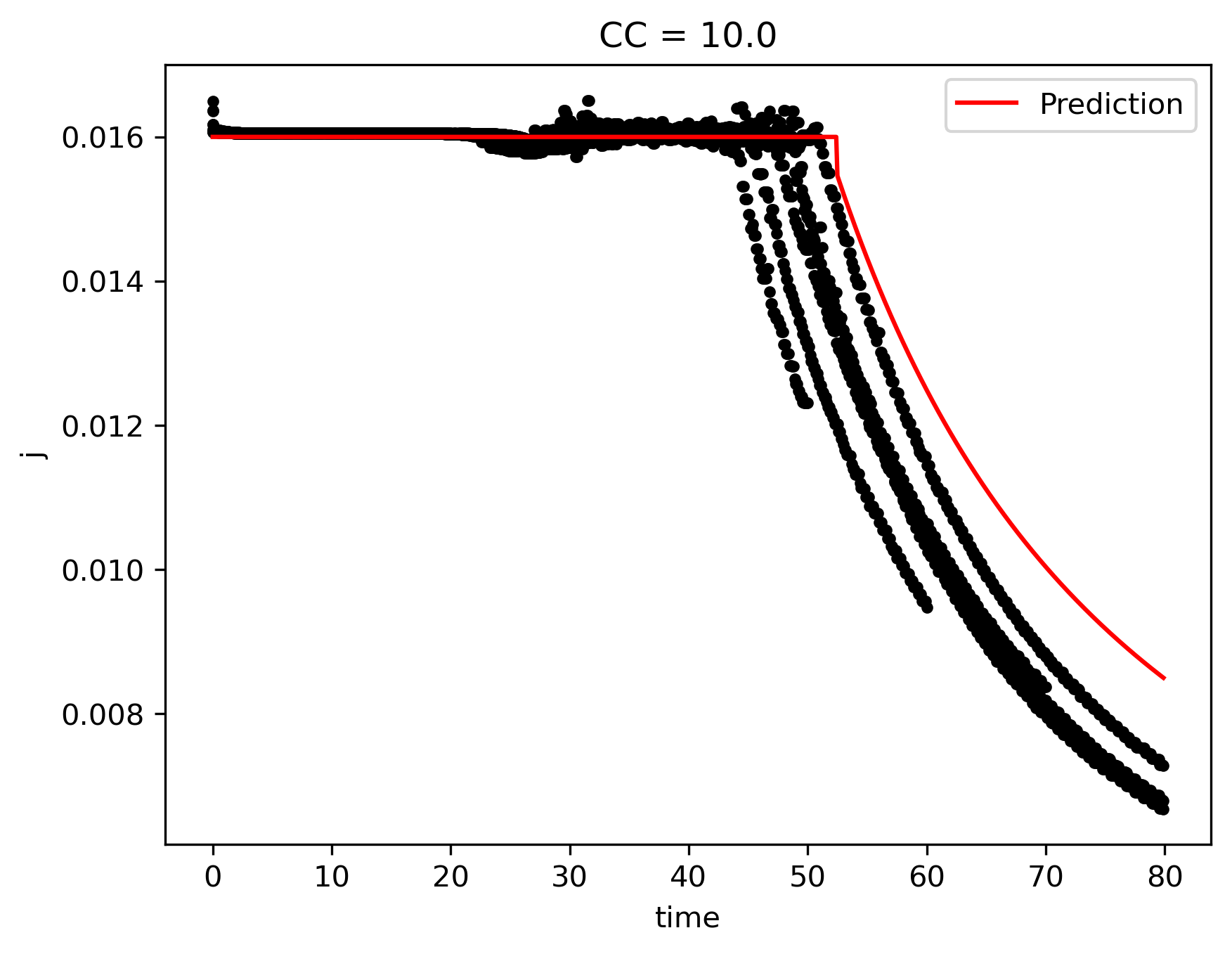}
        \caption{Constant current experiment prediction, $j_0=10.0mA$ (current)}
    \end{subfigure}
    \hfill
    \begin{subfigure}[b]{0.32\textwidth}
        \includegraphics[width=\linewidth]{TikzPictures/ecoat_neode_cur_pred_CC7.5_first_peak.png}
        \caption{Constant current experiment prediction, $j_0=7.5mA$ (current)}
    \end{subfigure}
    \hfill
    \begin{subfigure}[b]{0.32\textwidth}
        \includegraphics[width=\linewidth]{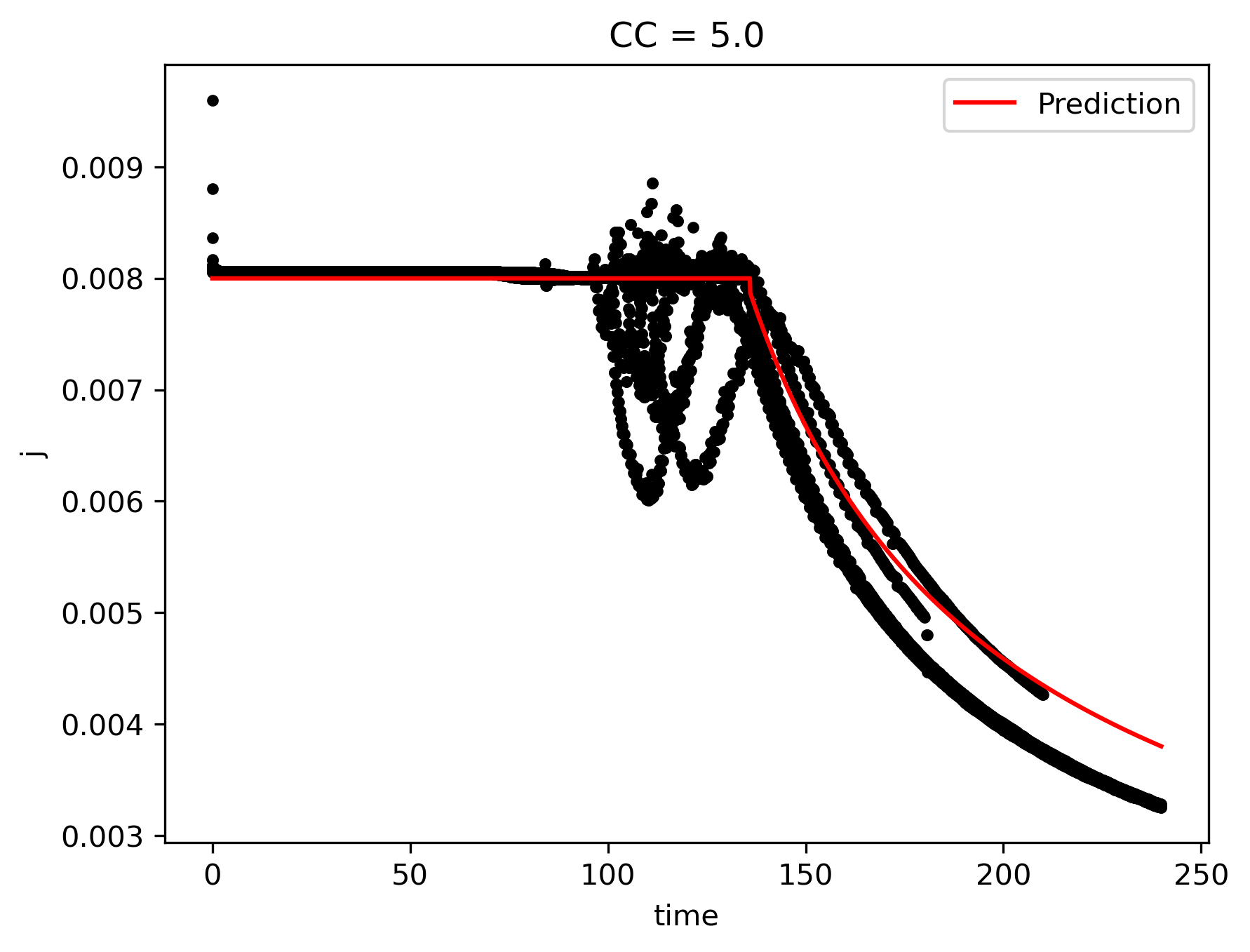}
        \caption{Constant current experiment prediction, $j_0=5.0mA$ (current)}
    \end{subfigure}
    \caption{ML-augmented model with first peak current predictions compared to experimental data.}
    \label{fig:app_current_first_peak}
\end{figure}

\begin{figure}[h!]
    \centering
    \begin{subfigure}[b]{0.32\textwidth}
        \includegraphics[width=\linewidth]{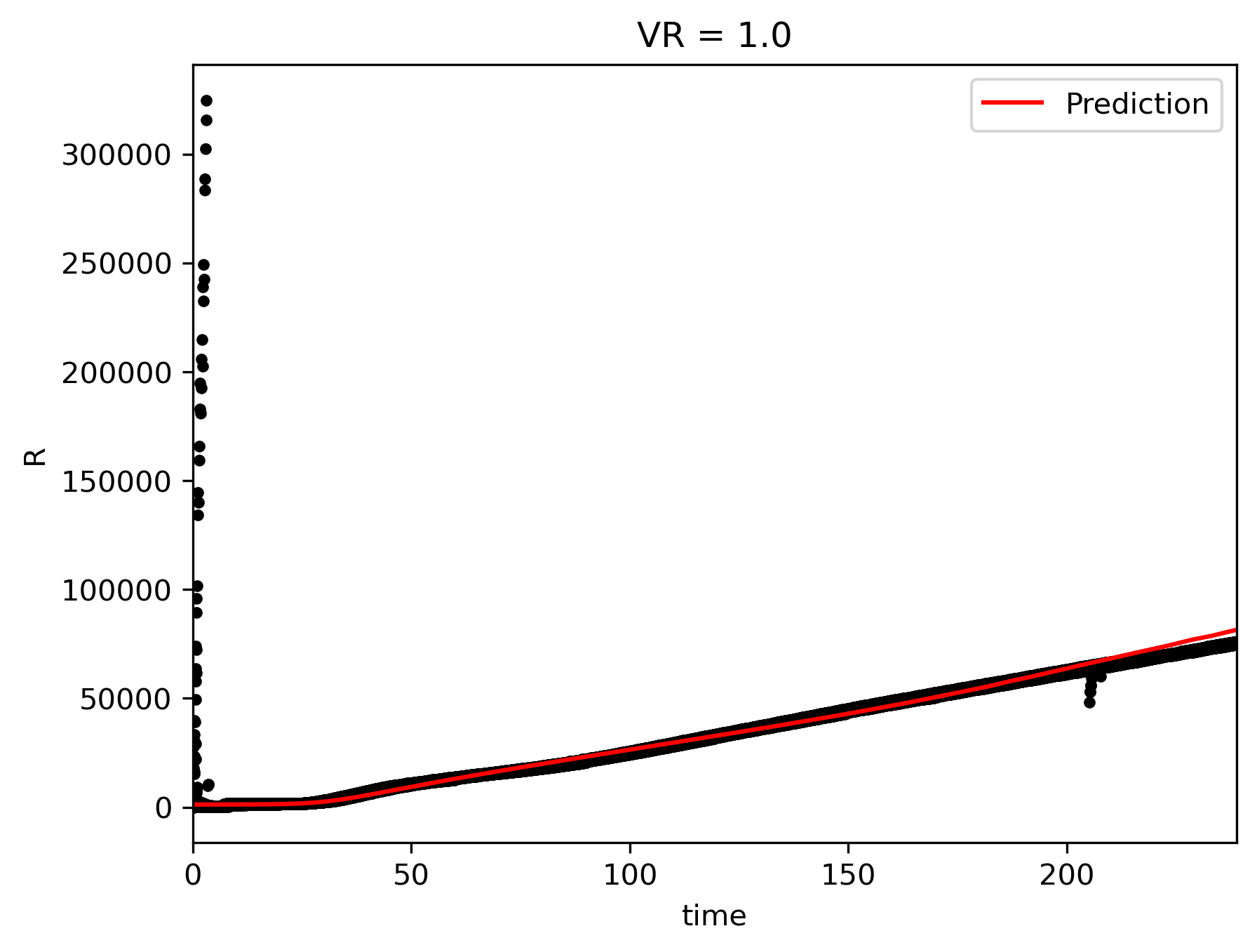}
        \caption{Voltage ramp experiment prediction, $V_R=1.0$ (resistance)}
    \end{subfigure}
    \hfill
    \begin{subfigure}[b]{0.32\textwidth}
        \includegraphics[width=\linewidth]{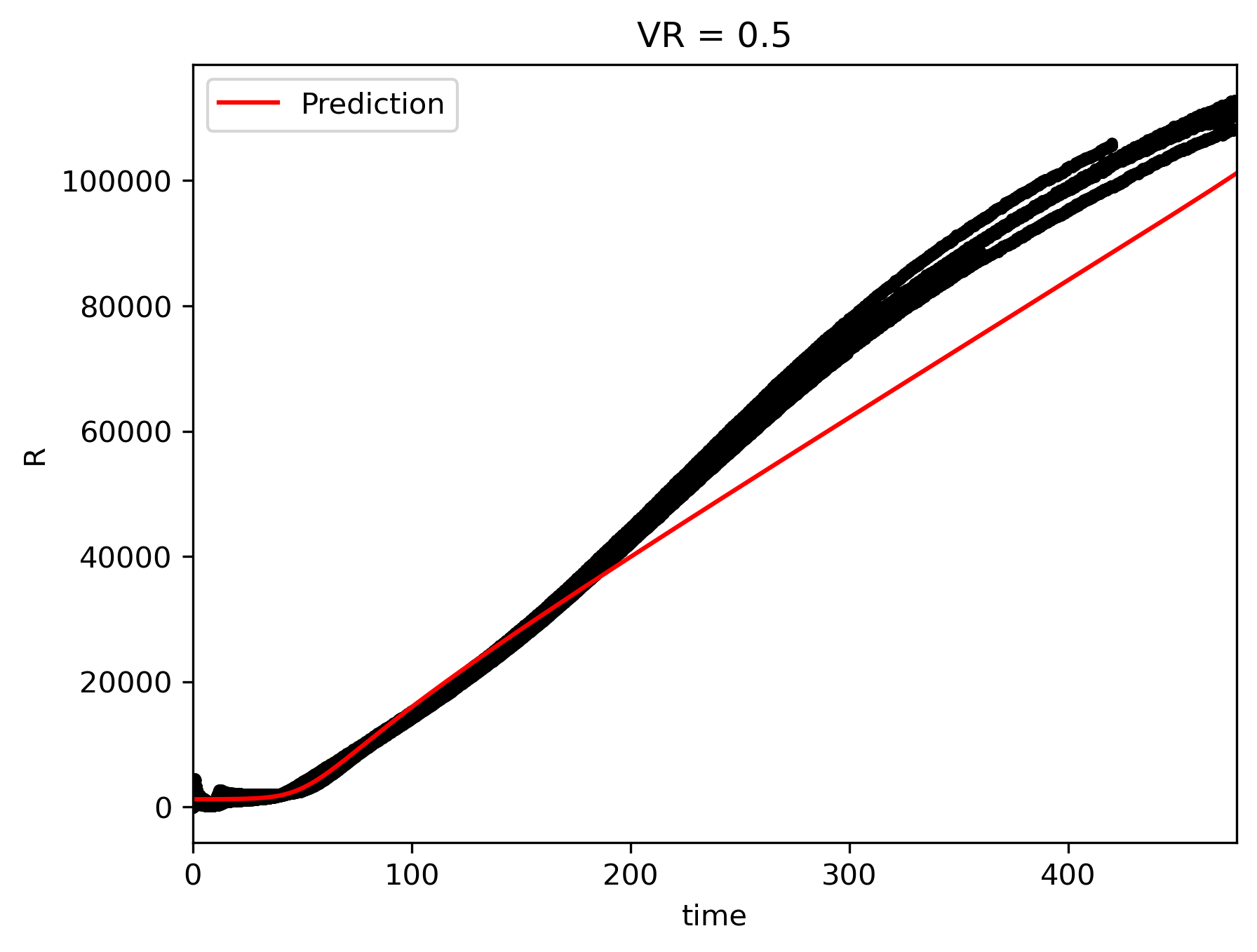}
        \caption{Voltage ramp experiment prediction, $V_R=0.5 V/s$ (resistance)}
    \end{subfigure}
    \hfill
    \begin{subfigure}[b]{0.32\textwidth}
        \includegraphics[width=\linewidth]{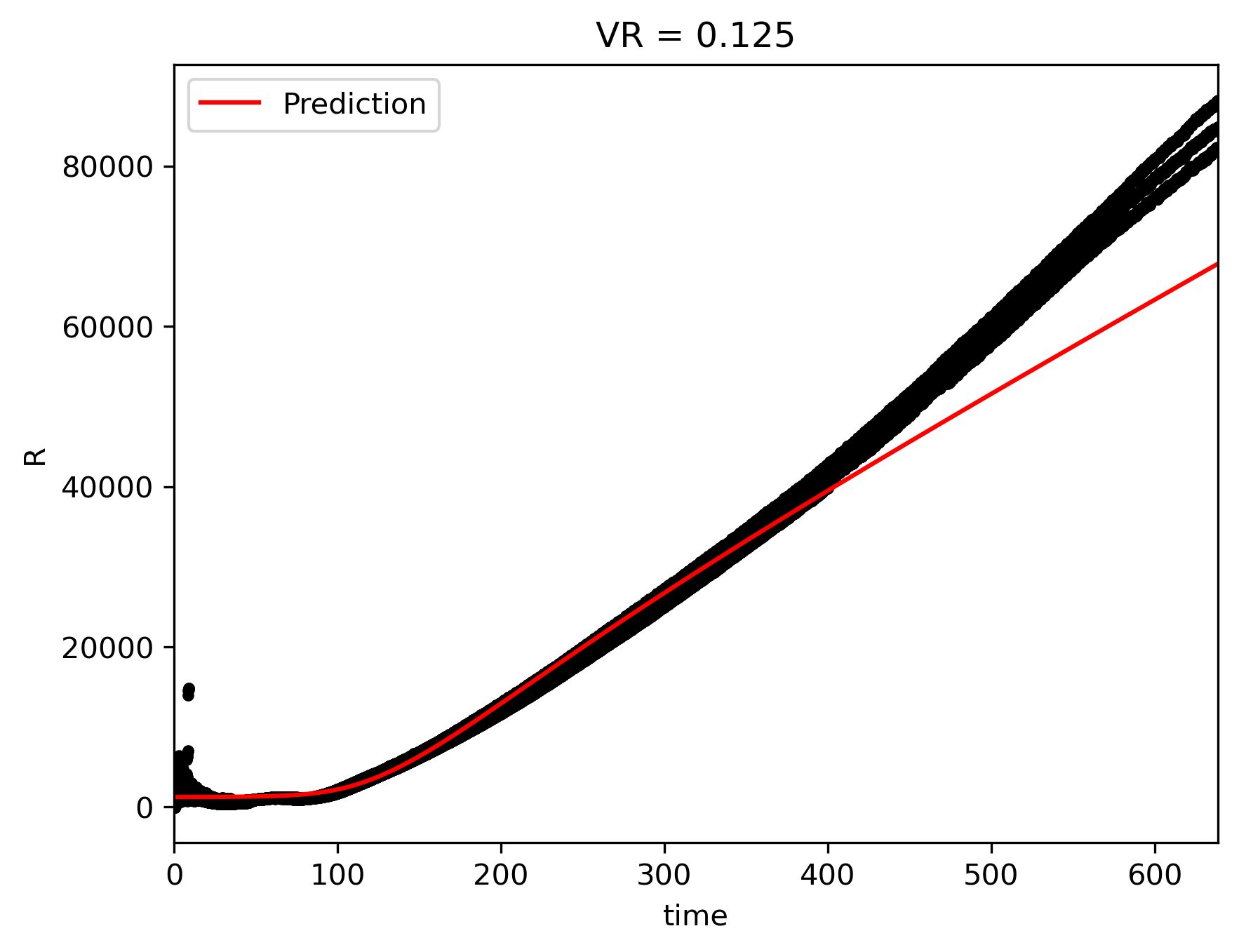}
        \caption{Voltage ramp experiment prediction, $V_R=0.125 V/s$ (resistance)}
    \end{subfigure}
    \hfill
    \begin{subfigure}[b]{0.32\textwidth}
        \includegraphics[width=\linewidth]{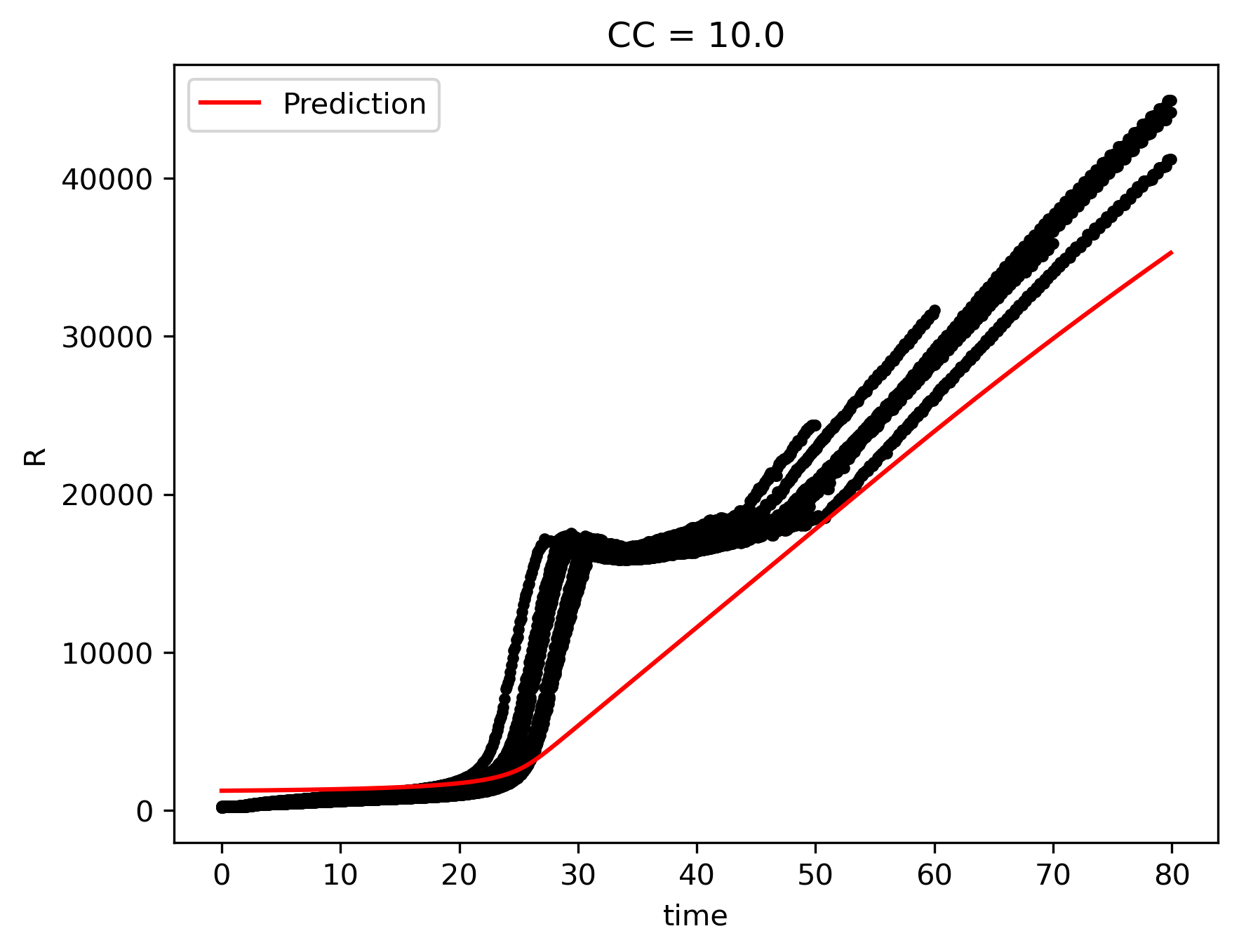}
        \caption{Constant resistance experiment prediction, $j_0=10.0mA$ (resistance)}
    \end{subfigure}
    \hfill
    \begin{subfigure}[b]{0.32\textwidth}
        \includegraphics[width=\linewidth]{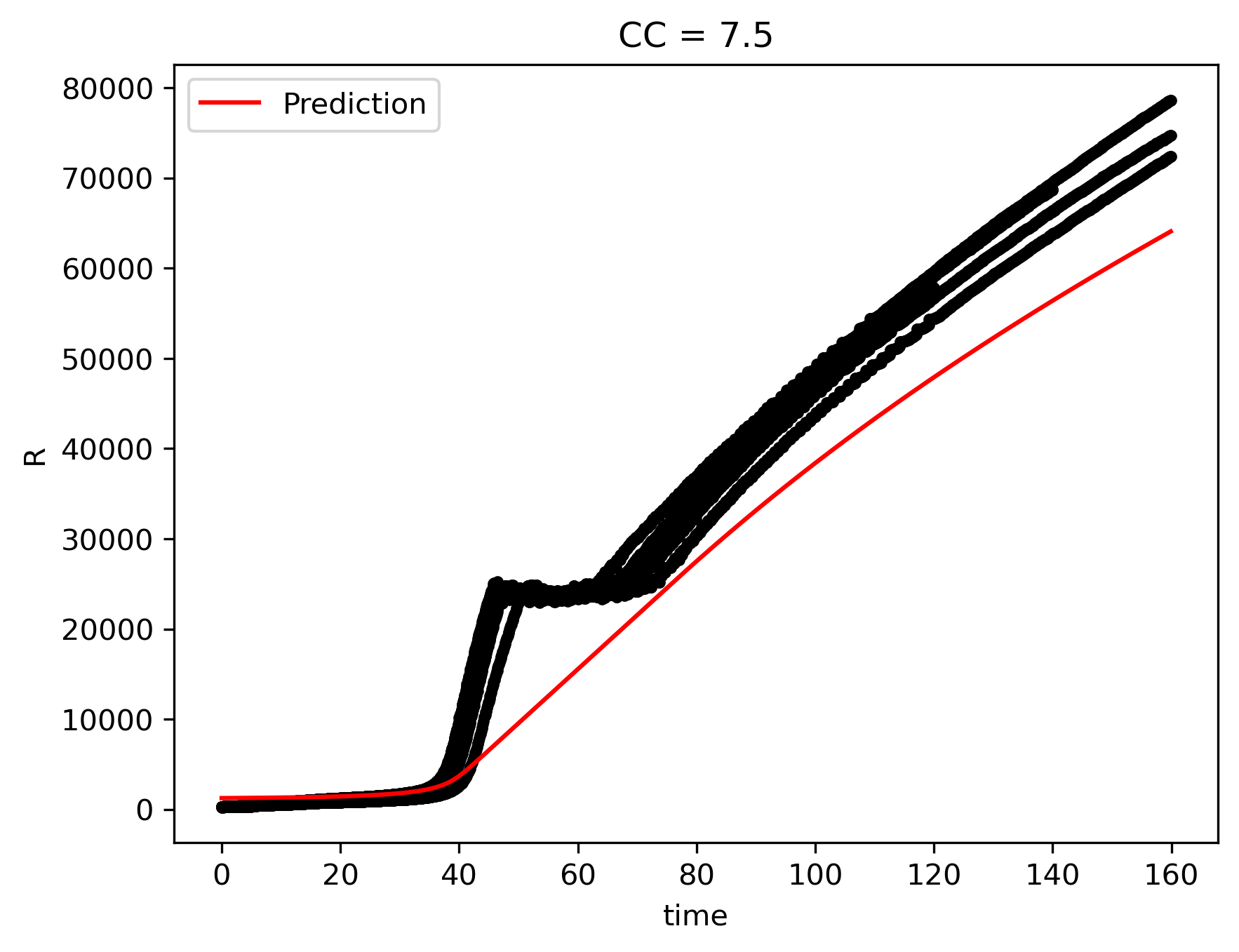}
        \caption{Constant current experiment prediction, $j_0=7.5mA$ (resistance)}
    \end{subfigure}
    \hfill
    \begin{subfigure}[b]{0.32\textwidth}
        \includegraphics[width=\linewidth]{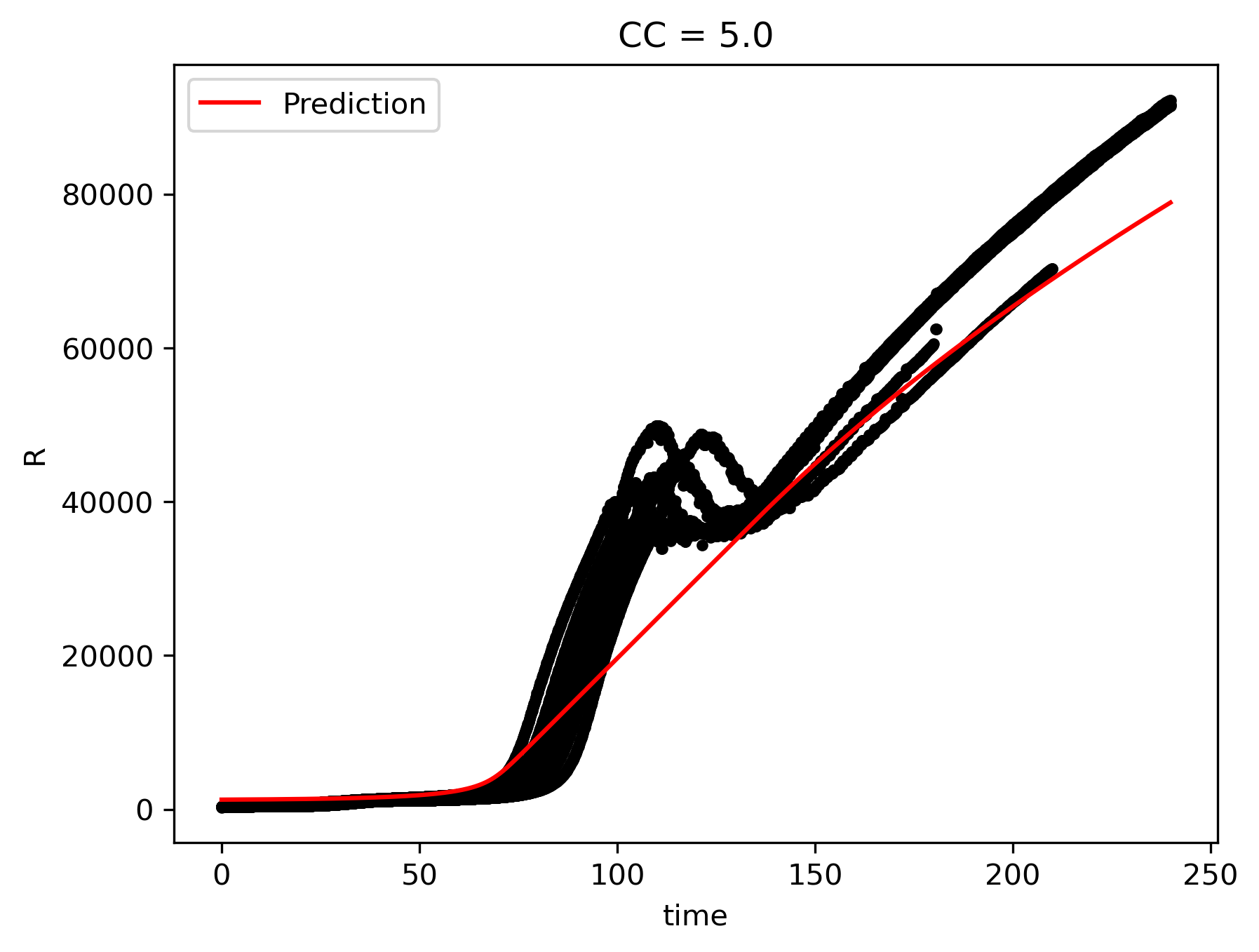}
        \caption{Constant current experiment prediction, $j_0=5.0mA$ (resistance)}
    \end{subfigure}
    \caption{ML-augmented model with first peak resistance predictions compared to experimental data.}
    \label{fig:app_resistance_first_peak}
\end{figure}

We present all prediction results on the ML-augmented model with first peak. Noteably, predictions are poor for VR experiments in the low $V_R$ regime, but accurate for all other experimental configurations. In particular, thickness prediction is much more accurate compared to the baseline and inference-informed models. Predictions for current and resistance are included in Figs.~\ref{fig:app_current_first_peak}-~\ref{fig:app_resistance_first_peak}, but the ML model is trained with current data for the 3 voltage ramp experiments only. 

\section{ML-augmented model without first-peak}\label{app:no_first_peak}
\begin{figure}[h!]
    \centering
    \begin{subfigure}[b]{0.32\textwidth}
        \includegraphics[width=\linewidth]{TikzPictures/ecoat_neode_cur_pred_VR1.0.png}
        \caption{Voltage ramp experiment prediction, $V_R=1.0 V/s$ (current)}
    \end{subfigure}
    \hfill
    \begin{subfigure}[b]{0.32\textwidth}
        \includegraphics[width=\linewidth]{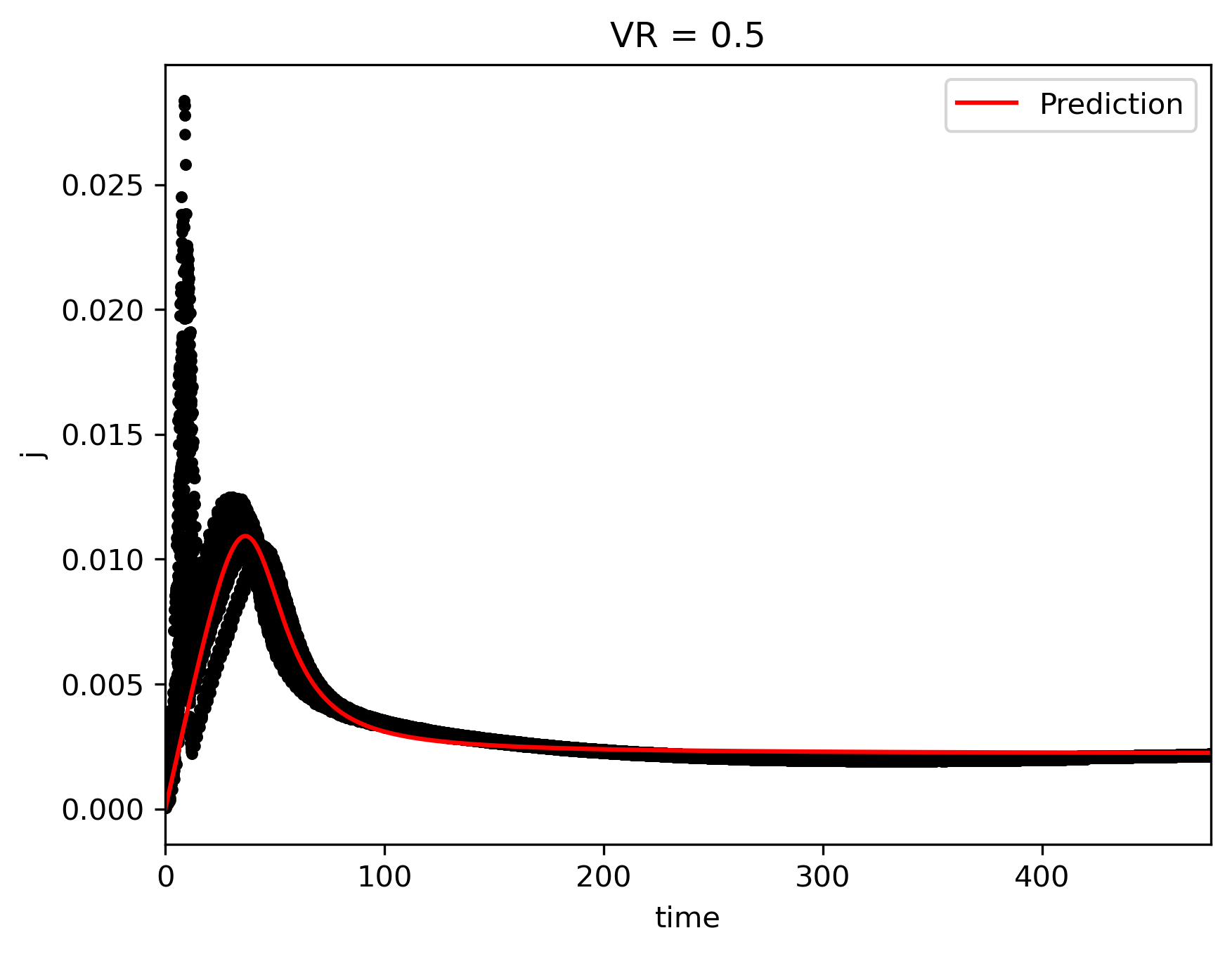}
        \caption{Voltage ramp experiment prediction, $V_R=0.5 V/s$ (current)}
    \end{subfigure}
    \hfill
    \begin{subfigure}[b]{0.32\textwidth}
        \includegraphics[width=\linewidth]{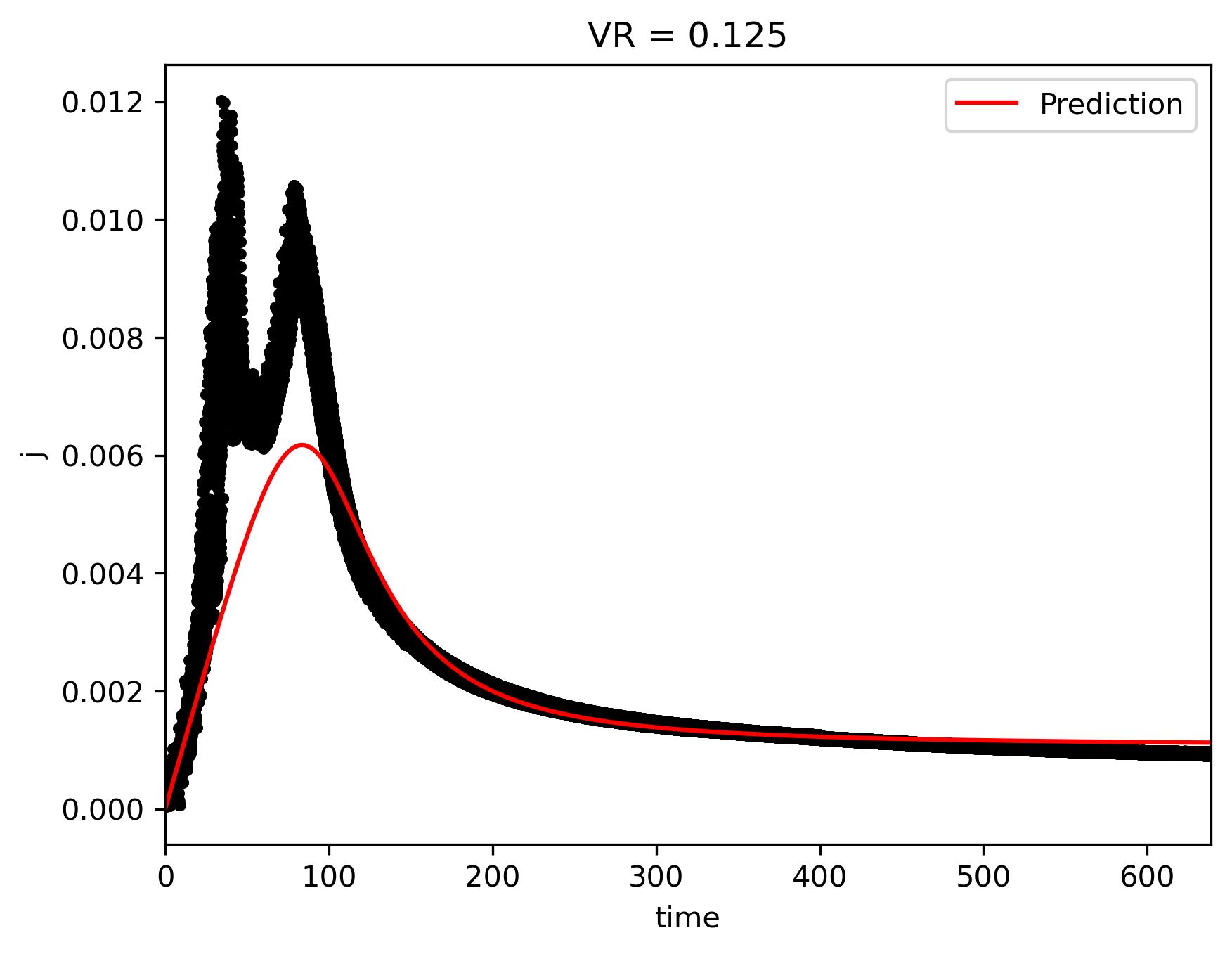}
        \caption{Voltage ramp experiment prediction, $V_R=0.125 V/s$ (current)}
    \end{subfigure}
    \hfill
    \begin{subfigure}[b]{0.32\textwidth}
        \includegraphics[width=\linewidth]{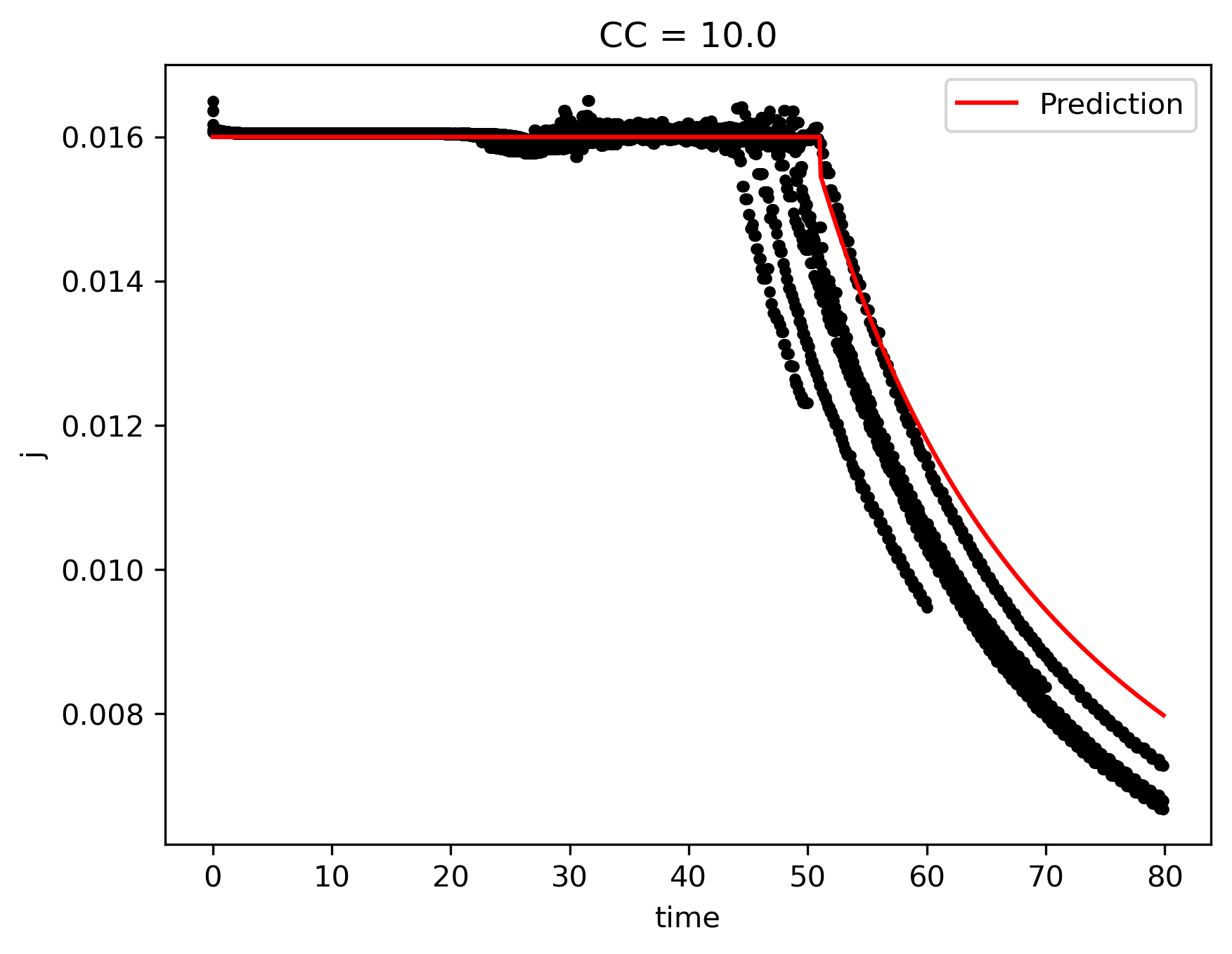}
        \caption{Constant current experiment prediction, $j_0=10.0mA$ (current)}
    \end{subfigure}
    \hfill
    \begin{subfigure}[b]{0.32\textwidth}
        \includegraphics[width=\linewidth]{TikzPictures/ecoat_neode_cur_pred_CC7.5.png}
        \caption{Constant current experiment prediction, $j_0=7.5mA$ (current)}
    \end{subfigure}
    \hfill
    \begin{subfigure}[b]{0.32\textwidth}
        \includegraphics[width=\linewidth]{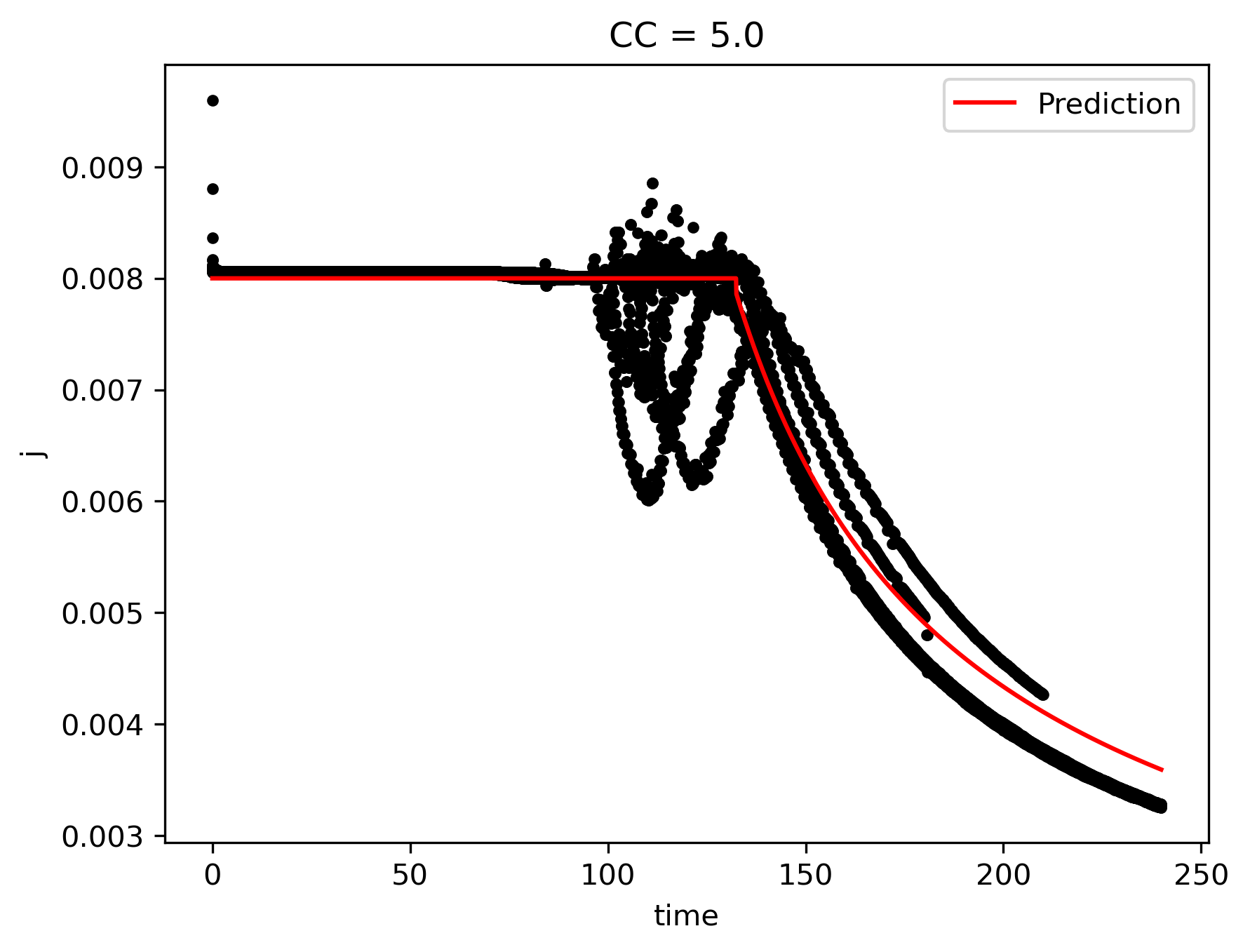}
        \caption{Constant current experiment prediction, $j_0=5.0mA$ (current)}
    \end{subfigure}
    \caption{ML-augmented model without first peak current predictions compared to experimental data.}
    \label{fig:app_current_no_first_peak}
\end{figure}

\begin{figure}[h!]
    \centering
    \begin{subfigure}[b]{0.32\textwidth}
        \includegraphics[width=\linewidth]{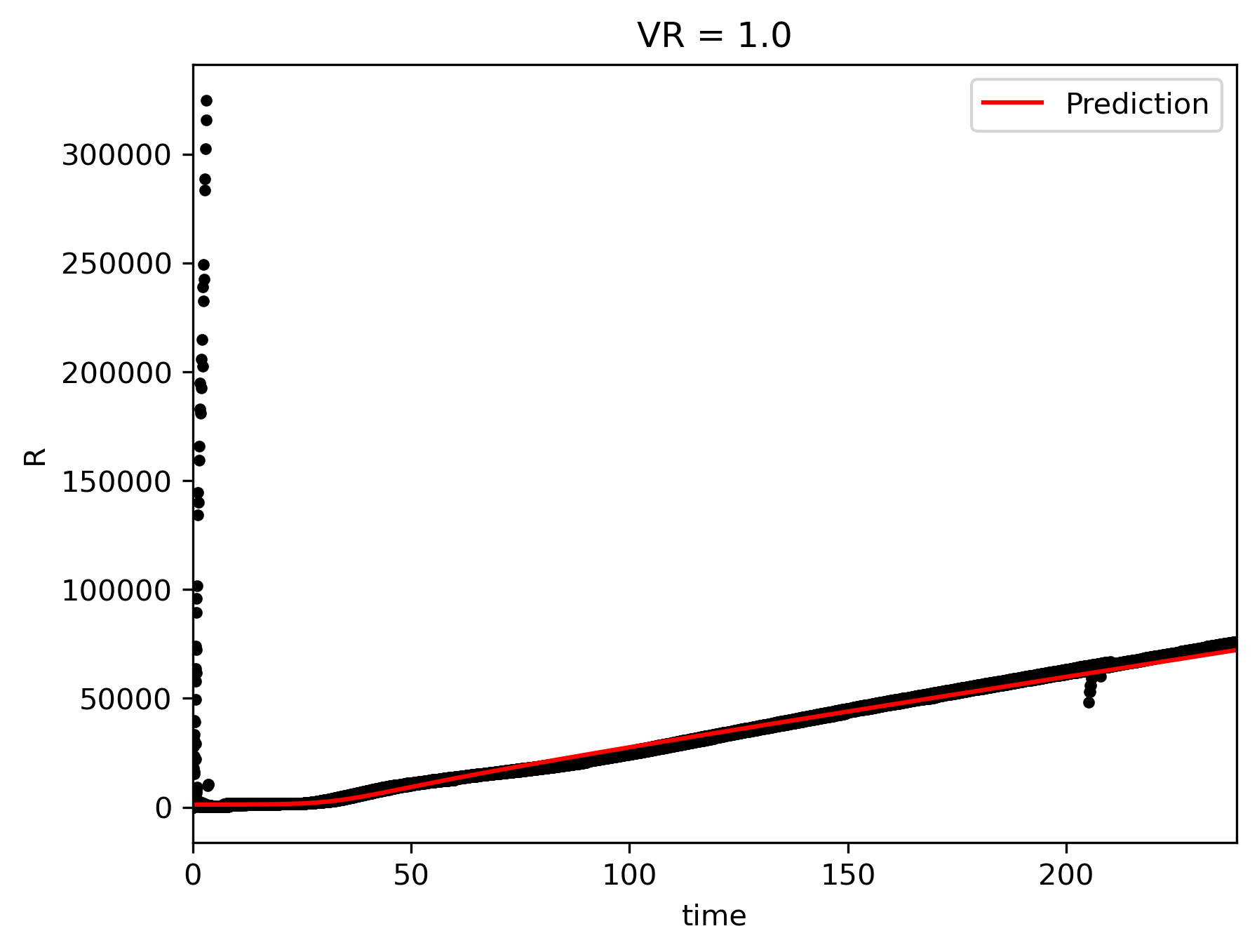}
        \caption{Voltage ramp experiment prediction, $V_R=1.0$ (resistance)}
    \end{subfigure}
    \hfill
    \begin{subfigure}[b]{0.32\textwidth}
        \includegraphics[width=\linewidth]{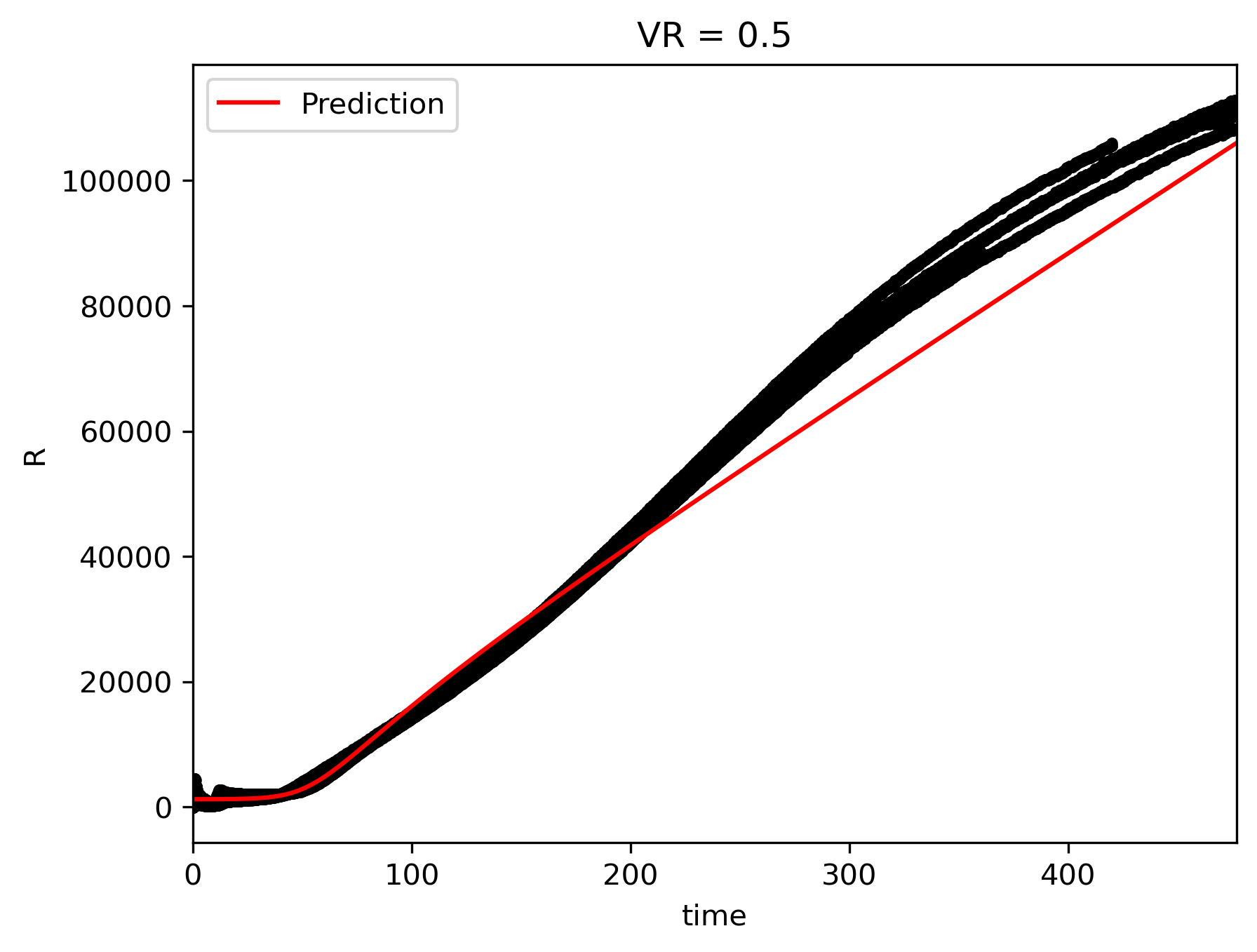}
        \caption{Voltage ramp experiment prediction, $V_R=0.5 V/s$ (resistance)}
    \end{subfigure}
    \hfill
    \begin{subfigure}[b]{0.32\textwidth}
        \includegraphics[width=\linewidth]{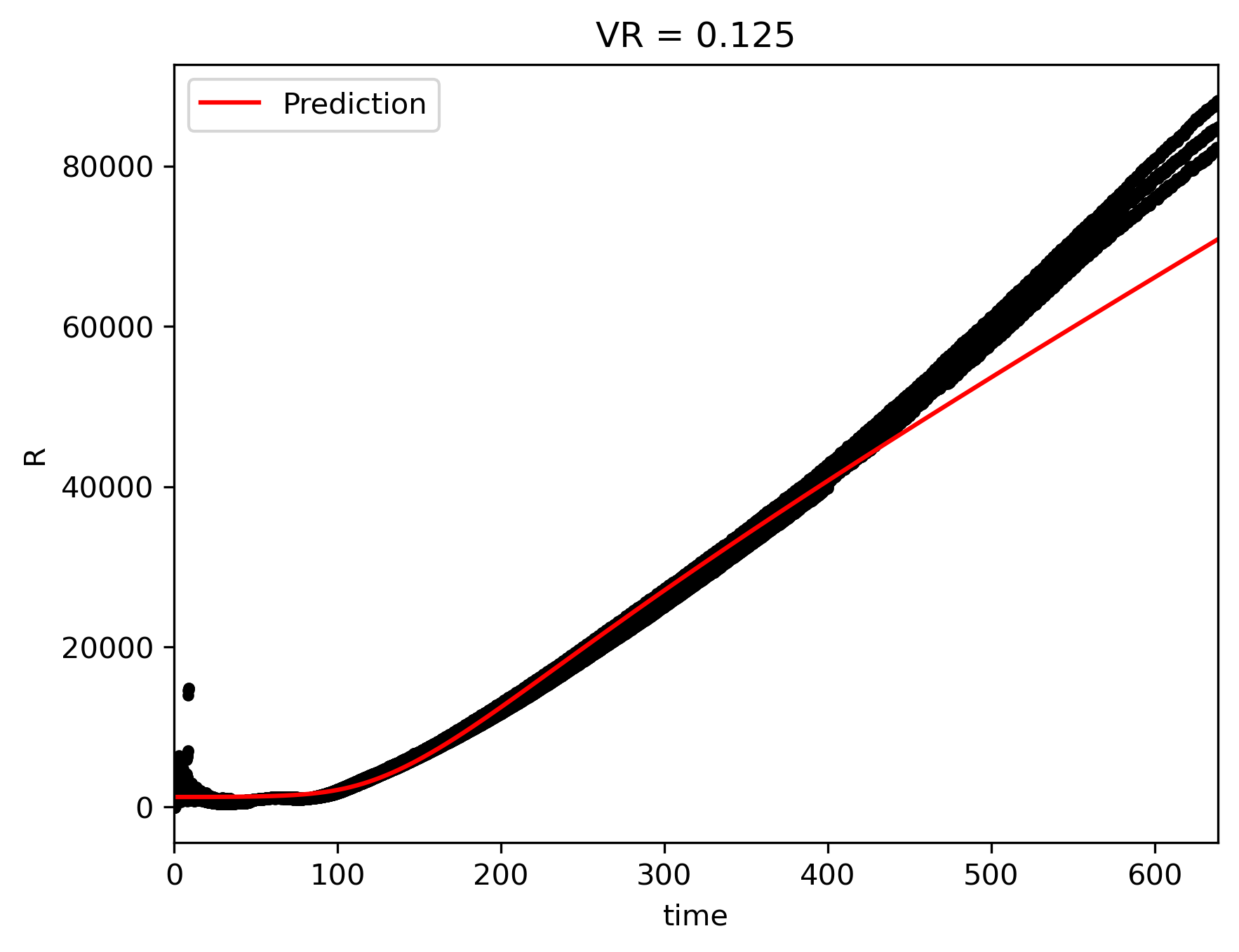}
        \caption{Voltage ramp experiment prediction, $V_R=0.125 V/s$ (resistance)}
    \end{subfigure}
    \hfill
    \begin{subfigure}[b]{0.32\textwidth}
        \includegraphics[width=\linewidth]{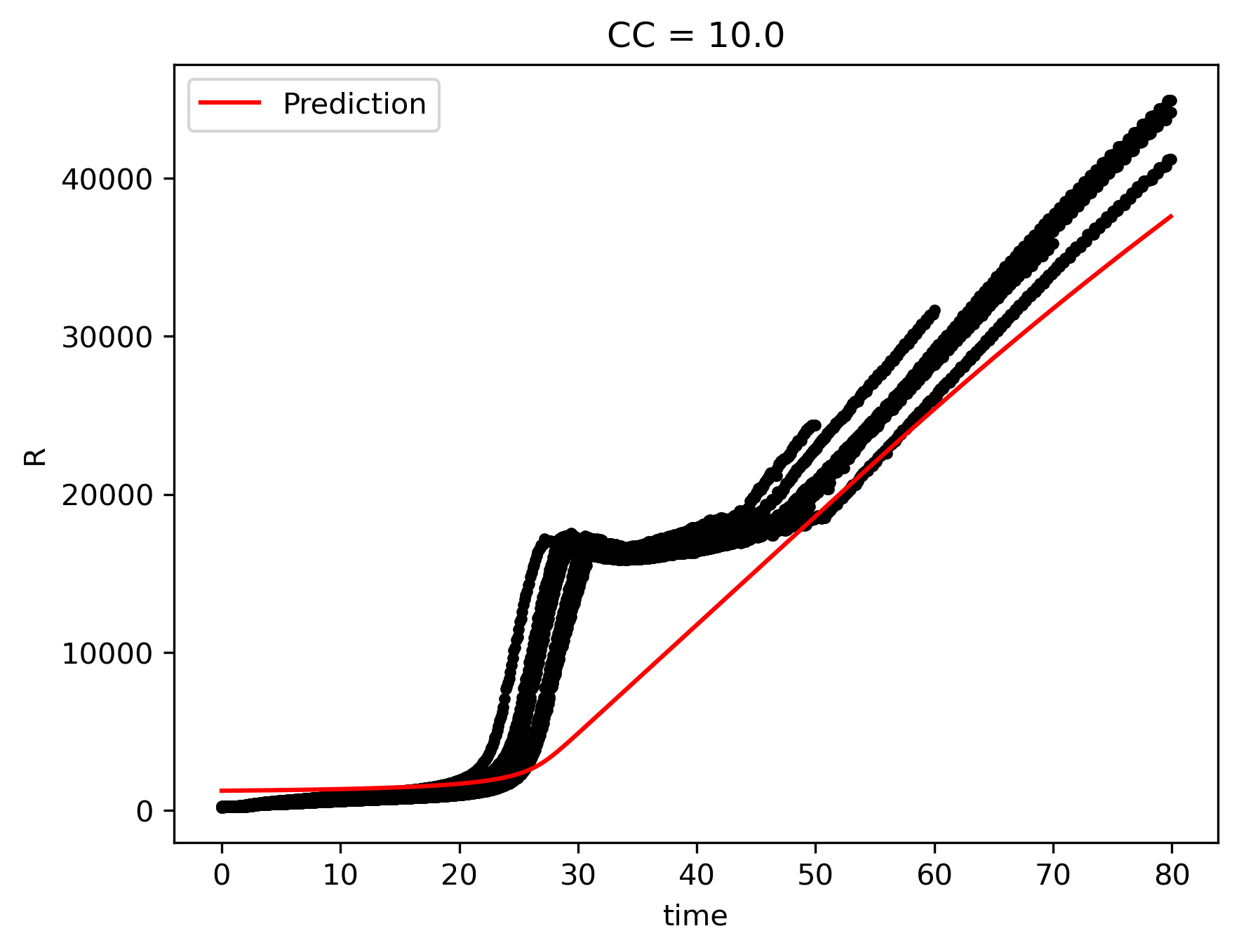}
        \caption{Constant resistance experiment prediction, $j_0=10.0mA$ (resistance)}
    \end{subfigure}
    \hfill
    \begin{subfigure}[b]{0.32\textwidth}
        \includegraphics[width=\linewidth]{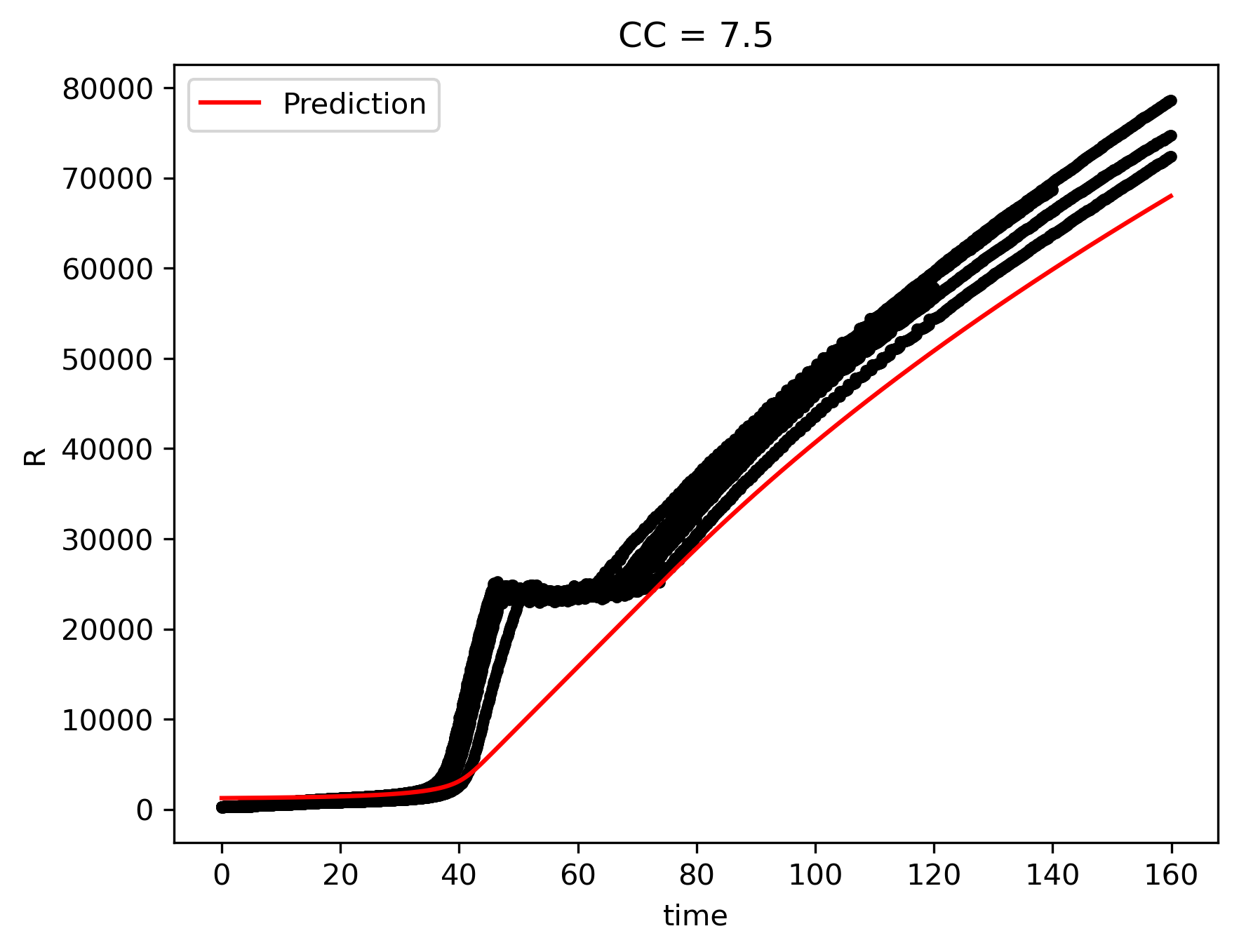}
        \caption{Constant current experiment prediction, $j_0=7.5mA$ (resistance)}
    \end{subfigure}
    \hfill
    \begin{subfigure}[b]{0.32\textwidth}
        \includegraphics[width=\linewidth]{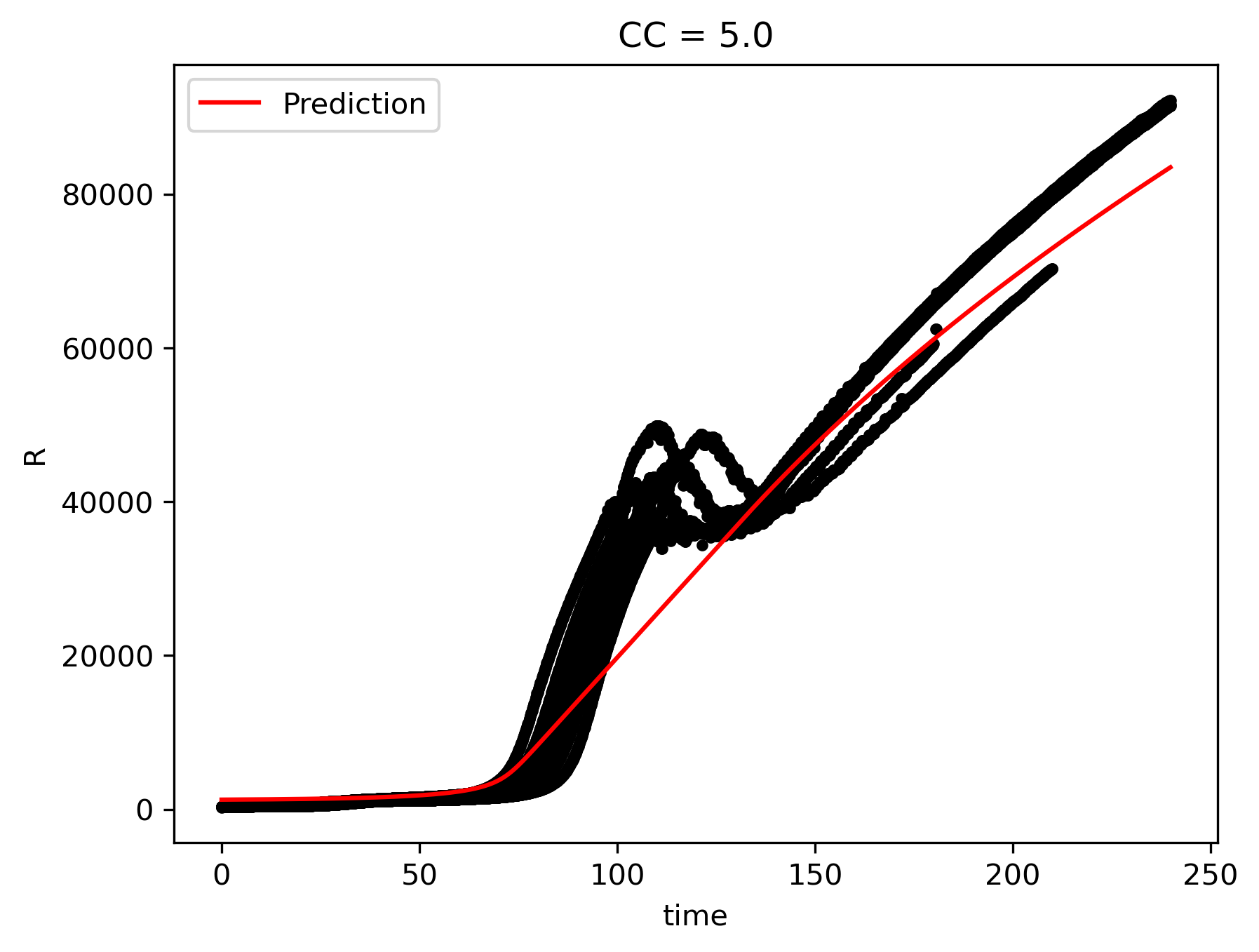}
        \caption{Constant current experiment prediction, $j_0=5.0mA$ (resistance)}
    \end{subfigure}
    \caption{ML-augmented model without first peak resistance predictions compared to experimental data.}
    \label{fig:app_resistance_no_first_peak}
\end{figure}

We present all prediction results on the ML-augmented model without the first peak model included. Removing the first peak model greatly improves model efficiency during prediction time (and training time). Predictions are still poor for VR experiments in the low $V_R$ regime, but improved over the baseline and inference-informed models. Again, thickness predictions are more accurate compared to previous models. Predictions for current, resistance, and thickness are included for all six experimental configurations in Figs.~\ref{fig:app_current_first_peak}-~\ref{fig:app_resistance_no_first_peak}, but the ML model is trained with current data for the 3 voltage ramp experiments only.

\bibliographystyle{elsarticle-num}
\bibliography{references}

\end{document}